\documentclass[useAMS,usenatbib]{mn2e}
\usepackage{graphicx} \usepackage{rotating}
\usepackage{txfonts}
\usepackage{url}

\def\nt/f{Nuclear Technology/Fusion}

\title[Carbon chemistry in GB PNe]
  {Carbon chemistry in Galactic Bulge Planetary Nebulae}
\author[L. Guzman-Ramirez et al.]
  {L. Guzman-Ramirez,$^1$\thanks{lizette.ramirez@postgrad.manchester.ac.uk}
  A.~A. Zijlstra,$^1$ R. N\'i Chuim\'in,$^1$ 
  \newauthor 
    K. Gesicki,$^2$
  E. Lagadec,$^3$
  T.~J. Millar$^4$
and Paul ~M.  Woods$^1$ \\
  $^1$Jodrell Bank Centre for Astrophysics, School of Physics \& Astronomy,
The University of Manchester, Manchester, M13
      9PL, UK \\
  $^2$Centrum Astronomii UMK, ul.Gagarina 11, 87-100  Torun,
      Poland\\ 
 $^3$European Southern Observatory, Karl-Schwarzschild-Str 2, 
       85748 Garching, Germany \\
 $^4$Astrophysics Research Centre, School of Mathematics and Physics,
      Queen's University Belfast, Belfast BT7 1NN, UK\\
}
\date{Released 2002 Xxxxx XX}

\pagerange{\pageref{firstpage}--\pageref{lastpage}} \pubyear{2002}

\def\LaTeX{L\kern-.36em\raise.3ex\hbox{a}\kern-.15em
    T\kern-.1667em\lower.7ex\hbox{E}\kern-.125emX}

\begin{document}

\date{Accepted 2010 November . Received 2010 November 04}

\pagerange{\pageref{firstpage}--\pageref{lastpage}} \pubyear{2010}

\maketitle

\label{firstpage}

\begin{abstract}
Galactic Bulge Planetary Nebulae show evidence of mixed chemistry with
emission from both silicate dust and PAHs. This mixed chemistry is
unlikely to be related to carbon dredge up, as third dredge-up is not expected
to occur in the low mass Bulge stars. We show that the phenomenon is
widespread, and is seen in 30 nebulae out of 40 of our sample, selected on
 the basis of their infrared flux. HST images and UVES spectra show that the mixed chemistry is
not related to the presence of emission-line stars, as it is in the Galactic
disk population. We also rule out interaction with the ISM as origin of the
PAHs. Instead, a strong correlation is found with morphology, and the presence
of a dense torus.  A chemical model is presented which shows that hydrocarbon
chains can form within oxygen-rich gas through gas-phase chemical reactions.
The model predicts two layers, one at $A_V\sim 1.5$ where small hydrocarbons
form from reactions with C$^+$, and one at $A_V\sim 4$, where larger chains
(and by implication, PAHs) form from reactions with neutral, atomic carbon.
These reactions take place in a mini-PDR.  We conclude that the mixed
chemistry phenomenon occurring in the Galactic Bulge Planetary Nebulae is best
explained through hydrocarbon chemistry in an UV-irradiated, dense torus.

\end{abstract}


\section{Introduction}

Planetary Nebulae (PNe) are remnants of the extreme mass loss
experienced by Asymptotic Giant Branch (AGB) stars, a phase
that occurs during the late stages of evolution of low and intermediate-mass stars,
between 1 and 8 M$_{\odot}$. The intense mass loss (from $10^{-6}$ to
$10^{-4}\,\rm M_{\odot}\, yr^{-1}$) leads to the formation of a
circumstellar envelope made of gas and dust (\citealp{kwok}). The dust
strongly emits in the infrared. After the termination of the AGB, the star
rapidly increases in temperature and photodissociates and photoionizes the
expanding ejecta.

During the AGB phase, a star may evolve from being oxygen-rich to being
carbon-rich, a change reflected in the chemical composition of the stellar
wind. The change occurs when carbon produced by He-burning is brought to
the surface by processes in the stellar interior thereby increasing the C/O
ratio until it exceeds unity and a carbon star is formed. This is the
so-called third dredge-up, which occurs at the end of flash burning in the
He-shell (\citealp{herwig}). The process depends on stellar mass: \cite{vassi}
showed the third dredge up occurs if the core mass of the star $M_{cs} >
1.5$\,M$_{\odot}$.

This predicts a clear distinction, with some (lower-mass) stars
showing oxygen-rich and some (higher-mass) stars carbon-rich
ejecta. In the molecular ejecta, the CO molecule locks away the less
abundant element, leaving the remaining free O or C to drive the
chemistry and dust formation.  Oxygen-rich shells are
characterized by silicate dust, while carbon-rich shells show
Polycyclic Aromatic Hydrocarbon (PAH) emission bands and carbonaceous
dust. This dichotomy is indeed by and large observed.

The first evidence for mixed-chemistry in post-AGB objects was found when one PN
(IRAS07027-7934) with strong PAH emission bands at 6.2, 7.7, 8.6, and
11.3 $\mu$m, was found to also show a 1.6GHz OH maser line
(\citealp{zijlstra91}).  Observations made with ISO uncovered several
further cases, where PAH emission in PNe occurred together with
emission bands of silicates at 23.5, 27.5, 33.8, and 10 $\mu$m,
usually found in O-rich shells (Waters et al. 1998a,b; Cohen et al.,
1999, 2002)\nocite{cohen99,cohen02}.

The mixed-chemistry phenomenon has also been seen in the Red Rectangle
(\citealp{waters}) and in PNe with late Wolf-Rayet type central stars
(\citealp{waters98}), where it is attributed to the presence of old
oxygen-rich material in a circumstellar disk, with the PAHs forming in
a more recent carbon-rich outflow. Weak PAH emission has however been
seen in some clearly  oxygen-rich objects, such as NGC 6302 and
Roberts 22 (\citealp{zijlstra01}). PAH bands have been detected in
S-type AGB stars, with C/O ratios close to, but below, unity
(\citealp{smolders}). \cite{cohenBarlow05} have shown that some PAH
emission can occur  at C/O ratios as low as 0.6.

\cite{perea} showed that the mixed-chemistry phenomenon is widespread among
the PNe in the Galactic Bulge (GB). Their Spitzer observations show that the
simultaneous presence of oxygen and carbon-rich dust features is common, and
is not restricted to objects with late/cool [WC] type stars. The traditional
explanation relating the mixed chemistry to a recent evolution towards
carbon-star is unlikely for the Bulge objects, as these old, low-mass stars
are not expected to show third dredge-up, and therefore should not show
enhanced C/O ratios. The few AGB carbon stars in the Bulge do not originate
from third dredge-up \citep{azzo}. Different explanations are needed.

In this paper, we investigate two alternative scenarios for the presence of
C-rich molecules: PAHs swept-up from the interstellar medium (ISM), and
in-situ PAH formation via chemical pathways in oxygen-rich environments.  This
is performed through the analysis of 40 PNe towards the Galactic
Bulge for which Spitzer spectroscopy is available. HST images are available for 22 of
these.  In Section 2 we present the observations. In Section 3 we present a
correlation between the silicate emission and the presence of a torus in the
PNe from HST images. In Section 4 we present a chemical model that forms long
C-chains in an O-rich environment and in Section 5 we discuss these results and
present our main conclusions.

\section{Observations}

\cite{perea} uncovered the mixed chemistry in 21 Galactic Bulge PNe, out of a
total sample of 26, using Spitzer Space Telescope spectra.  We retrieved and
re-reduced these 21 Spitzer spectra. We also included the PN H2-20 from \cite{guten},
and a further 18 PNe in the GB for which we found spectroscopic data in the
Spitzer archive, yielding a total sample of 40.  

All the spectra were taken with the Infrared Spectrograph
(IRS) (\citealp{houck}). The spectra cover a range between 5.2-37.2 $\mu$m by
using the Short-Low (SL: 5.2-14.5 $\mu$m; 64$<$R$<$128), Short-High (SH:
9.9-19.6 $\mu$m; R$\sim$600) and Long-High (LH: 18.7-37.2 $\mu$m; R$\sim$600)
modules.  The data were retrieved from the Spitzer Science Center Data Archive
using Leopard.  The Spitzer IRS Custom Extractor (SPICE) was used to do the
extraction of the spectra for each nod position from the 2D images.  All
spectra were cleaned for bad data points, spurious jumps and glitches, and
then they were combined and merged to create a single 5--37$\mu$m spectrum.
We did not correct for flux offsets between different wavelengths ranges but
these are minor. Offsets can be caused by the fact that the different
wavelength ranges use different slits, and our objects have sizes similar to
or larger than the slit widths.

Twenty-two of the objects were included in an HST SNAP-shot survey of
37 Galactic Bulge PNe (proposal 9356, PI Zijlstra). The WFPC2 images were
taken in three different filters, H$\alpha$, V, and [OIII], with
typical exposure times of $2\times 100$ sec, 60 sec and $2 \times 80$ sec, respectively.  The objects were placed on the CCD of the
Planetary Camera (PC) with a pixel scale of $0\farcs0455$.
Pipe-line reduced images were retrieved. Only the  H$\alpha$ images
are used here.

The target selection for the original HST observation was based on a
catalogued diameter less than 5$^{\prime\prime}$. It included a few objects without a
known size. Other than this, the target selection was random among the known
PNe in the direction of the Bulge. The fact that so many of the Spitzer sample
were included in the HST observations indicate that the Spitzer target
selection indirectly also made use of the diameter. The Spitzer samples were
mostly selected based on 12$\mu$m flux. The overlap in selection reflects 
the brighter infrared emission from compact nebulae.  

VLT UVES echelle spectra were obtained for these 22 objects, using
an exposure time of 600 sec. The spectra cover the wavelength range
from 3300 to 6600\AA, with a resolution of 80,000. These spectra were
used to investigate the nature of the central stars, based on the
presence of stellar emission lines.

\section{Results}

\subsection{External causes: ISM interaction}

As argued above, PAHs could be present in an oxygen-rich environment
if they are acquired from the ISM. This can be done by gathering interstellar PAHs through the
``wall'' where the expanding nebula collides with the ISM, creating a
stationary shock (\citealp{chris}), in which the interaction region is
dominated by ISM material. As the Bulge PNe have a high velocity
dispersion with respect to the ISM, the ``wall'' is expected to be close
to the nebula, and one-sided.

To test this model, we checked the spatial location of the PAH emission bands,
at 6.2 and 11.3$\mu$m, and compared these to that of the 12.8$\mu$m [Ne\,II]
line. The 12.8$\mu$m [Ne\,II] emission line comes from the ionized region, so the position of the line traces the position of the PN and is not affected by extinction. On the other hand, if the PAHs have been swept up from the ISM, then their physical position would not be at the same position as the PN (or the [Ne\,II] for this case) so there should be an offset in their positions. This was done for 40 objects in the sample. The peak of the [Ne\,II] line
and the 11.3$\mu$m PAH feature were measured using the SH part of the
spectrum.

 In a few cases we found evidence for an offset. In PN Hb\,4, the faint
 11.3$\mu$m feature is offset by about 6$^{\prime\prime}$ from the [Ne II] 12.80$\mu$m
 emission. In M2-14, the faint 11.3$\mu$m emission is offset by 4$^{\prime\prime}$ from
 the [Ne II] line. We also found an offset of 2$^{\prime\prime}$ between the [NeII] line
 and the PAH bands at 11.3$\mu$m in the case of H1-16, H1-50, H2-20, M1-40 and
 Cn1-5. These smaller offsets may not be real, as the resolution at the
 position of the peak of the line and the feature is about 2$^{\prime\prime}$. Hb\,4
 provides the strongest case for a spatial offset. However, as shown below,
it has at best weak evidence for mixed chemistry.

Thus, there are only a few potential cases where  interaction with the
ISM, or other forms of ISM confusion, may contribute to the PAH
features.  In the large majority of cases, the dust and PAH
features are co-located and centered on the nebula. In these cases
the mixed chemistry is likely to have an internal origin, related solely to
the star and its ejecta.
 
\subsection{Internal causes: star and nebula}

\begin{table*}
\caption{Galactic Bulge PNe. Column 3 and 4 list the identified bands in the
  IRS spectra. Col. 5 lists the peak line flux of the 7.7$\mu$m PAH
  band. Col. 6 indicate whether the star has emission lines [WC] or is of weak emission-line (wels)
  type. The last column indicates whether an HST image exists. A star in
  Col. 3 indicates that the PAH bands are very faint and may be in some
  doubt. Mixed-chemistry objects are those unstarred objects which do not have
  ``none'' in col. 4 (30 in total). }
\label{GBPNe}
\centering
\begin{tabular}{lllllll} 
\hline\hline PN G Name & Name &  PAHs($\mu$m)  & Silicates ($\mu$m) & 7.7$\mu$m (mJy) & Central star & HST image \\
 \hline
008.3-01.1 & M 1-40    &  6.2, 7.7, 11.3             & 27.5, 33.8       & 400
& wels  & no  \\ 
002.2-09.4 & Cn 1-5    &  6.2, 7.7, 11.3             & 27.5, 33.8       & 280
& [WC4] & no  \\  
356.8+03.3 & Th 3-12   &  6.2, 7.7, 8.6, 11.3        & 33.8             & 180 &       & yes \\
006.5-03.1 & H 1-61    &  6.2, 7.7, 11.3             & 27.5, 33.8       & 150 & wels  & no  \\ 
006.4+02.0 & M 1-31    &  6.2, 7.7, 11.3             & 10, 27.5, 33.8   & 140 & wels  & yes \\
357.1-04.7 & H 1-43    &  6.2, 7.7, 8.6, 11.3        & 10, 27.5, 33.8   & 130 & [WC11]& yes \\
007.2+01.8 & Hb 6      &  6.2, 7.7, 11.3             & 27.5, 33.8       & 120
&       & no  \\ 
358.7+05.2 & M 3-40    &  6.2, 7.7, 8.6, 11.3        & 27.5, 33.8       & 111 &       & yes \\
356.5-02.3 & M 1-27    &  6.2, 7.7, 11.3             & 27.5             & 100 & [WC11]& no  \\
004.9+04.9 & M 1-25    &  6.2, 7.7, 11.3             & 27.5, 33.8       &  90 & [WC6] & no  \\
359.3-01.8 & M 3-44    &  6.2, 7.7, 11.3             & 27.5, 33.8       &  80 & [WC11]& no  \\
358.9+03.4 & H 1-19    &  6.2, 7.7, 8.6, 11.3        & 27.5, 33.8       &  80 &       & yes \\
000.1+04.3 & H 1-16    &  6.2, 7.7, 11.3             & 27.5, 33.8       &  80 &       & no  \\
352.6+03.0 & H 1-8     &  6.2, 7.7, 8.6, 11.3        & 27.5, 33.8       &  70 &       & yes \\
003.1+03.4 & Hen 2-263 &  6.2, 7.7, 8.6, 11.3        & 23.5, 27.5, 33.8 &  60 &       & yes \\
354.5+03.3 & Th 3-4    &  6.2, 7.7, 8.6, 11.3        & 10, 27.5, 33.8   &  60 &       & yes \\
359.9-04.5 & M 2-27    &  6.2, 7.7, 11.3             & 27.5, 33.8       &  60 & wels  & no  \\
000.0-06.8 & Hen 2-367 &  6.2, 7.7, 11.3             & 27.5,33.8        &  60 &       & no  \\
351.9-01.9 & K 5-4     &  6.2, 7.7, 8.6, 11.3        & none             &  50 &       & yes \\
356.9+04.4 & M 3-38    &  6.2, 7.7, 11.3             & 27.5, 33.8       &  50 &       & yes \\
002.8-01.7 & H 2-20    &  6.2, 7.7, 11.3             & 27.5, 33.8       &  40 &       & yes \\ 
006.8+04.1 & M 3-15    &  6.2, 7.7, 11.3             & 27.5, 33.8       &  20 & [WC5] & yes \\
007.5+04.3 & Th 4-1    &  7.7, faint 11.3            & 10, 27.5, 33.8   &  20 &       & yes \\
358.5-04.2 & H 1-46    &  7.7, 11.3                  & 10, 27.5, 33.8   &  18 &       & yes \\
004.0-03.0 & M 2-29    &  7.7, 8.6, 11.3             & none             &  10 &       & no  \\
355.9+03.6 & H 1-9     &  7.7, 8.6                   & none             &  10 &       & no  \\
357.2+02.0 & H 2-13    &  faint 7.7, 11.3            & 33.8             &   8 &       & yes \\
001.2+02.1 & Hen 2-262 &  faint 6.2, 7.7, 8.6, 11.3  & 33.8             &   8 &       & no  \\               
004.1-03.8 & KFL 11    &  faint 7.7, 8.6, 11.3       & 33.8             &   5 &       & yes \\
003.1+02.9 & Hb 4      &  faint 7.7 and 11.3  (*)        & 27.5, 33.8       &   5 & [WC4] & no  \\
002.9-03.9 & H 2-39    &  faint 7.7  (*)                 & 23.5, 27.5, 33.8 &   1 &       & yes \\
006.3+04.4 & H 2-18    &  11.3                       & 33.8             &   0 &       & yes \\
003.6+03.1 & M 2-14    &  faint 11.3 (*)             & 27.5, 33.8       &   0 & wels  & yes \\
356.5-03.6 & H 2-27    &  faint 11.3 (*)                 & none             &   0 & [WC11]& yes \\
358.5+02.9 & Al 2-F    &  faint 11.3 (*)                 & 33.8             &   0 &       & yes \\
008.6-02.6 & MaC 1-11  &  faint 8.6, 11.3            & 33.8             &   0 &       & yes \\
006.1+08.3 & M 1-20    &  6.2, 8.6, 11.3             & none             &   0 & [WC8] & no  \\
005.9-02.6 & MaC 1-10  &  11.3                       & 33.8             &   0 &       & no  \\ 
008.2+06.8 & Hen 2-260 &  8.6                        & none             &   0 &       & no  \\
359.7-02.6 & H 1-40    &  6.2, 11.3                  & 10, 27.5, 33.8   &   0 &       & no  \\
 \hline
  \end{tabular}
\end{table*}

\subsubsection{Spitzer spectra}

Table \ref{GBPNe} lists the 40 PNe we analyzed. We identified both PAH and
silicate bands, as listed in the Table. PAHs bands were seen in all objects,
except for 5 objects where only very faint 11.3$\mu$m/7.7$\mu$m bands and no other
features were seen. Silicate emission was seen in 34
PNe. We found evidence for mixed chemistry (both PAH and silicate emission) in
30 PNe,  where we exclude those with dubious PAH bands.

Table \ref{GBPNe} is in order of PAH strength, where the PAH strength is defined as
the peak line flux of the 7.7$\mu$m PAH band.  From H2-18 onward, no 7.7$\mu$m
emission feature is detected. The spectra are shown in the left-most panels of
Figs. \ref{correl1}, \ref{correl2}, \ref{correl3} and \ref{correl4}. We only
show the short wavelength region (SL and HL) to cover the 10$\mu$m silicate
feature and the PAH bands. The vertical red lines indicate the wavelength of
PAH bands: 6.2, 7.7, 8.6 and 11.3$\mu$m . Offsets between the different
spectral segments are visible in three cases.

Th 3-4 (PN G354.5+03.3), H 1-46 (PN G358.5-04.2), and Th 4-1 (PN G 007.5+04.3)
show a strong broad band at 10$\mu$m, interpreted as silicate emission. In
all three cases, the peak is shifted to longer wavelengths, beyond 10$\mu$m.
In the classification scheme of Sloan \&\ Price (1998)\nocite{sloan}, they are
of type SE4/5. The 13$\mu$m feature typically associated with the reddest
silicates (SE1-4) in AGB stars, is not present in our objects.  \cite{perea}
show three more objects with silicate emission without PAH features -- these also
have reddish silicates, and are of type SE5-6, without a 13$\mu$m feature. The
reddening of the silicate emission may be attributed to large dust grains.

\subsubsection{Emission-line stars}

The right-most panels of these figures show the UVES spectra located around the
strong stellar emission C IV-5800/5811\AA\ located. This line is an
indicator for [WC]-type stars. The y-axis scale is the same for all panels, except
for two objects with strong emission lines. In the case of H1-43 (PN
G357.1-04.7) we show the stellar broad line C III at 5696\AA\ together with
the nebular narrow line of [NII] at 5755\AA. The UVES spectra confirm the
classification of \cite{gorny04} but do not reveal any further emission-line
stars.

Table \ref{GBPNe} shows that the mixed chemistry phenomenon is not limited to
emission-line objects (cf. \citealp{gorny}). In our sample of 30 PNe, 10 have
emission-line stars and 20 do not. In the rejected sample of 10 PNe, 4 have
emission-line stars and 6 do not. This confirms that in the Bulge, there is no
clear relation between mixed chemistry and class of central star. However,
among the mixed chemistry objects, there is a strong tendency for the emission
line stars to have the strongest 7.7$\mu$m band. Thus, although the mixed
chemistry is itself independent of the stellar class, it is amplified by
emission-line stars.

\subsubsection{Morphology}

We use HST images of 22 of our objects to study their
morphology. The H$\alpha$ HST images are shown in the middle panels of
Figs. \ref{correl1}, \ref{correl2}, \ref{correl3} and \ref{correl4}, again in
the order listed above (of decreasing PAH strength).  The arrows indicate the
orientation of the image and the bar indicates the spatial scale.  The images
are displayed using an ``asinh'' scale to enhance the contrast and to better
see the faintest structures. (Absolute intensities are not used in this
paper.) Note that the images of different objects are not to the same spatial
or intensity scale.

A few further images are available in the literature. M1-40 has a diameter of
20$^{\prime\prime}$ \citep{schwarz} and is therefore likely a foreground nebula. Cn1-5 has a
similar-sized halo \citep{corradi03} and also is probably not in the Bulge. M1-25 was imaged
by \citet{aaquist} who find a torus with diameter 1.7$^{\prime\prime}$. Hb 6 and Hb 4 are also not
expected to be in the Bulge as their radio fluxes are too high \citep{zijlstra89,aaquist}.

\begin{figure*}
     \hbox{\includegraphics[width=50mm]{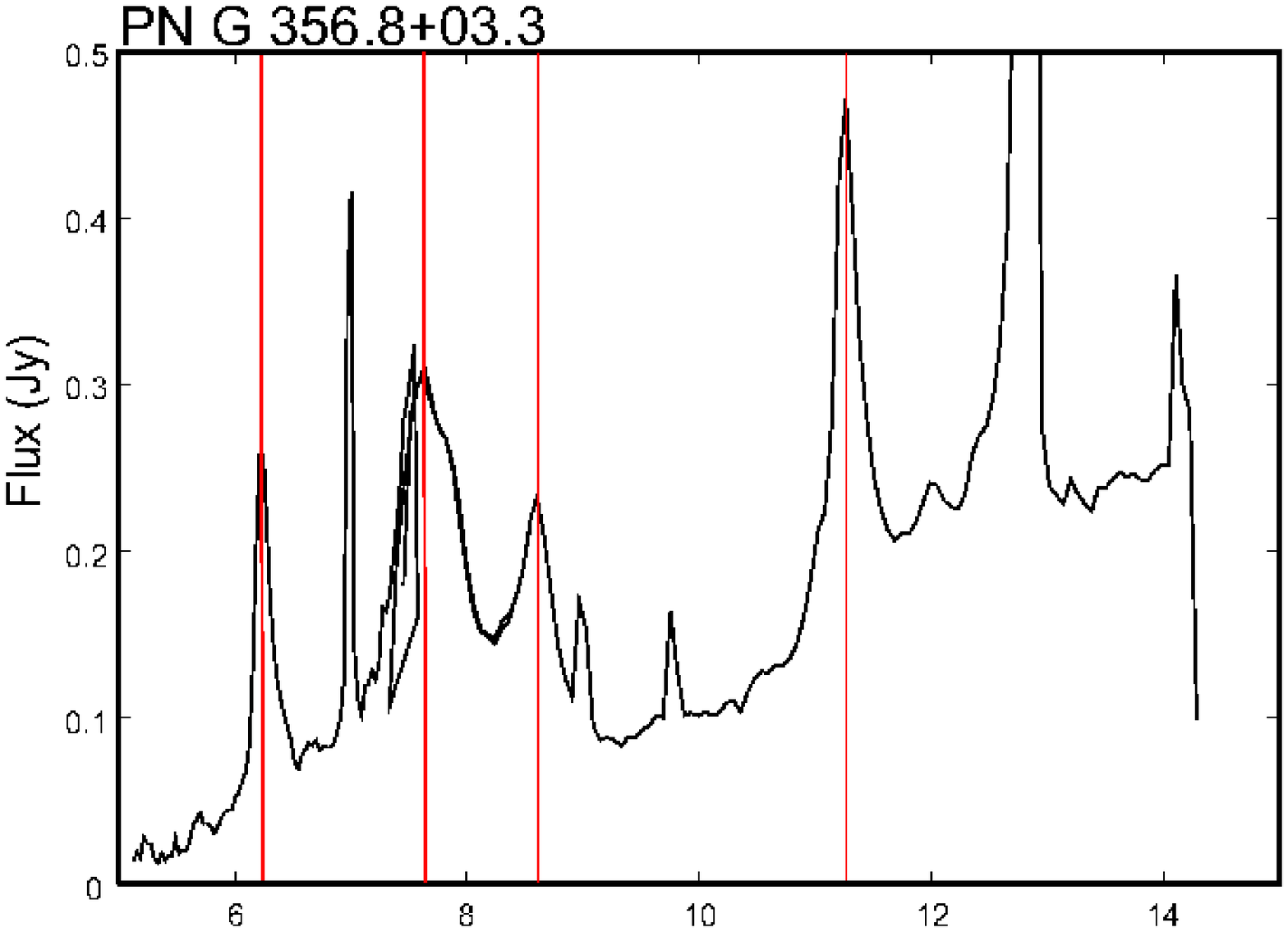}
          \hspace{3mm}
          \includegraphics[width=40mm]{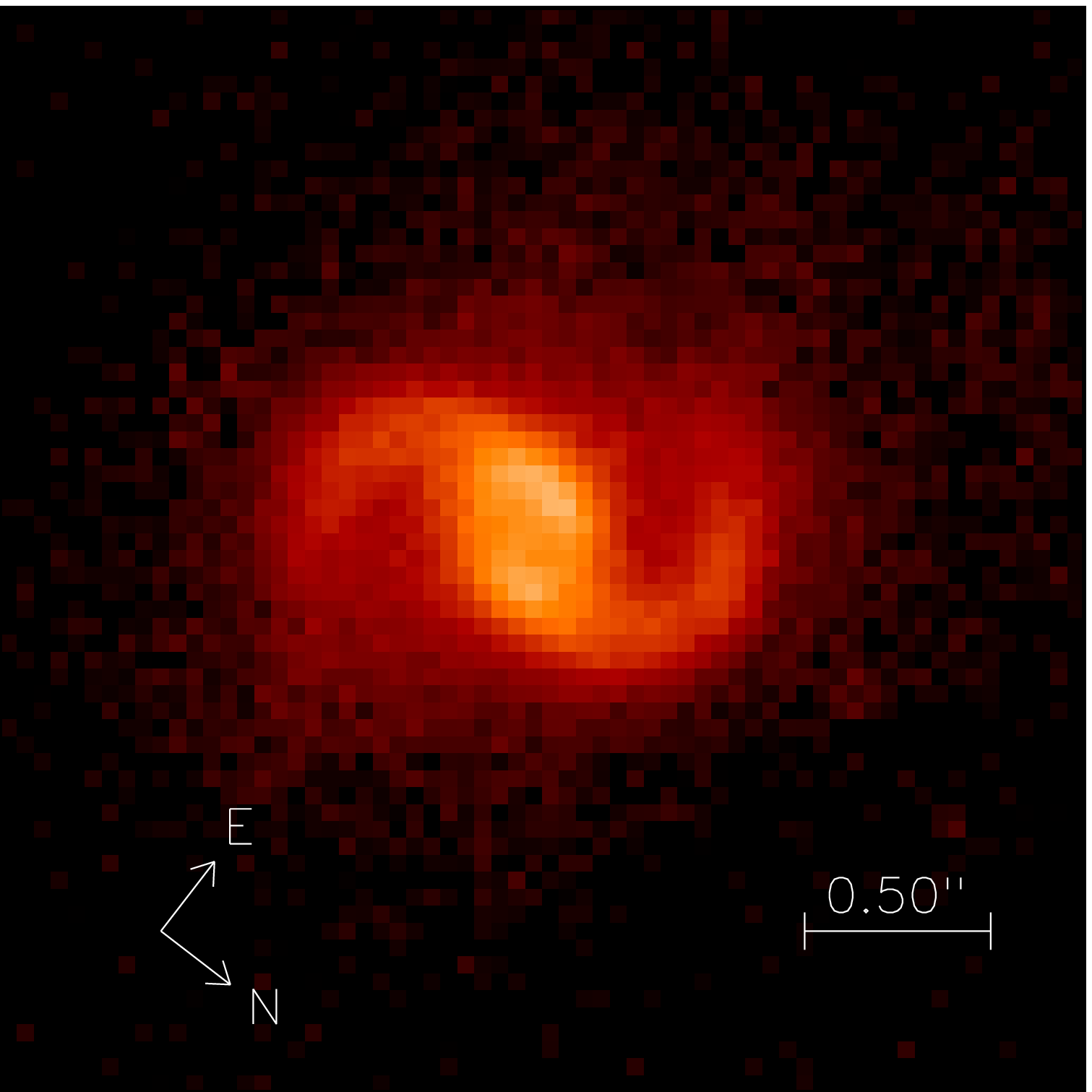}
          \hspace{6mm}
          \includegraphics[width=55mm]{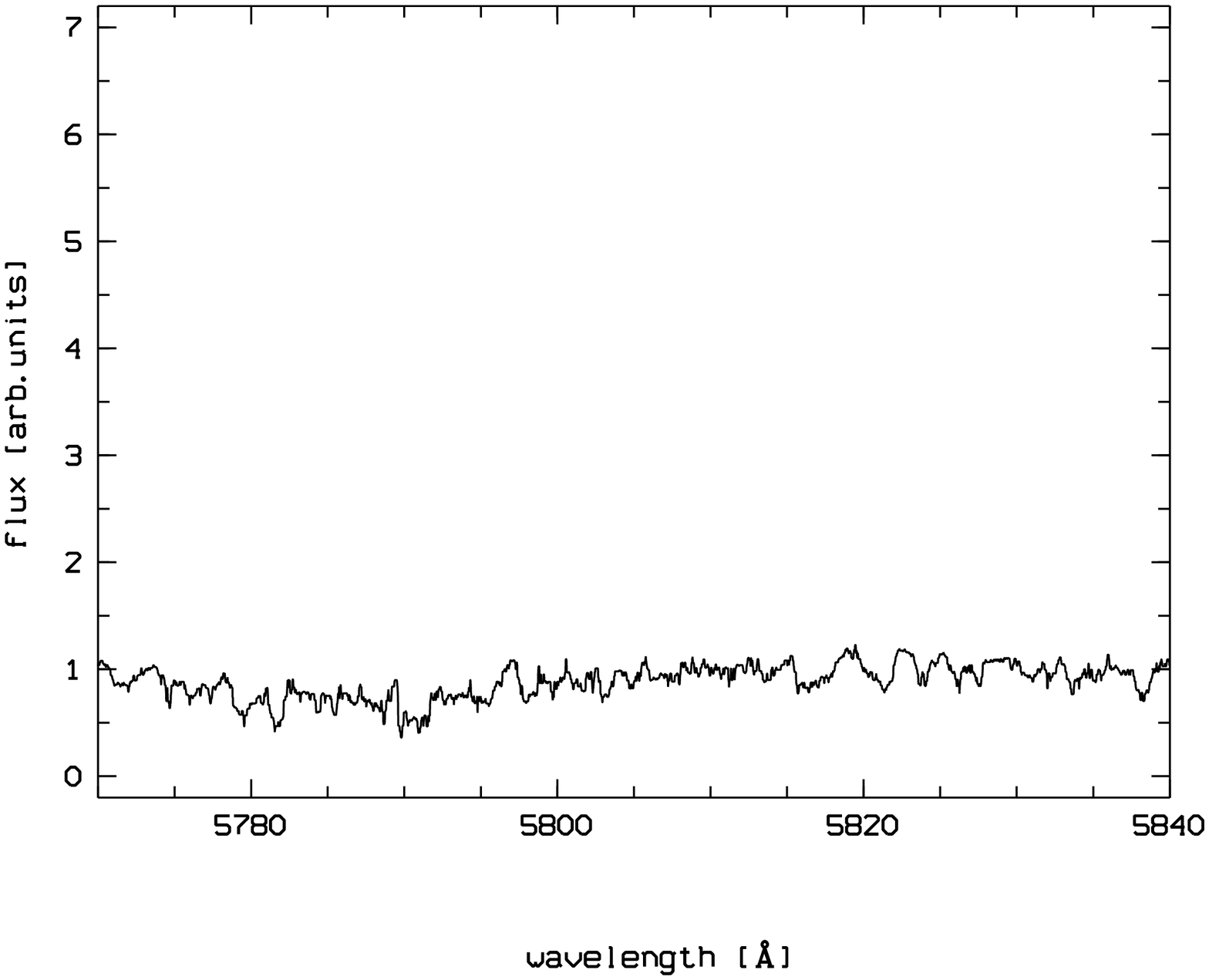}} 
        \vspace{1mm}
      \hbox{\includegraphics[width=50mm]{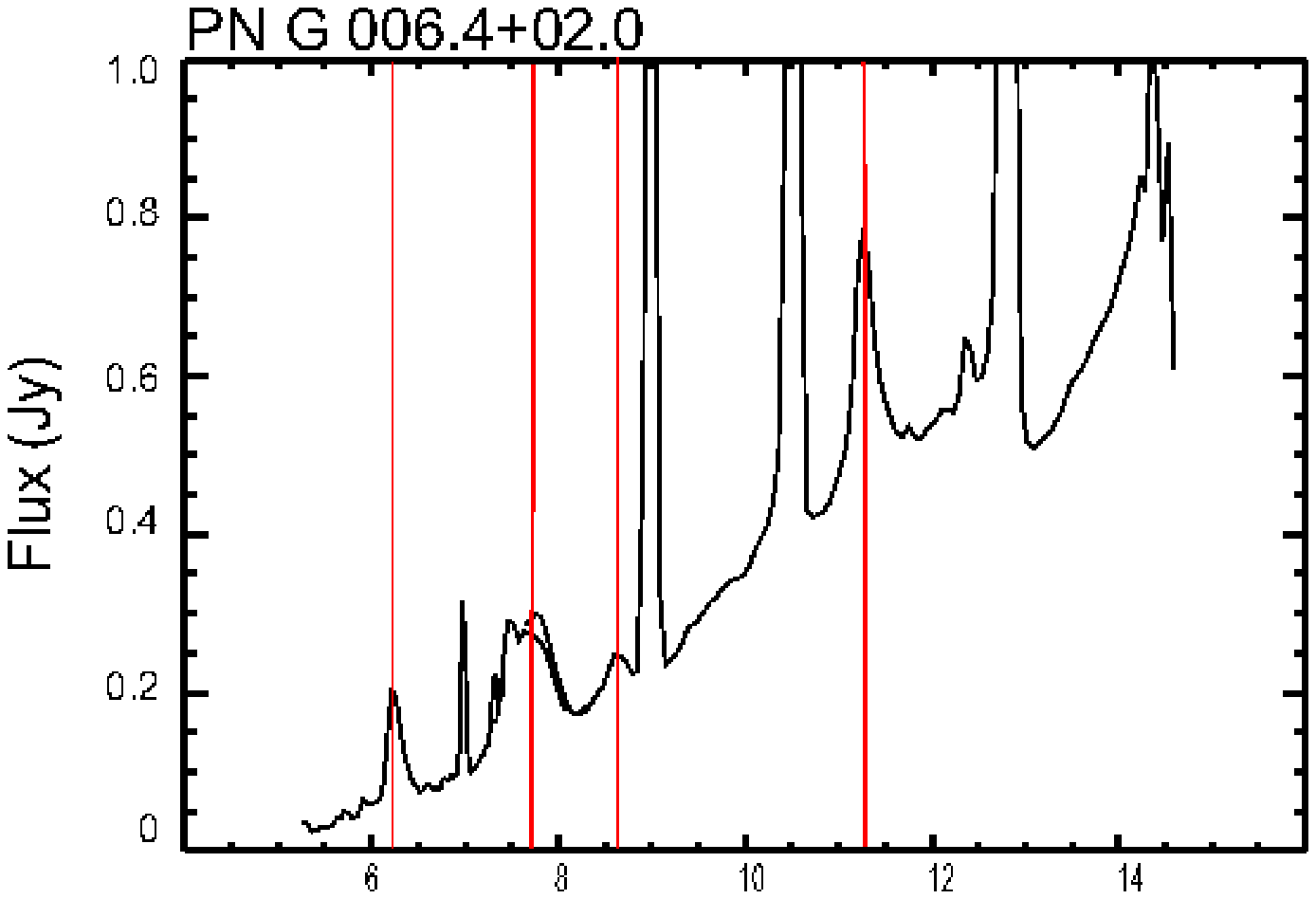}
          \hspace{3mm}
           \includegraphics[width=40mm]{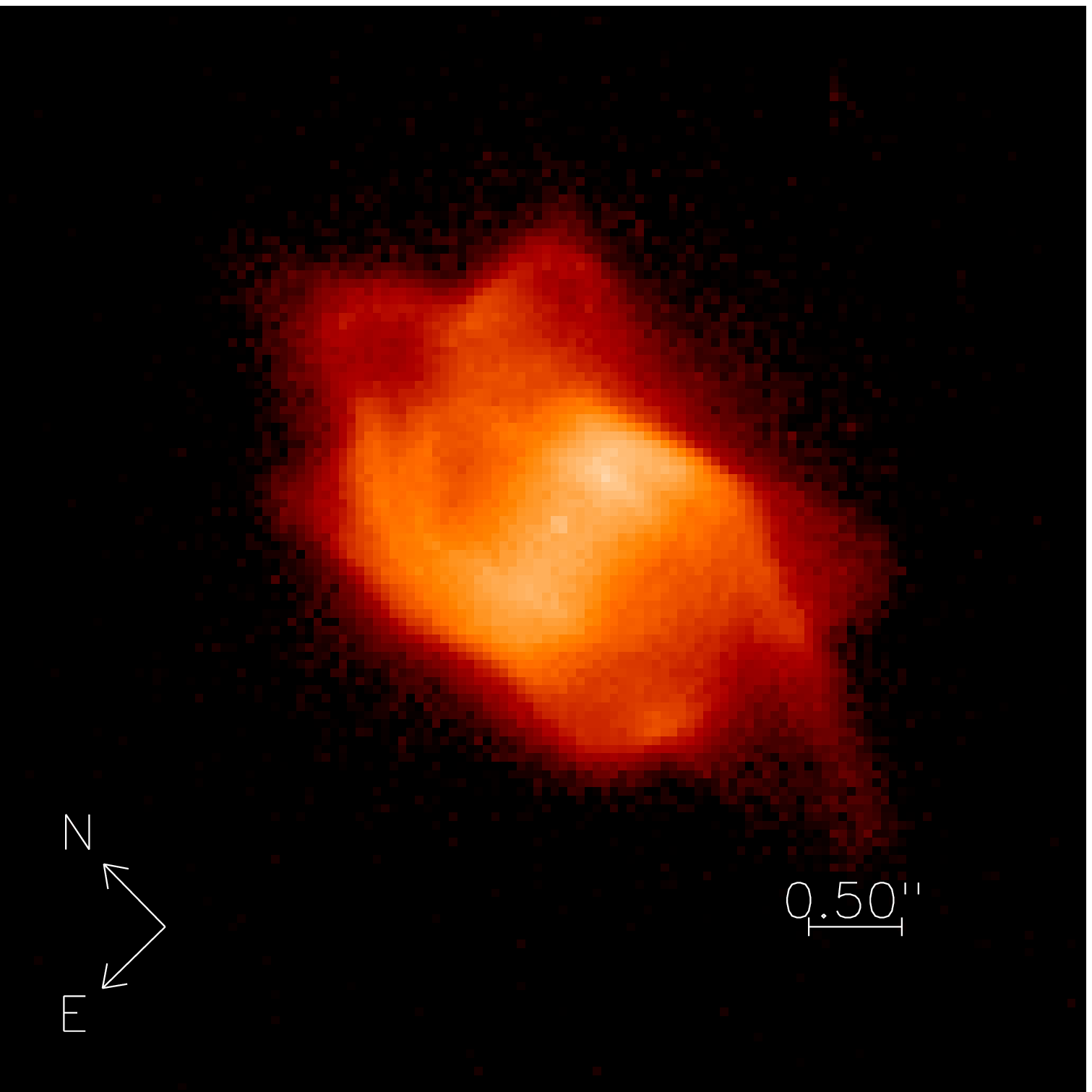}
          \hspace{3mm}
          \includegraphics[width=55mm]{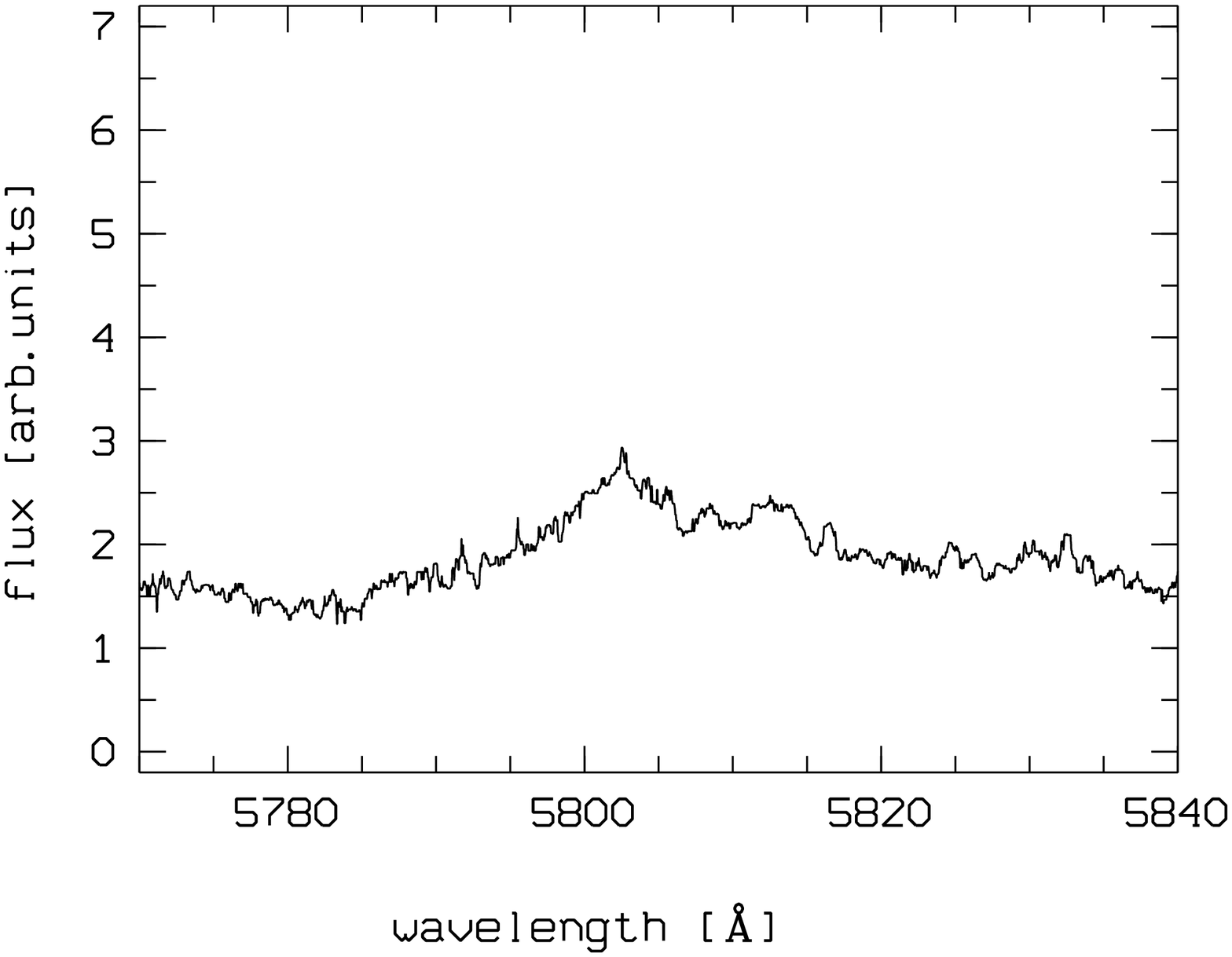}}
      \vspace{1mm}
     \hbox{\includegraphics[width=50mm]{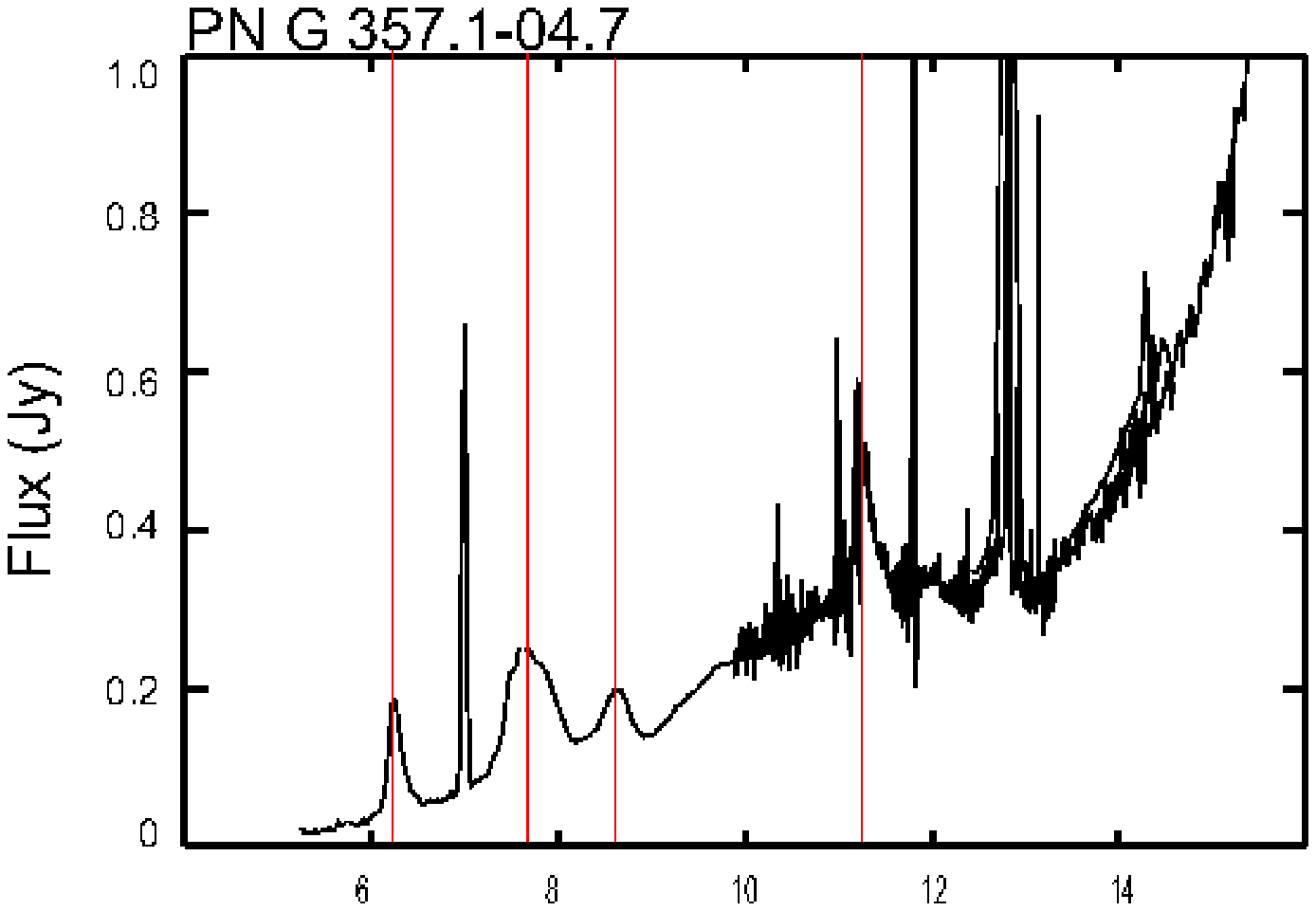}
          \hspace{3mm}
          \includegraphics[width=40mm]{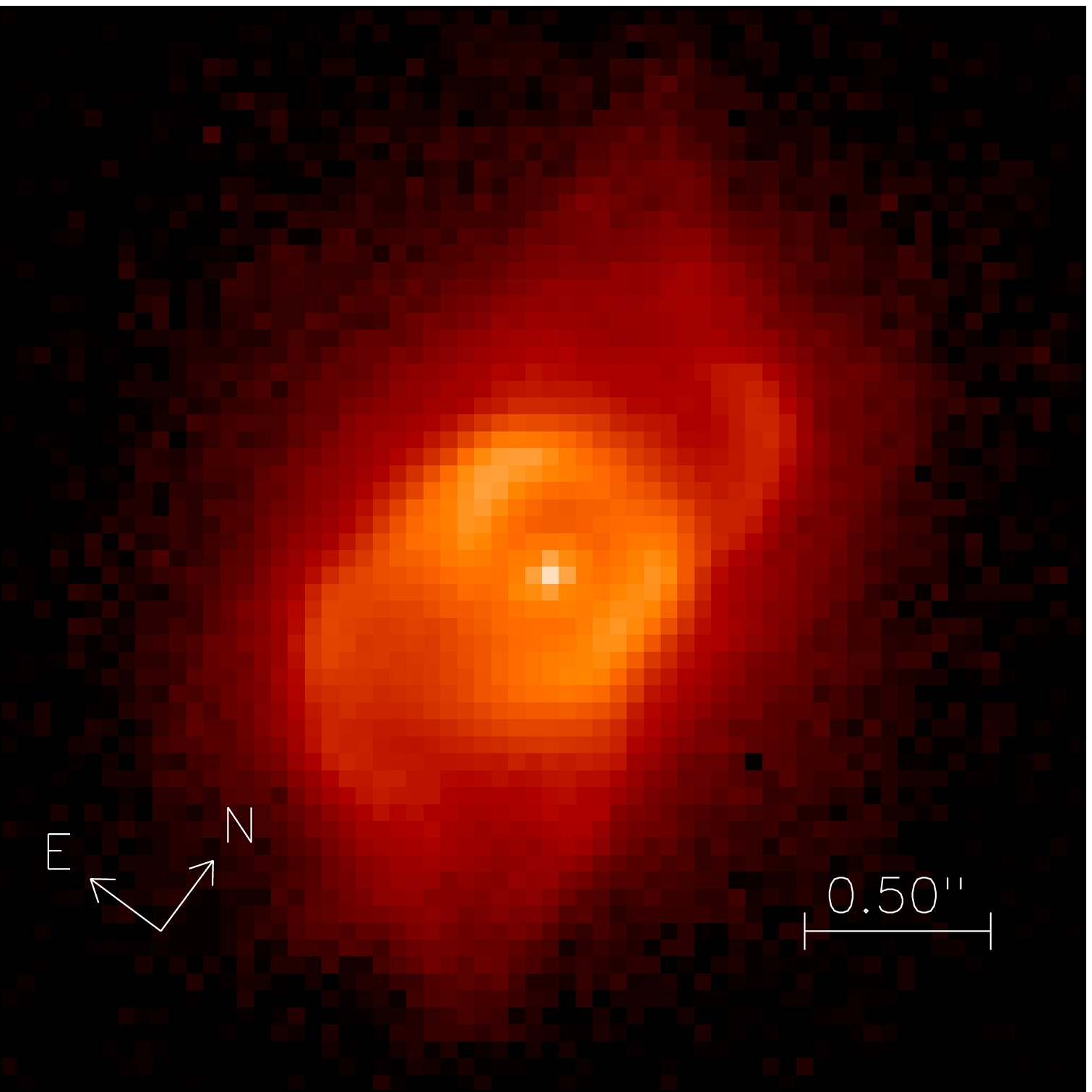}
          \hspace{3mm}
          \includegraphics[width=55mm]{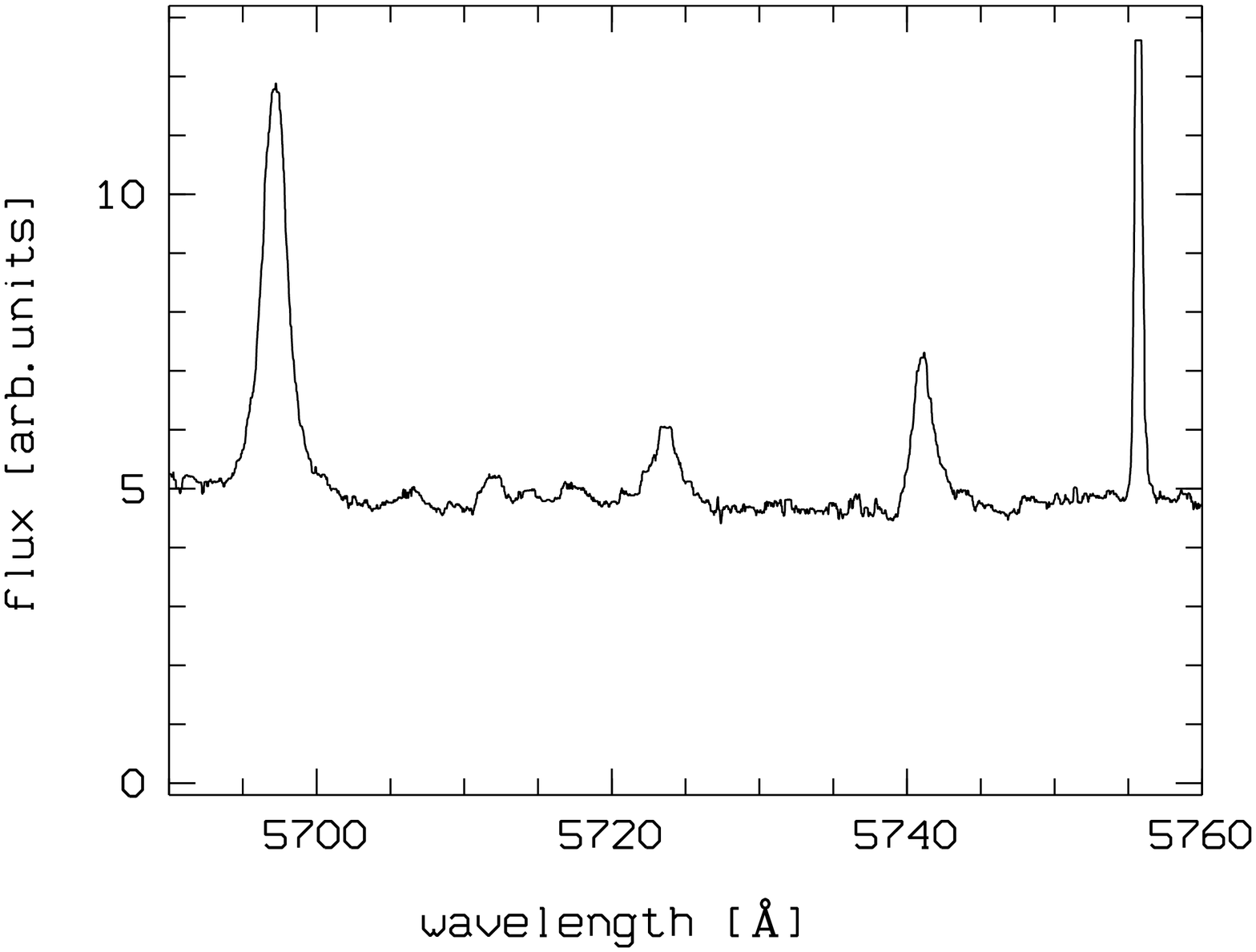}}
        \vspace{1mm}
     \hbox{\includegraphics[width=50mm]{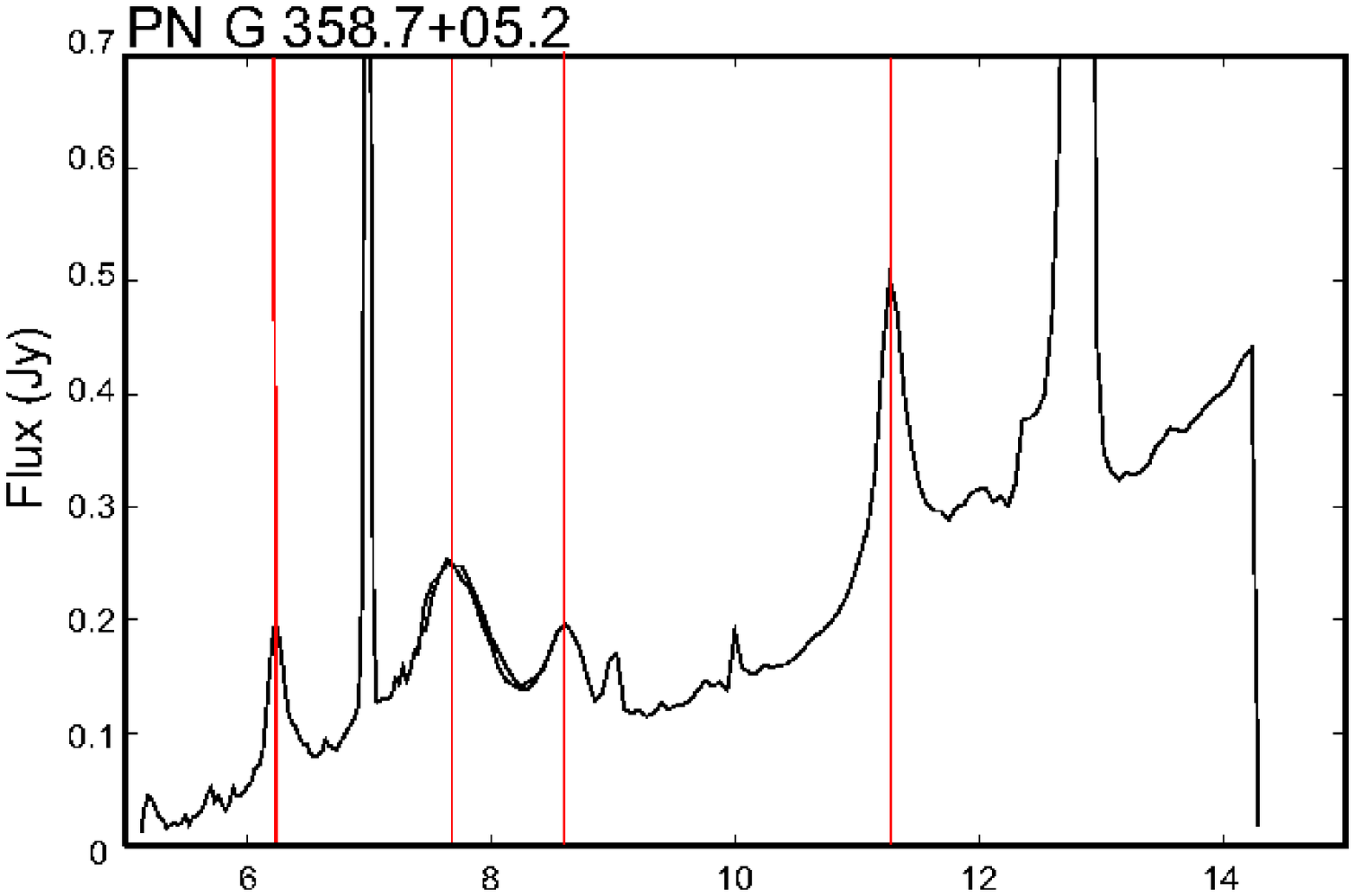}
          \hspace{3mm}
          \includegraphics[width=40mm]{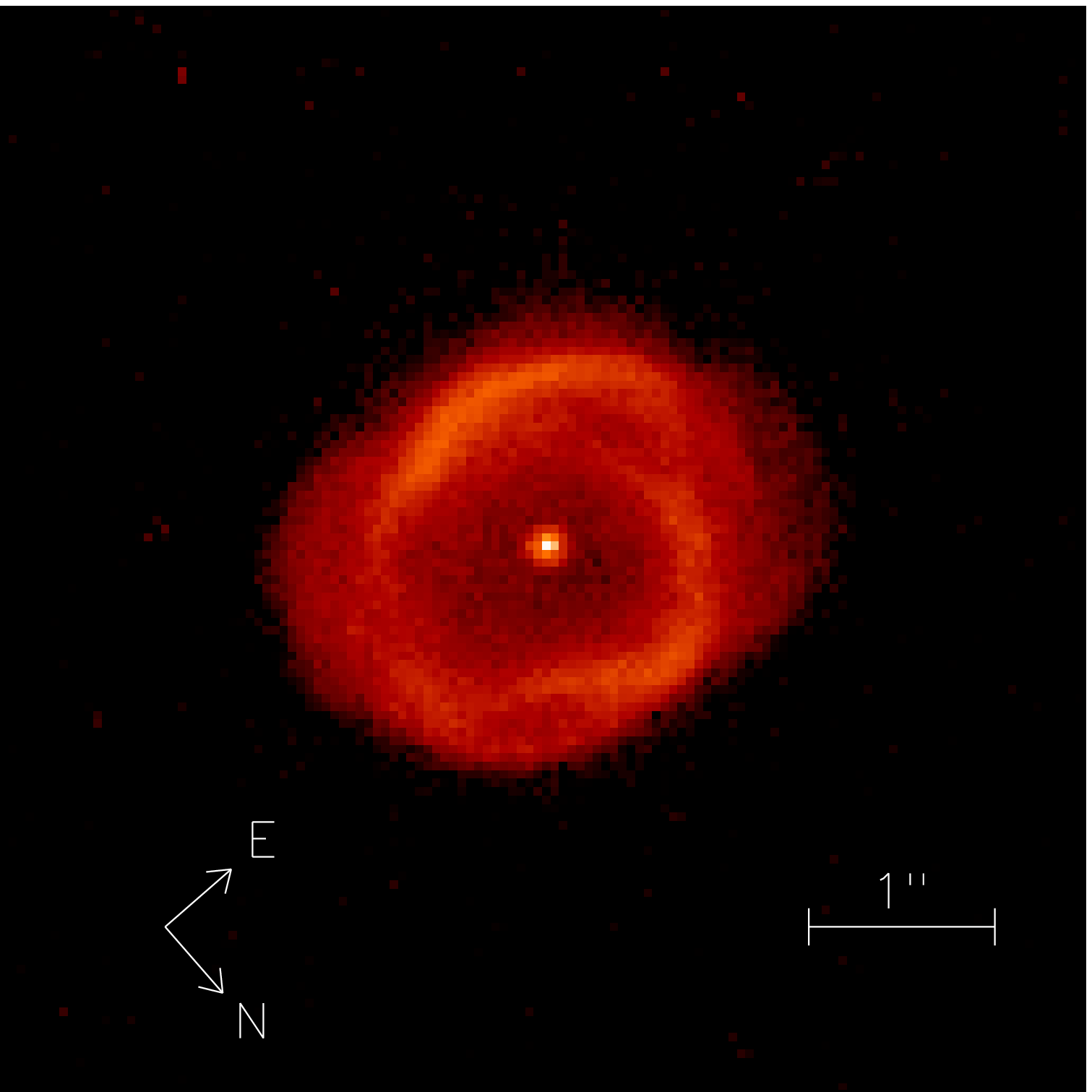}
          \hspace{6mm}
          \includegraphics[width=55mm]{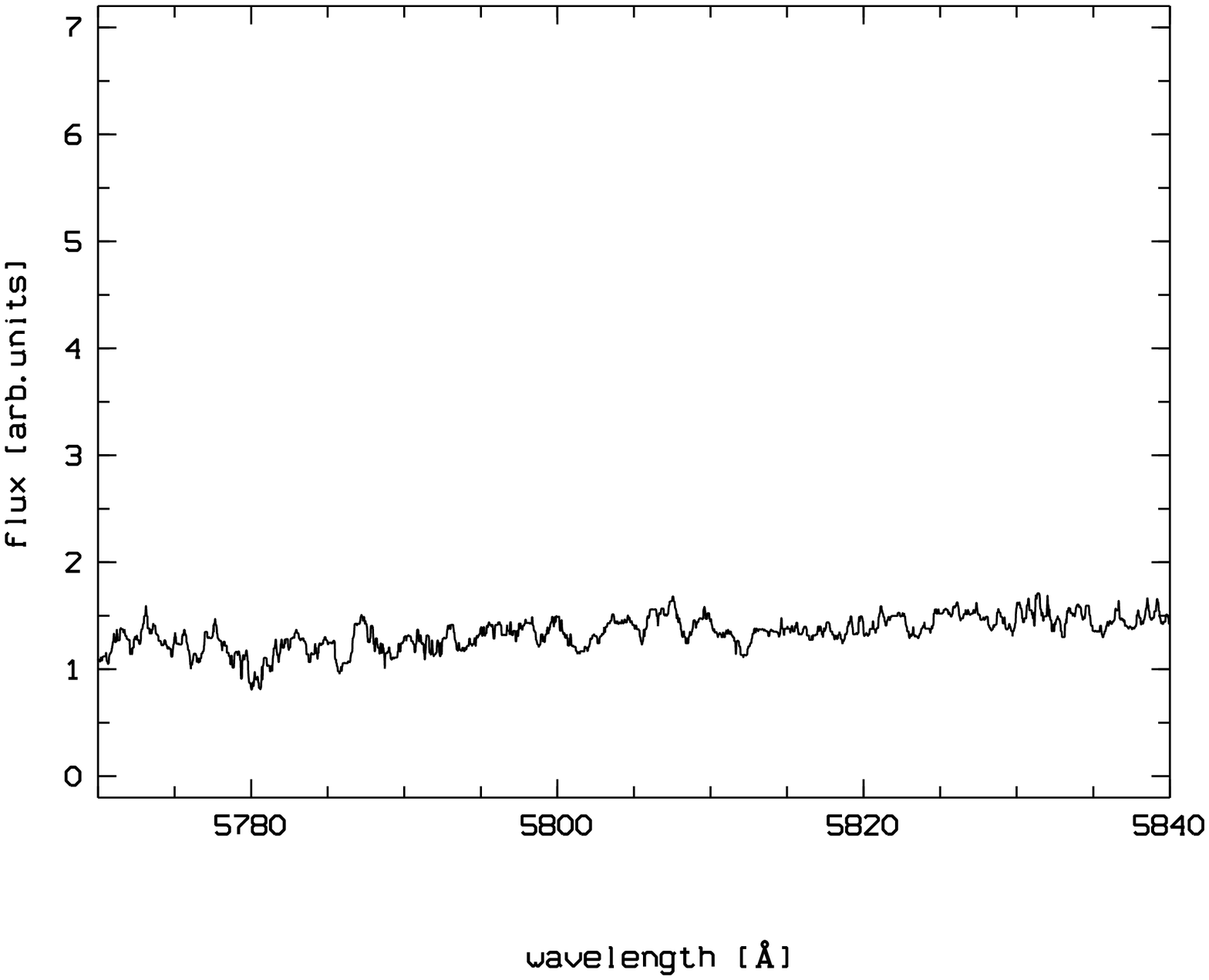}}
       \vspace{1mm}
        \hbox{\includegraphics[width=50mm]{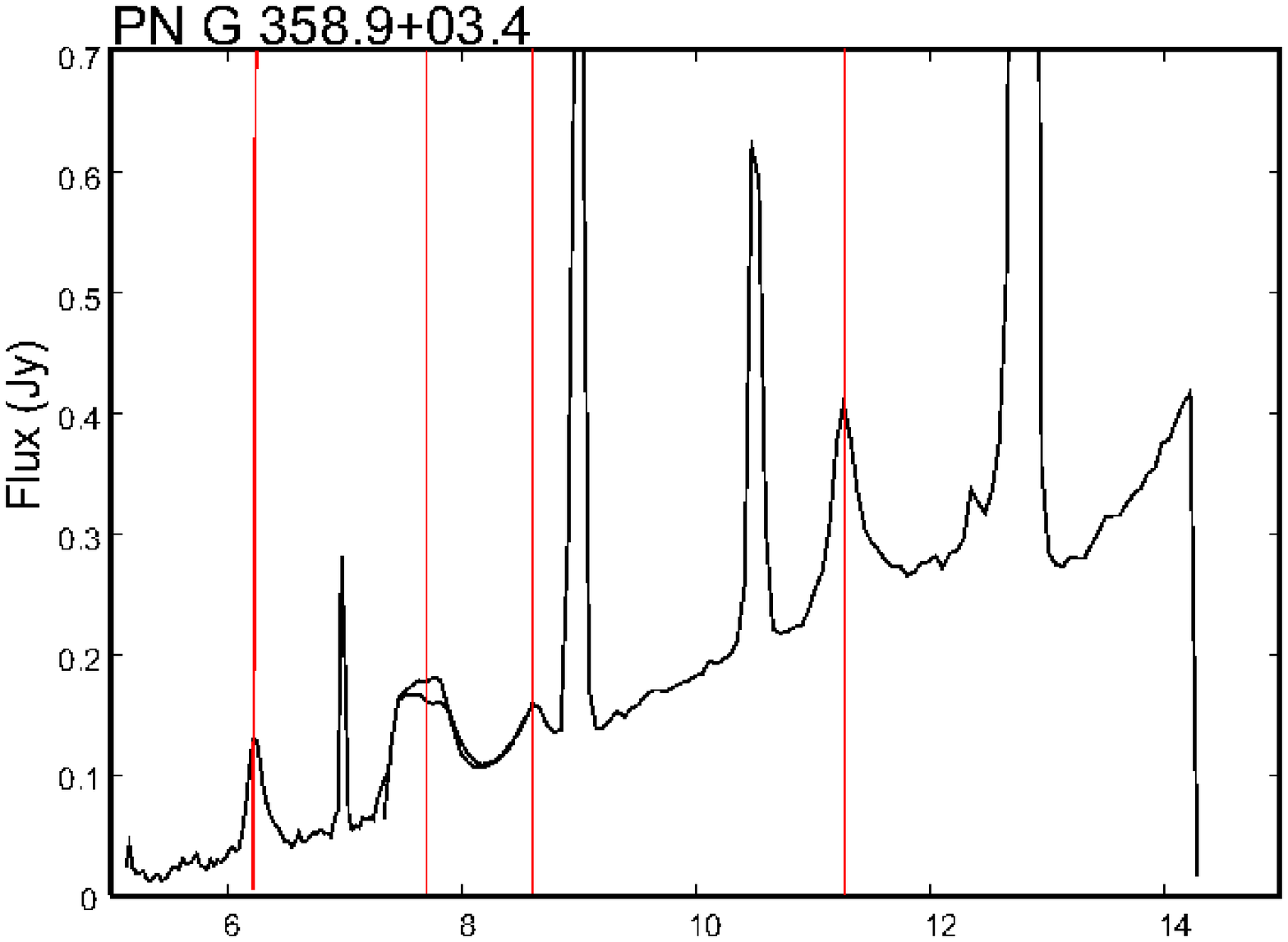}
          \hspace{3mm}
          \includegraphics[width=40mm]{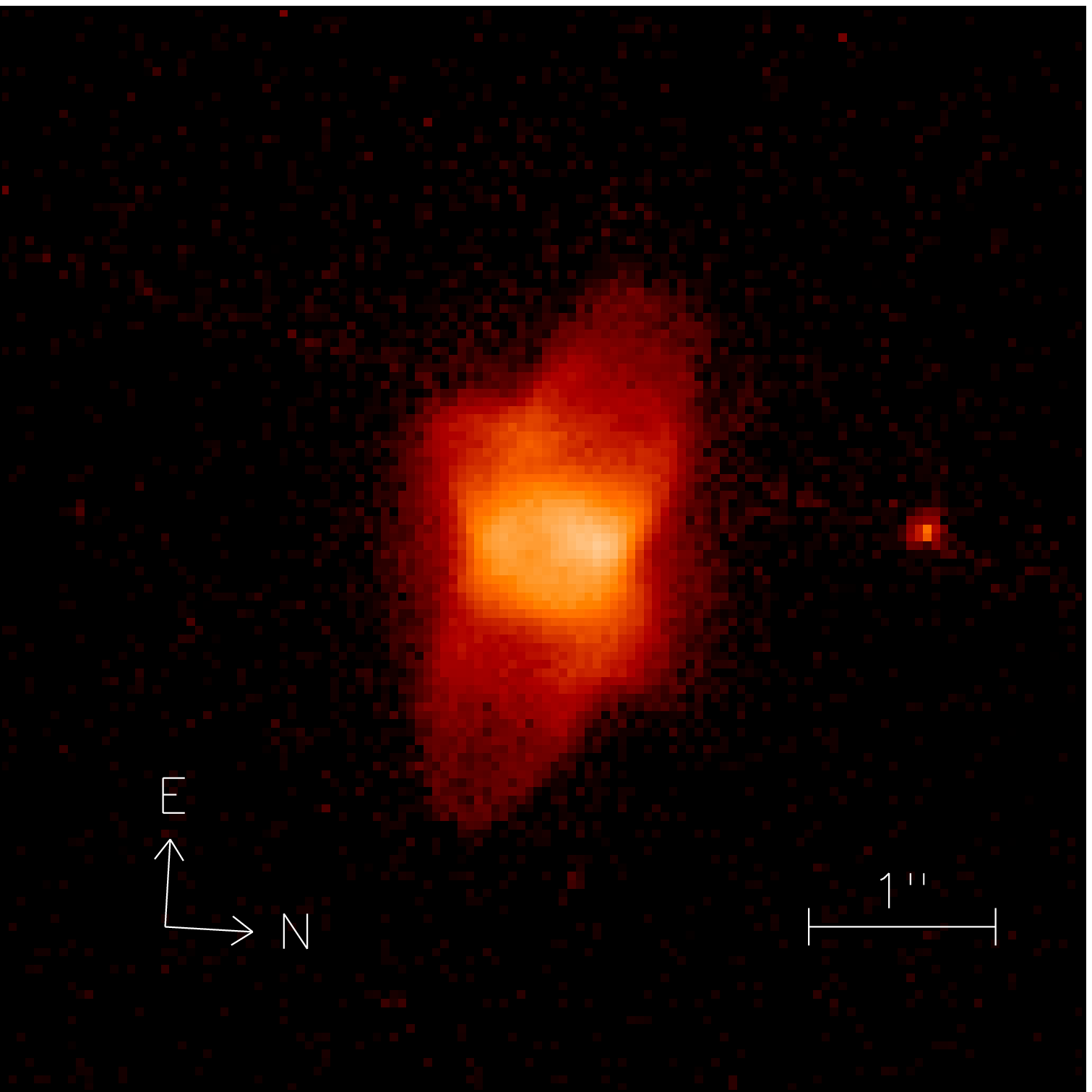}
          \hspace{6mm}
          \includegraphics[width=55mm]{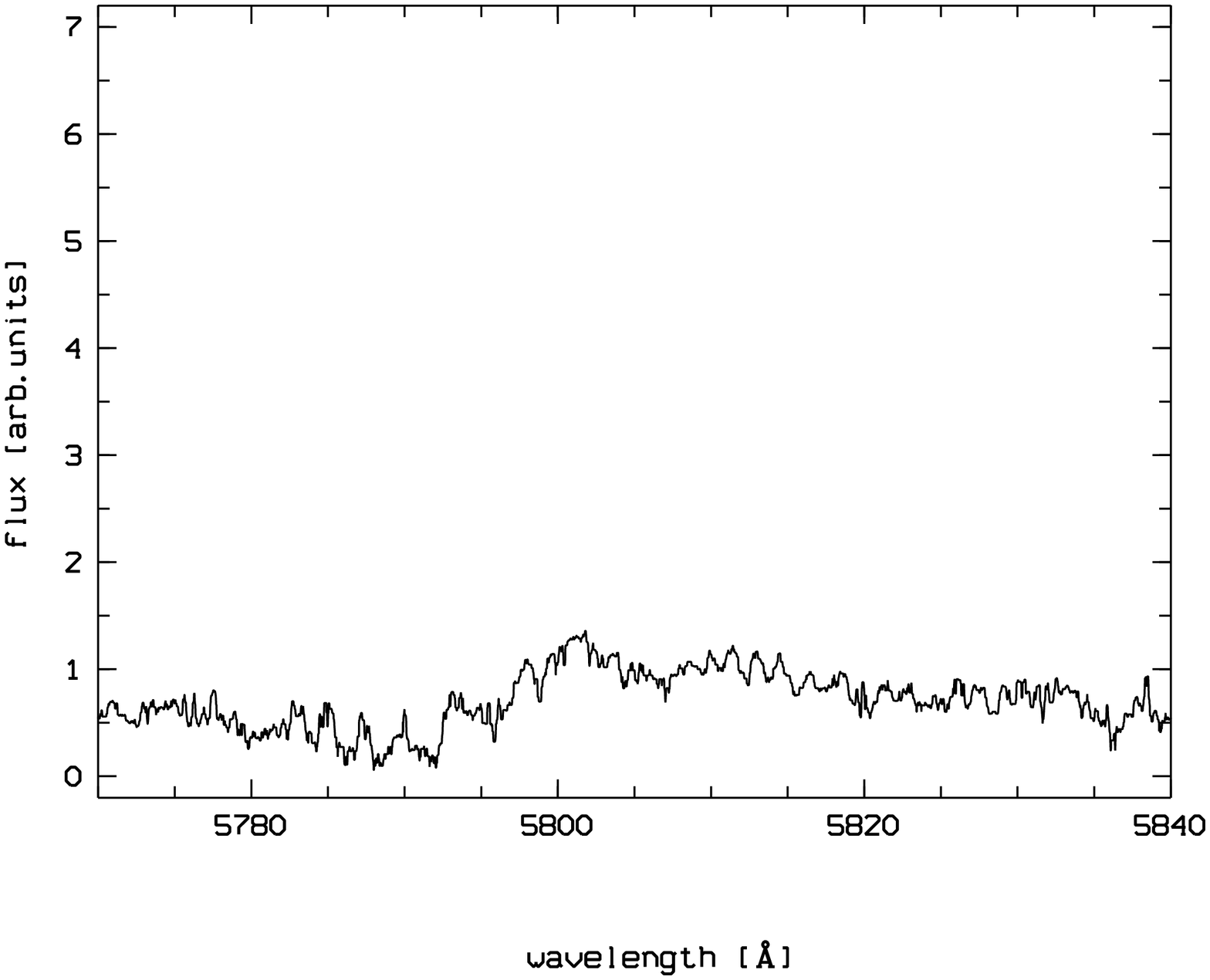}}
     \caption{Correlation image showing in the first column the IR
       Spitzer spectrum, showing the short wavelength region, the
       vertical red lines show the PAH bands. In the second column is the
       corresponding HST image and in the third column we show a part
       of the UVES spectrum. }
     \label{correl1}
\end{figure*}

\begin{figure*}
      \vspace{1mm}
     \hbox{\includegraphics[width=50mm]{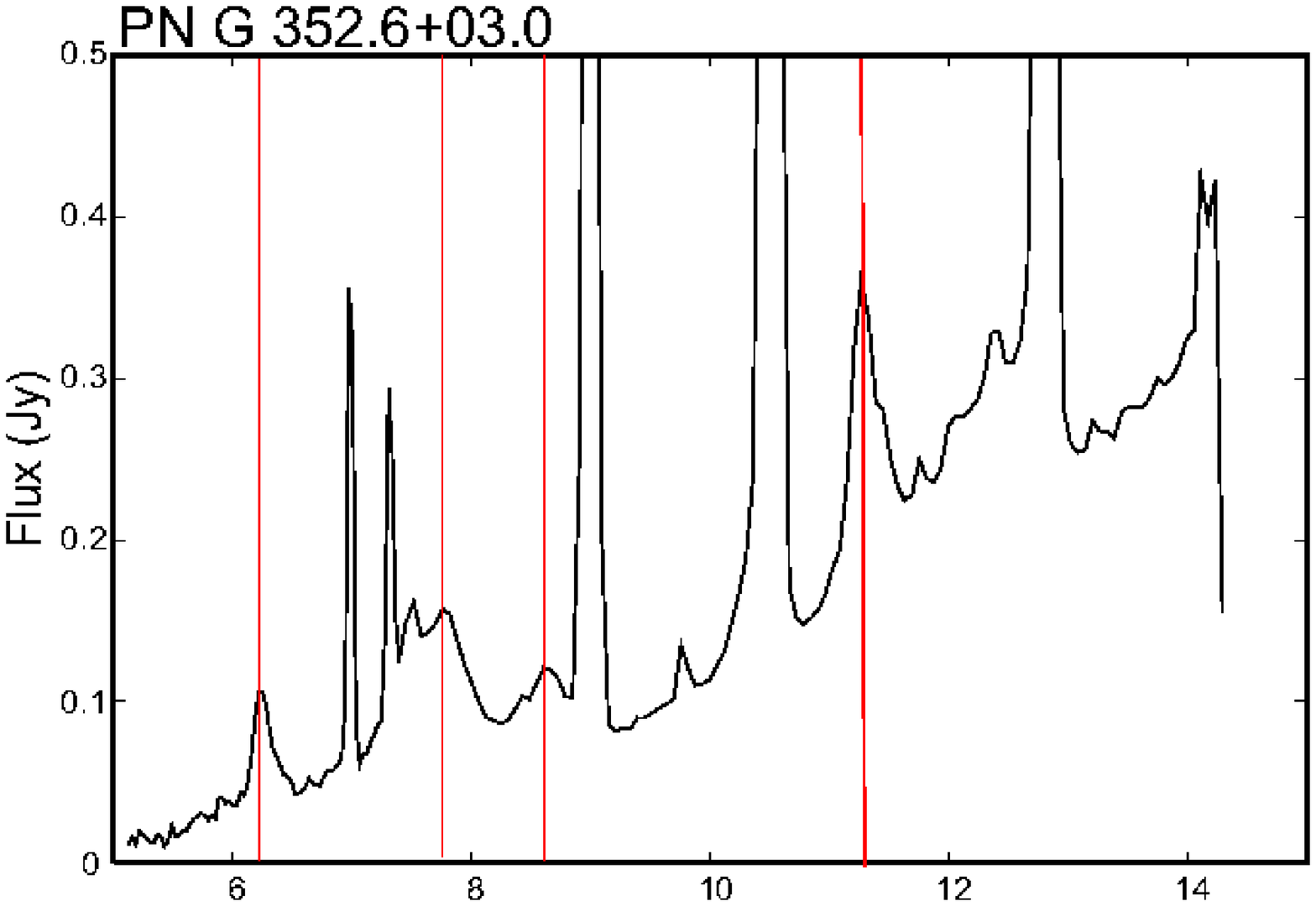}
          \hspace{3mm}
          \includegraphics[width=40mm]{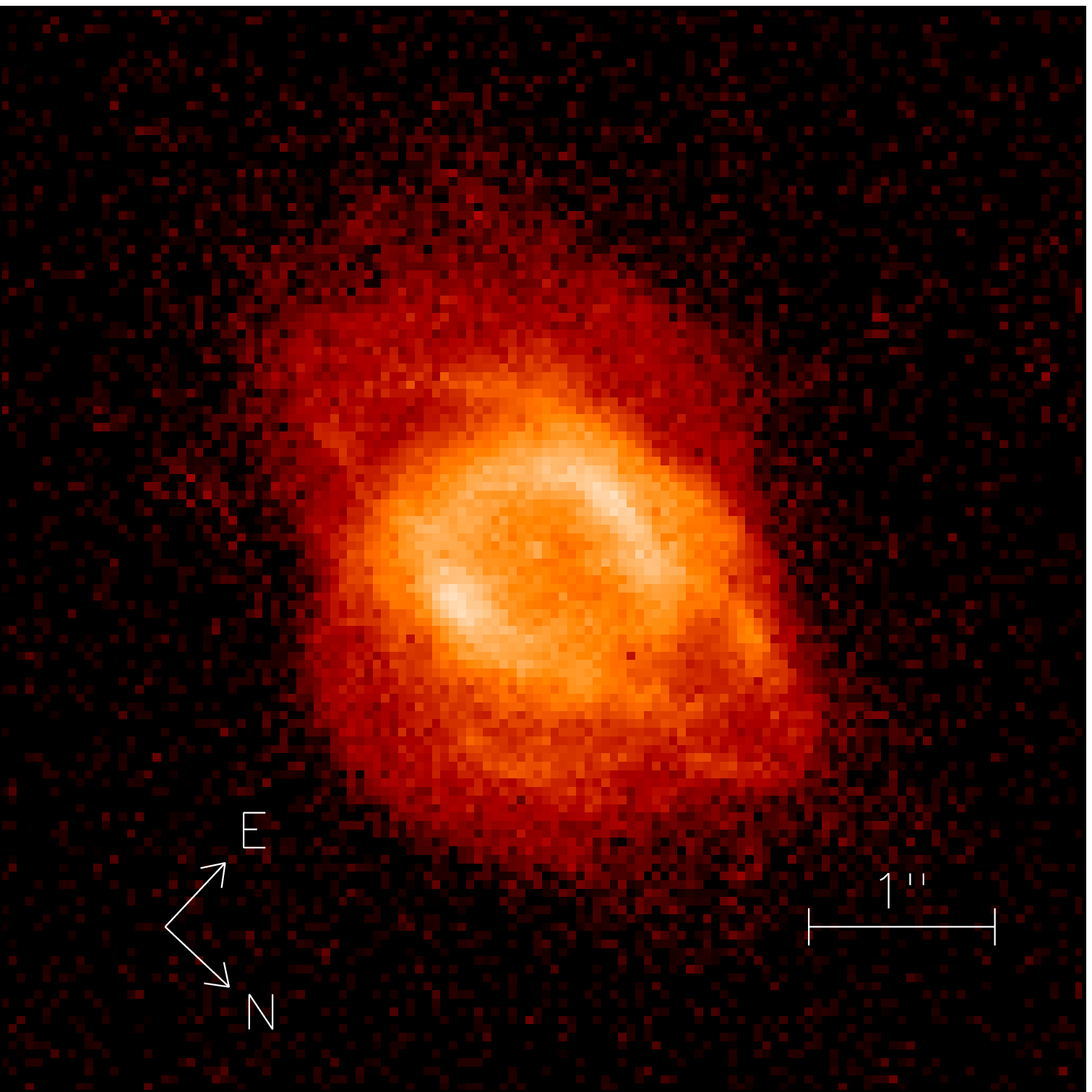}
          \hspace{6mm}
          \includegraphics[width=55mm]{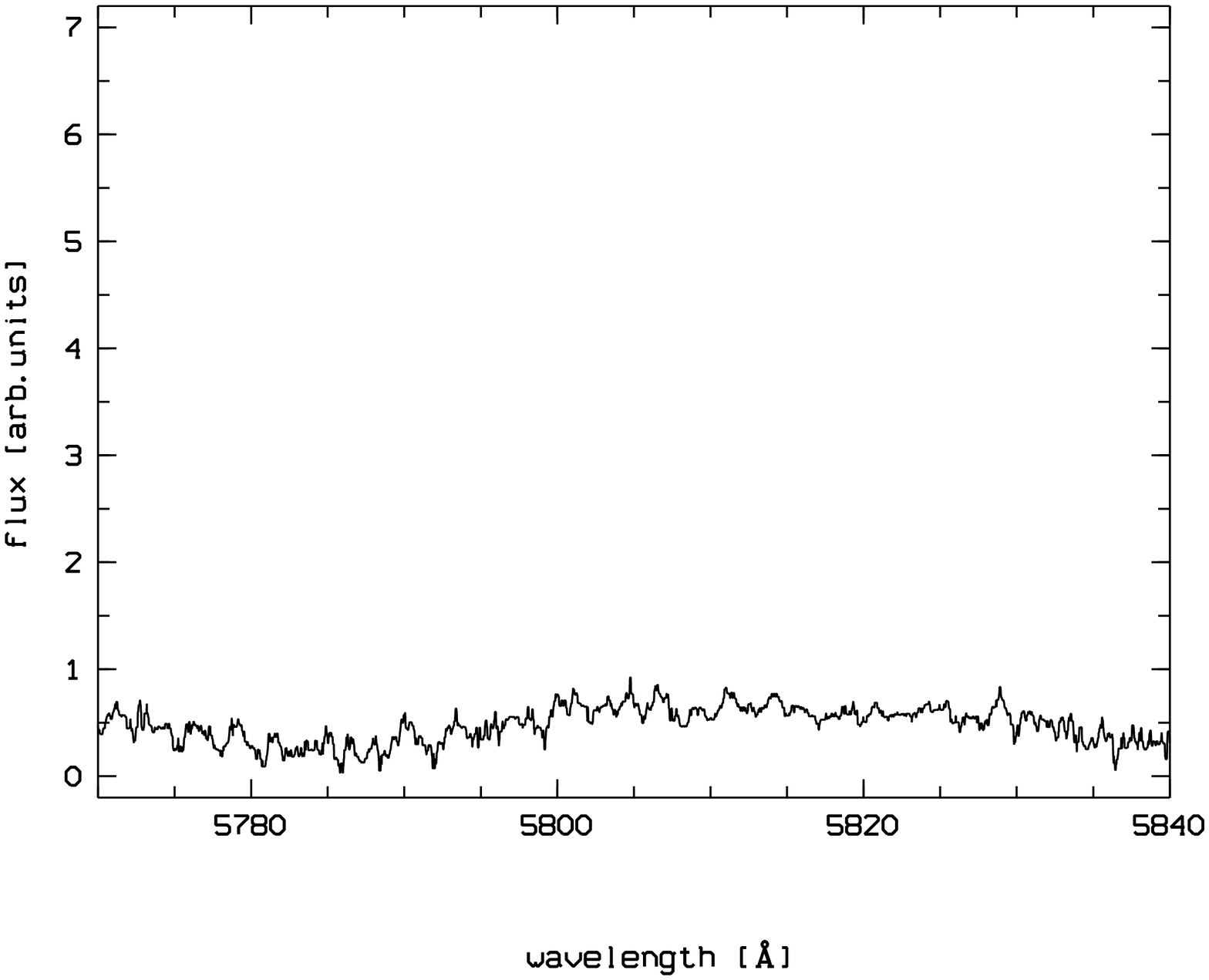}}
      \vspace{1mm}
     \hbox{\includegraphics[width=50mm]{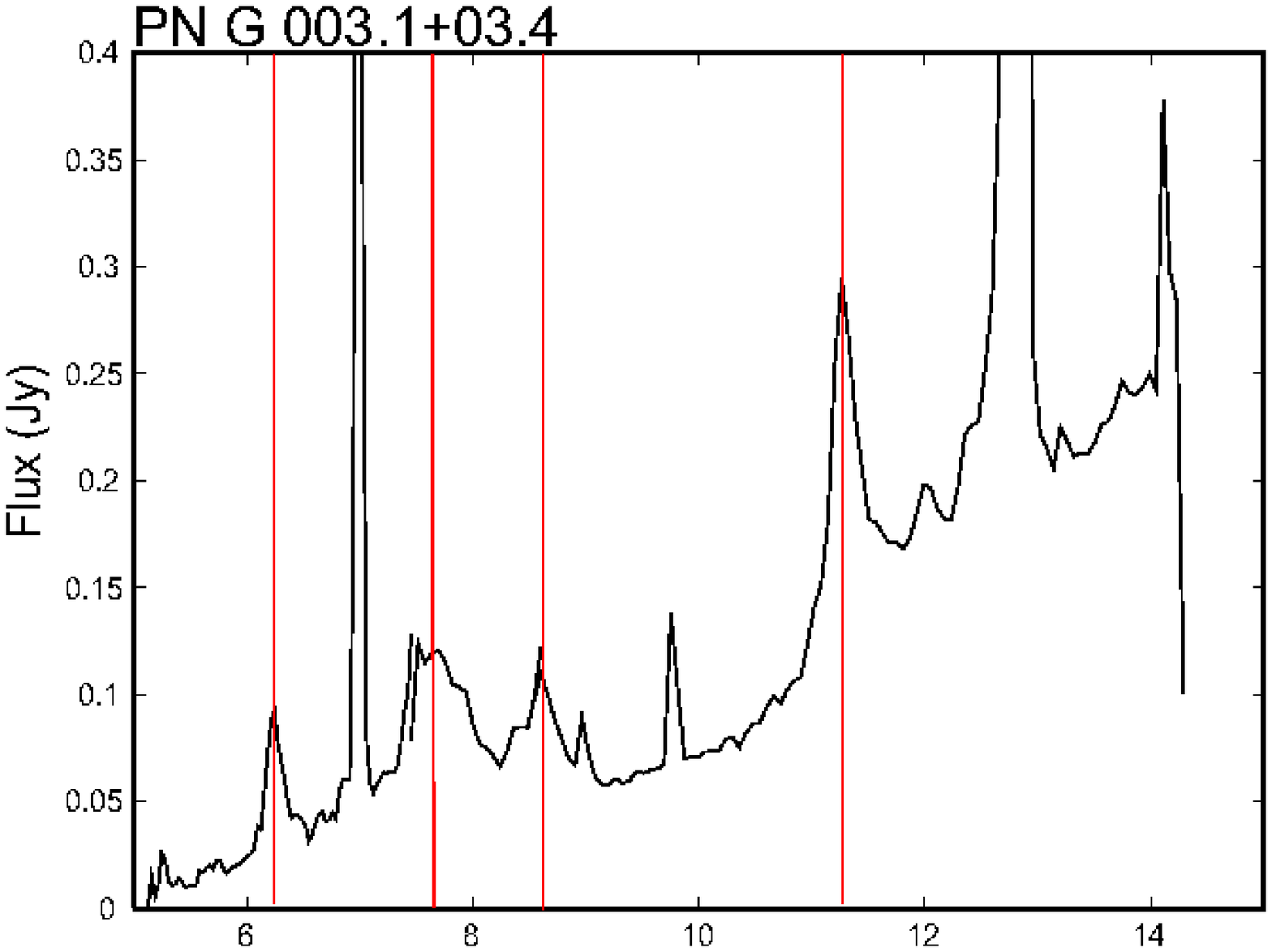}
          \hspace{3mm}
          \includegraphics[width=40mm]{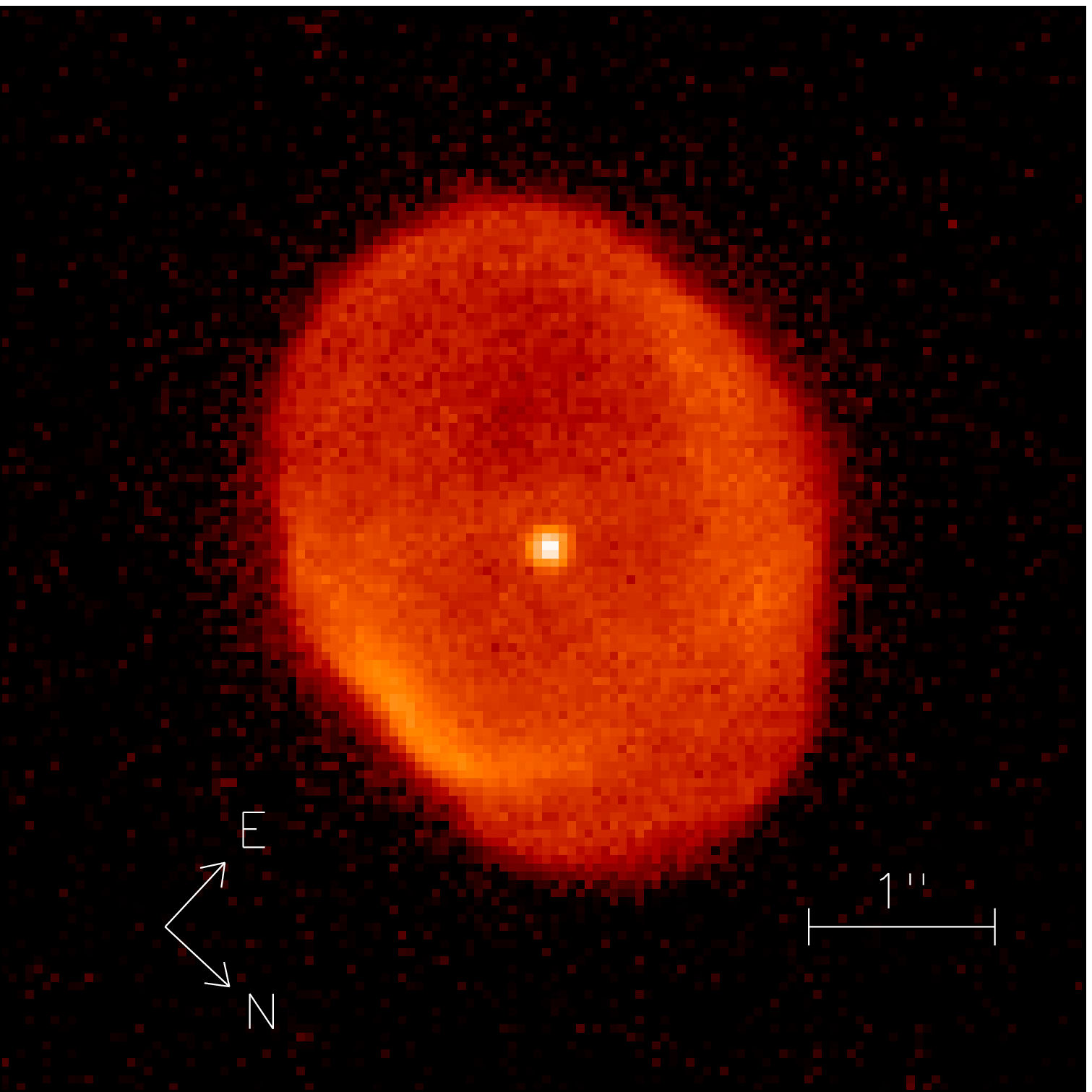}
          \hspace{6mm}
          \includegraphics[width=55mm]{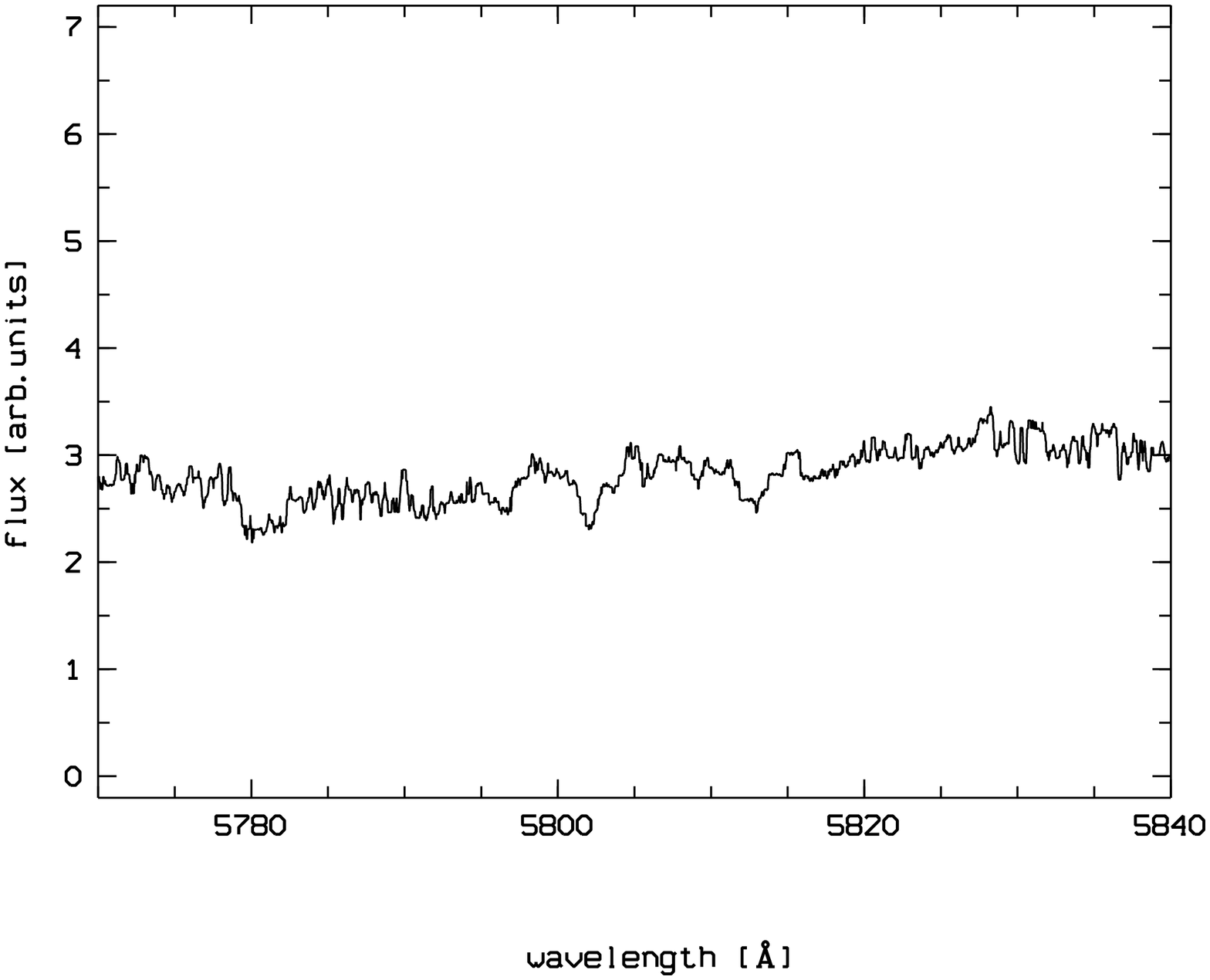}}
      \vspace{1mm}
      \hbox{\includegraphics[width=50mm]{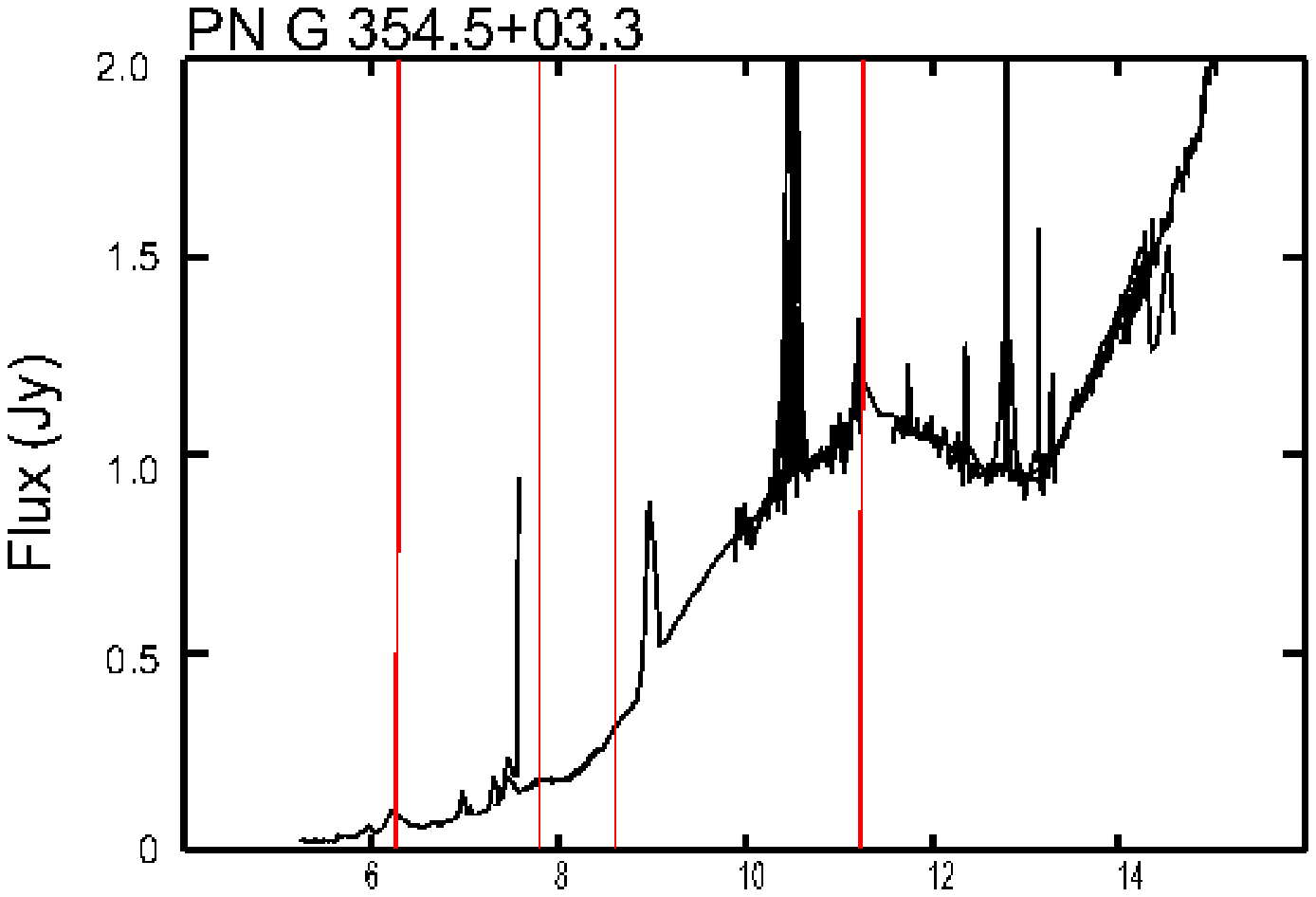}
          \hspace{3mm}
          \includegraphics[width=40mm]{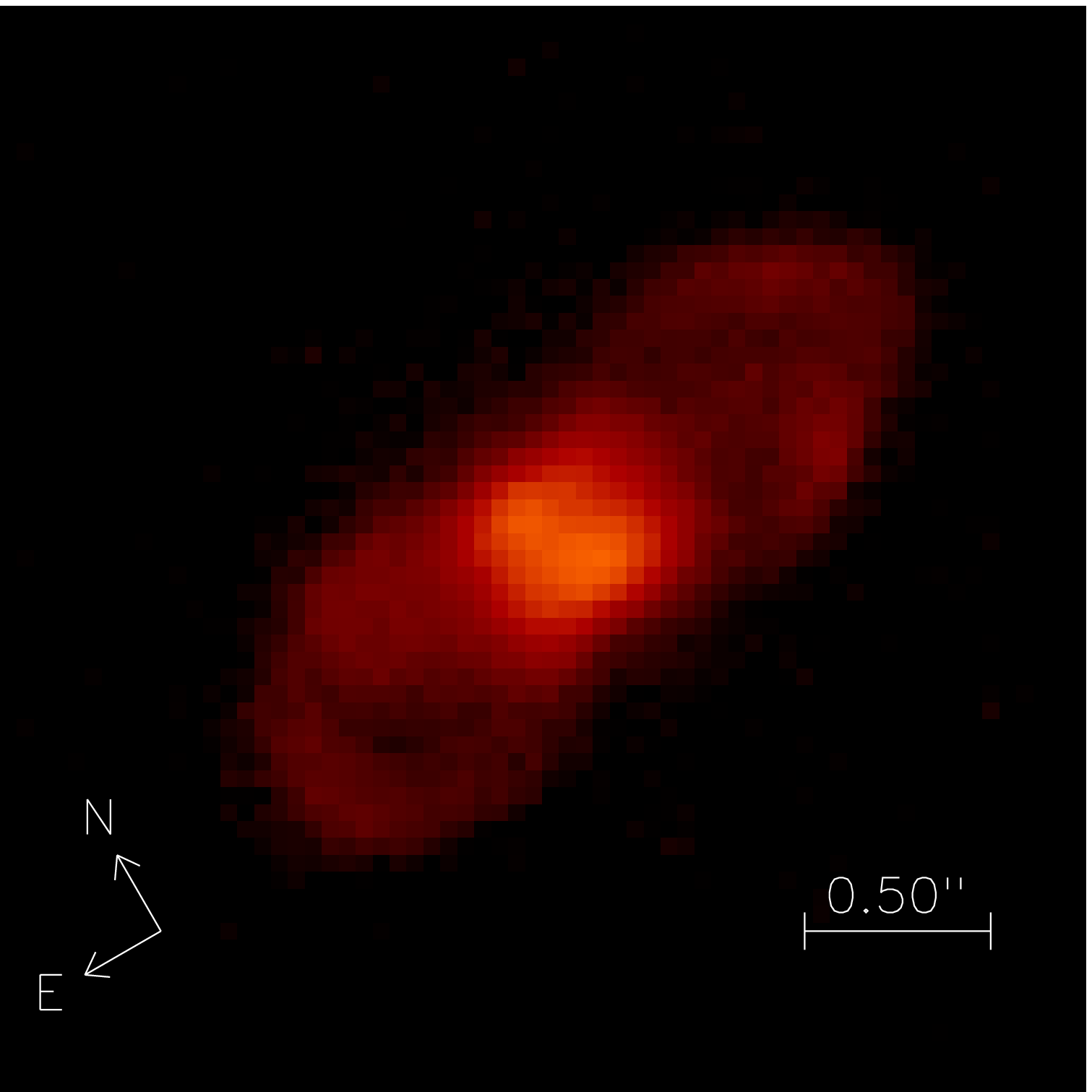}
          \hspace{3mm}
          \includegraphics[width=55mm]{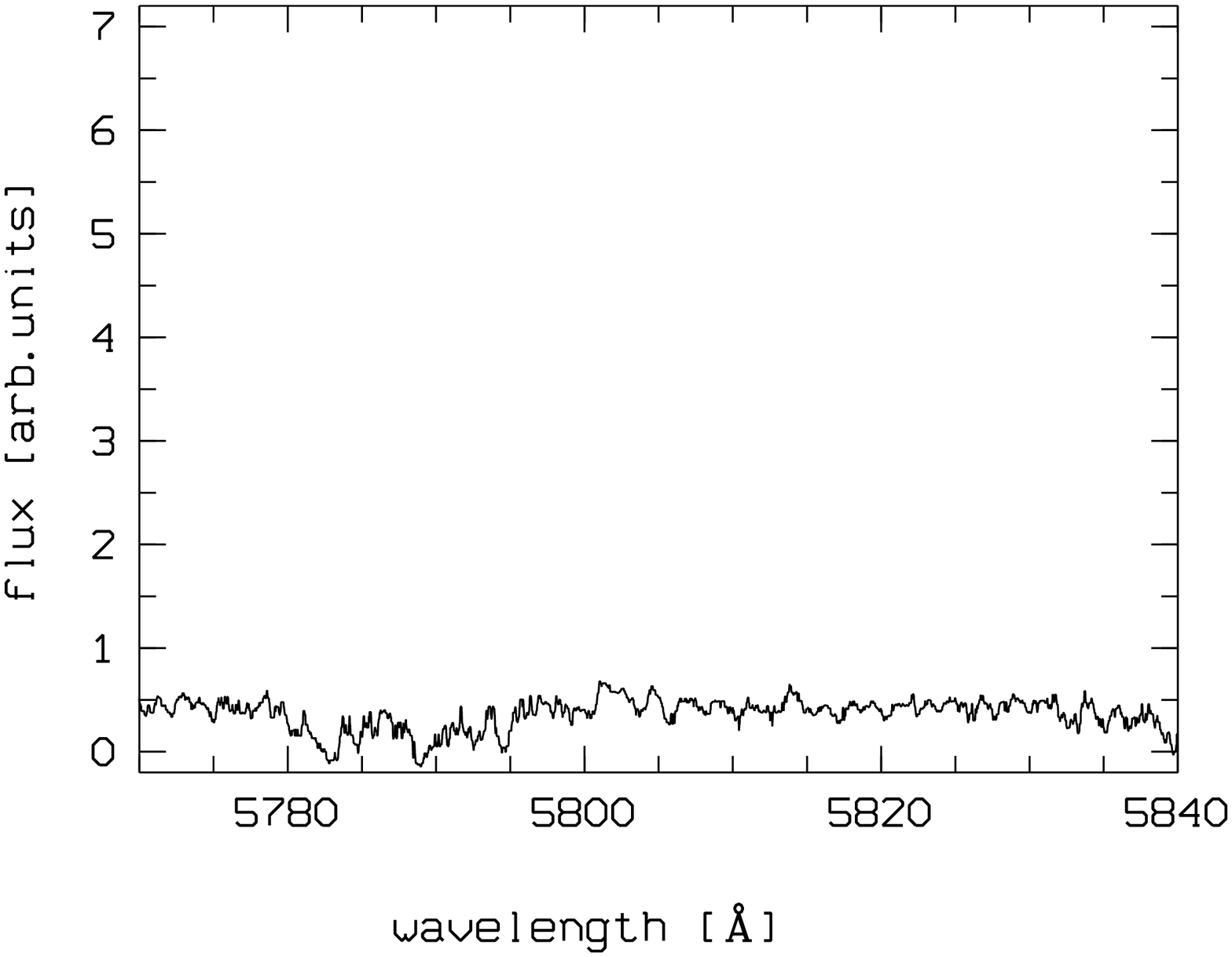}}
      \vspace{1mm}
      \hbox{\includegraphics[width=50mm]{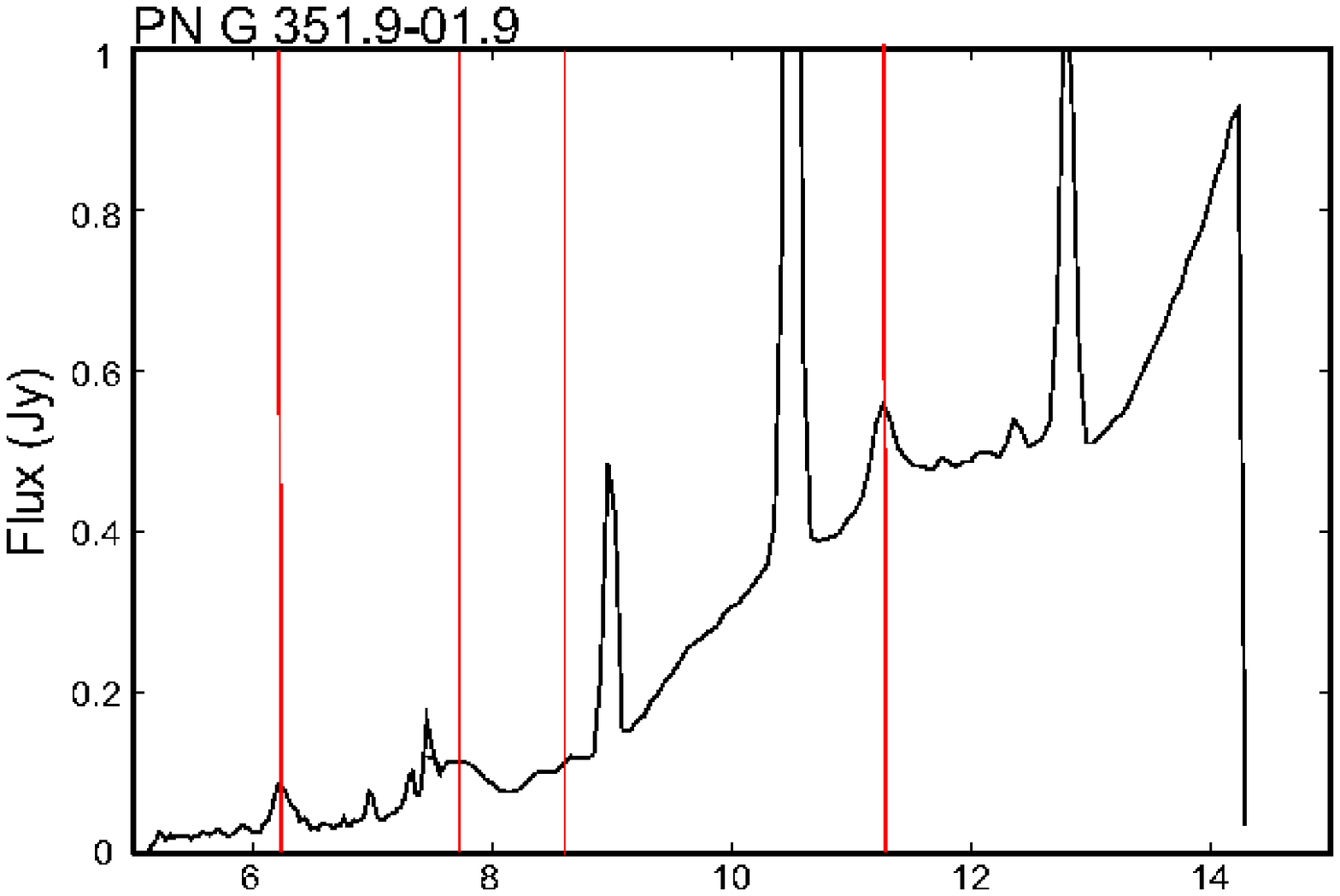}
          \hspace{3mm}
          \includegraphics[width=40mm]{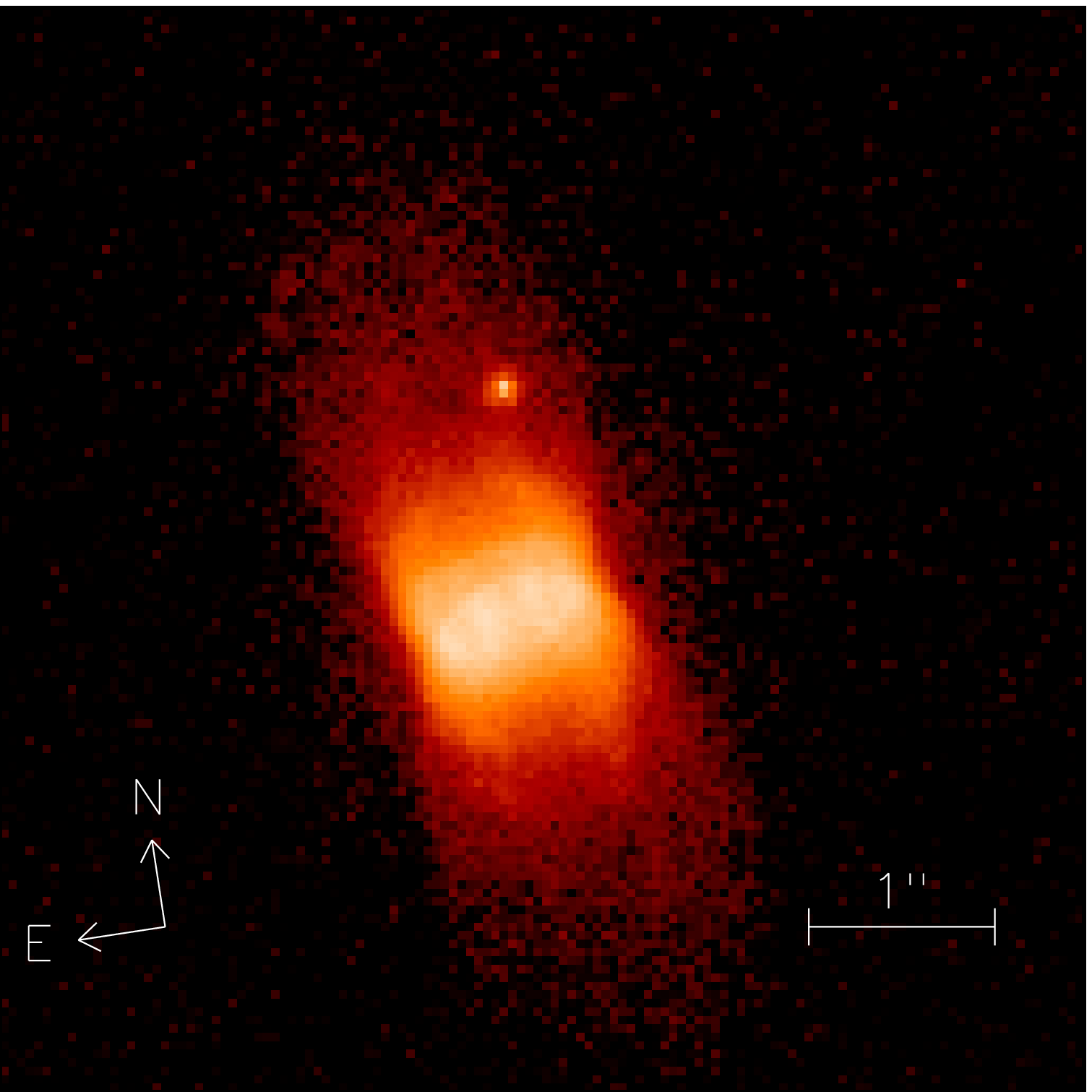}
          \hspace{6mm}
          \includegraphics[width=55mm]{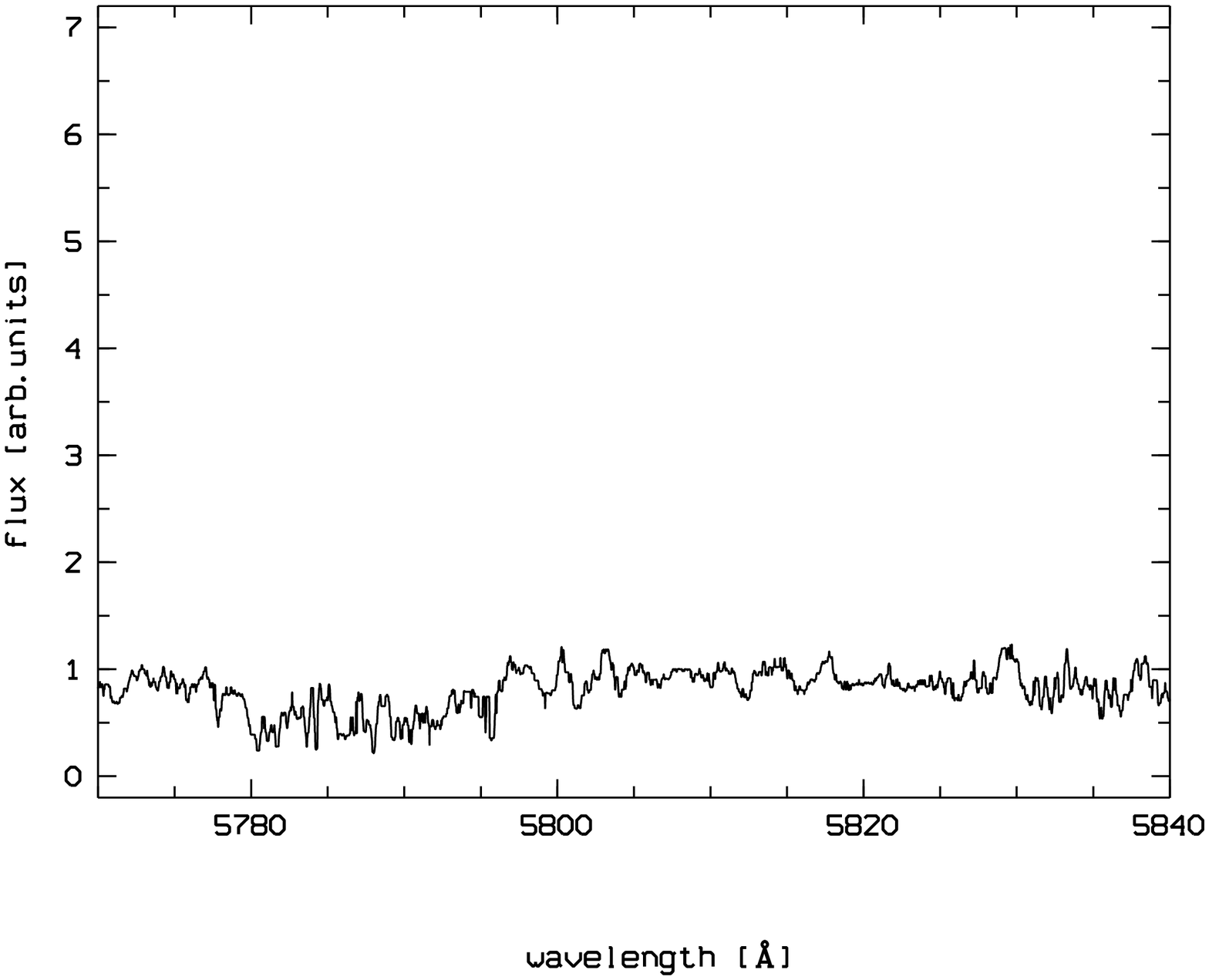}}
      \vspace{1mm}
    \hbox{\includegraphics[width=50mm]{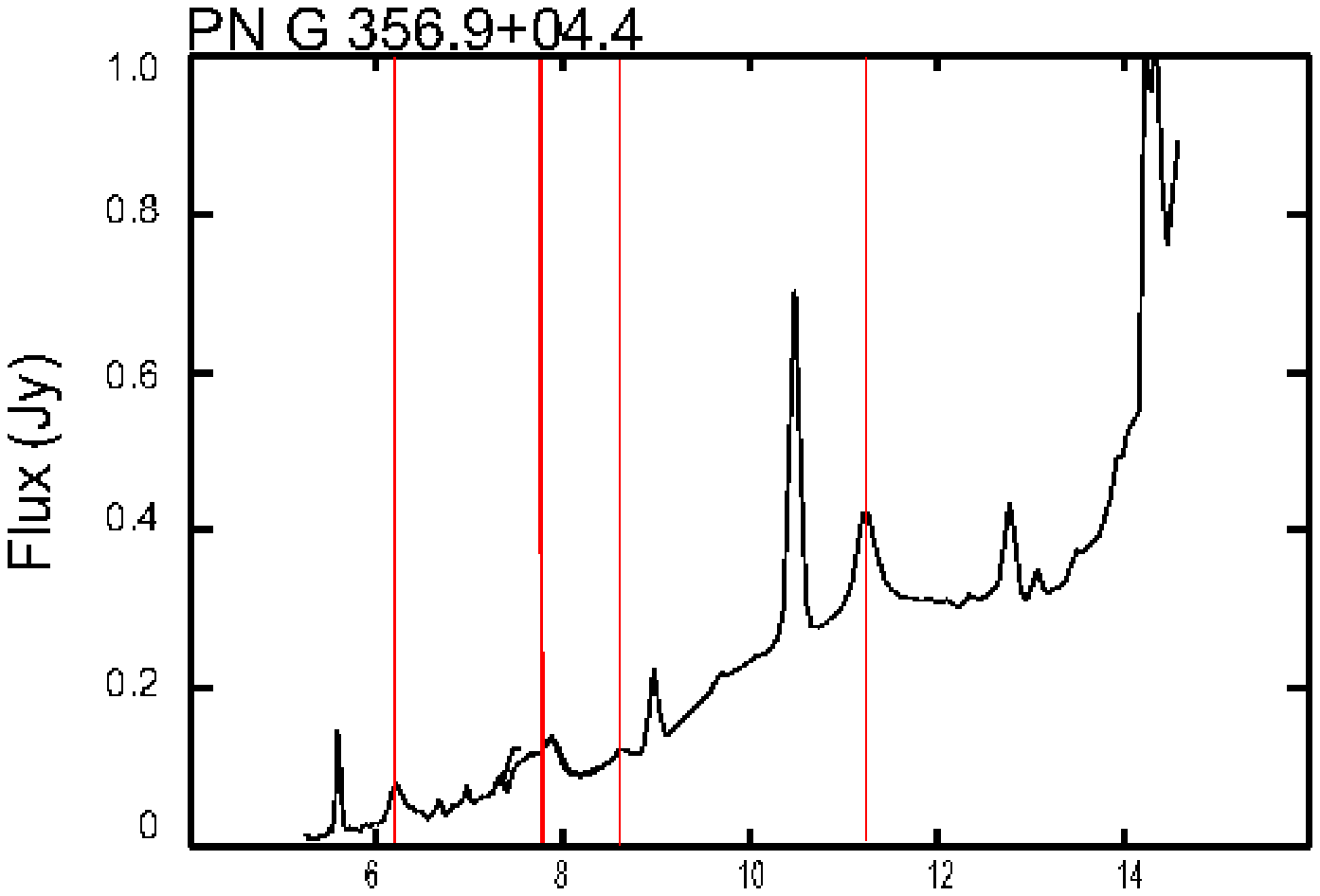}
          \hspace{3mm}
           \includegraphics[width=40mm]{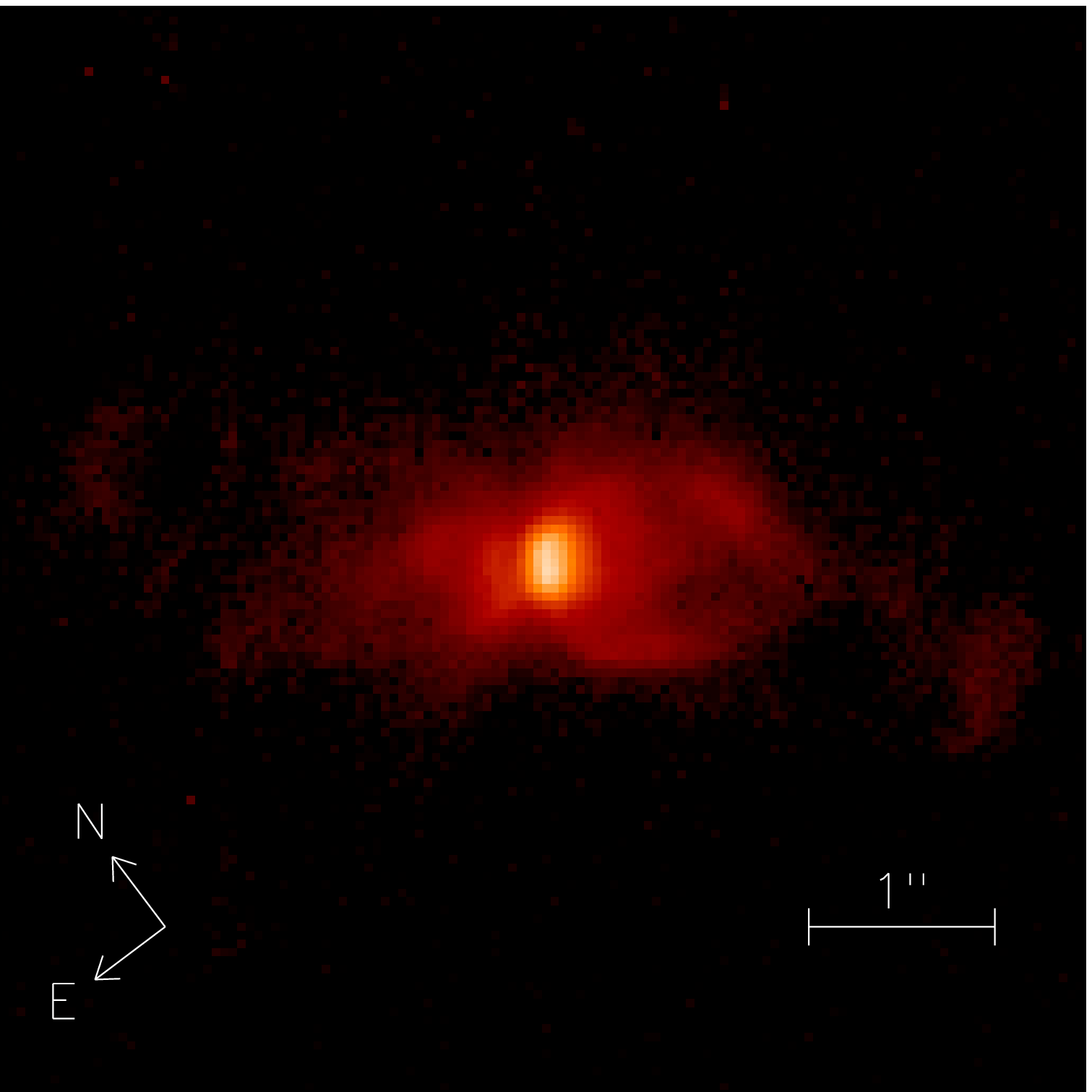}
          \hspace{3mm}
          \includegraphics[width=55mm]{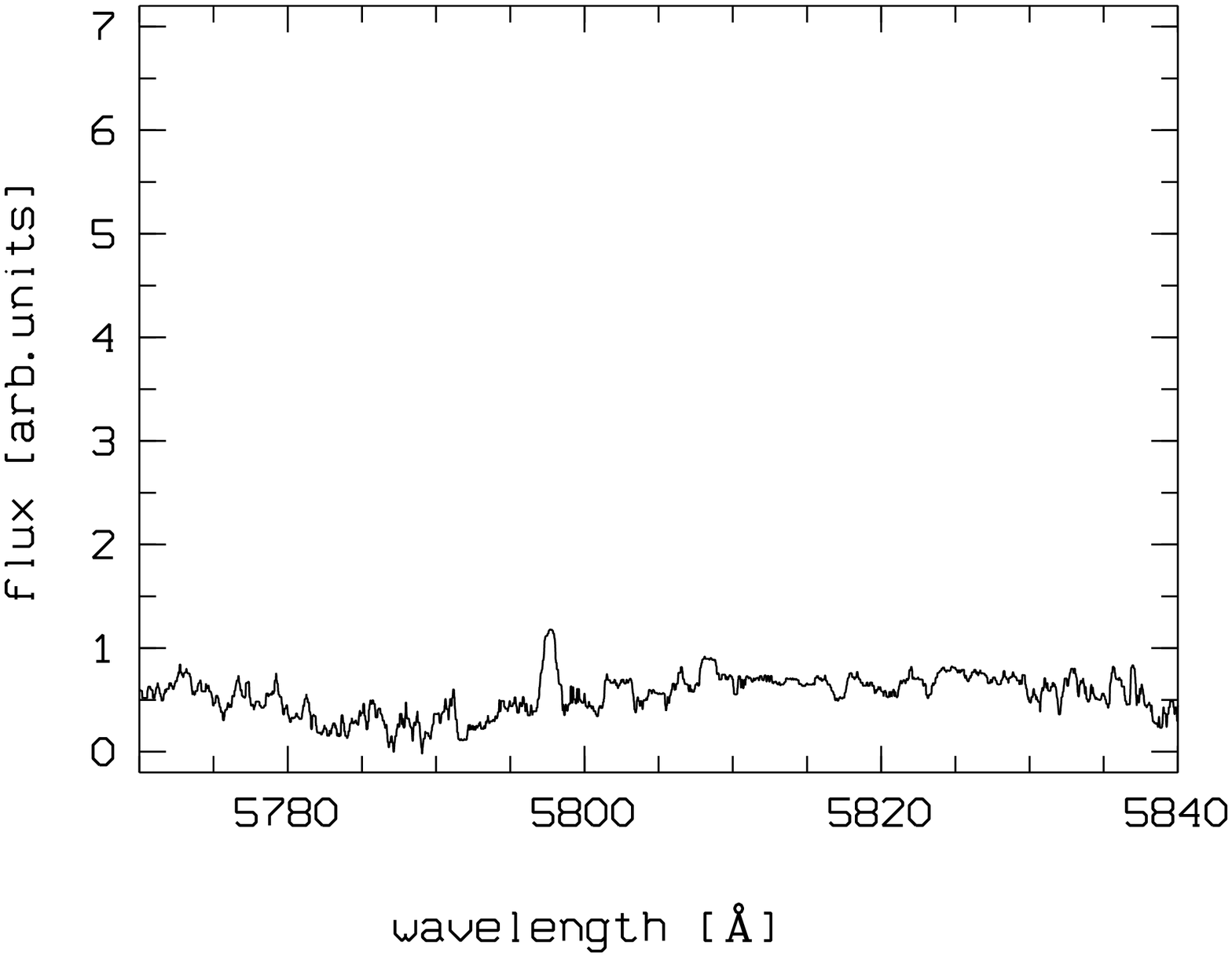}}
       \vspace{1mm}
     \hbox{\includegraphics[width=50mm]{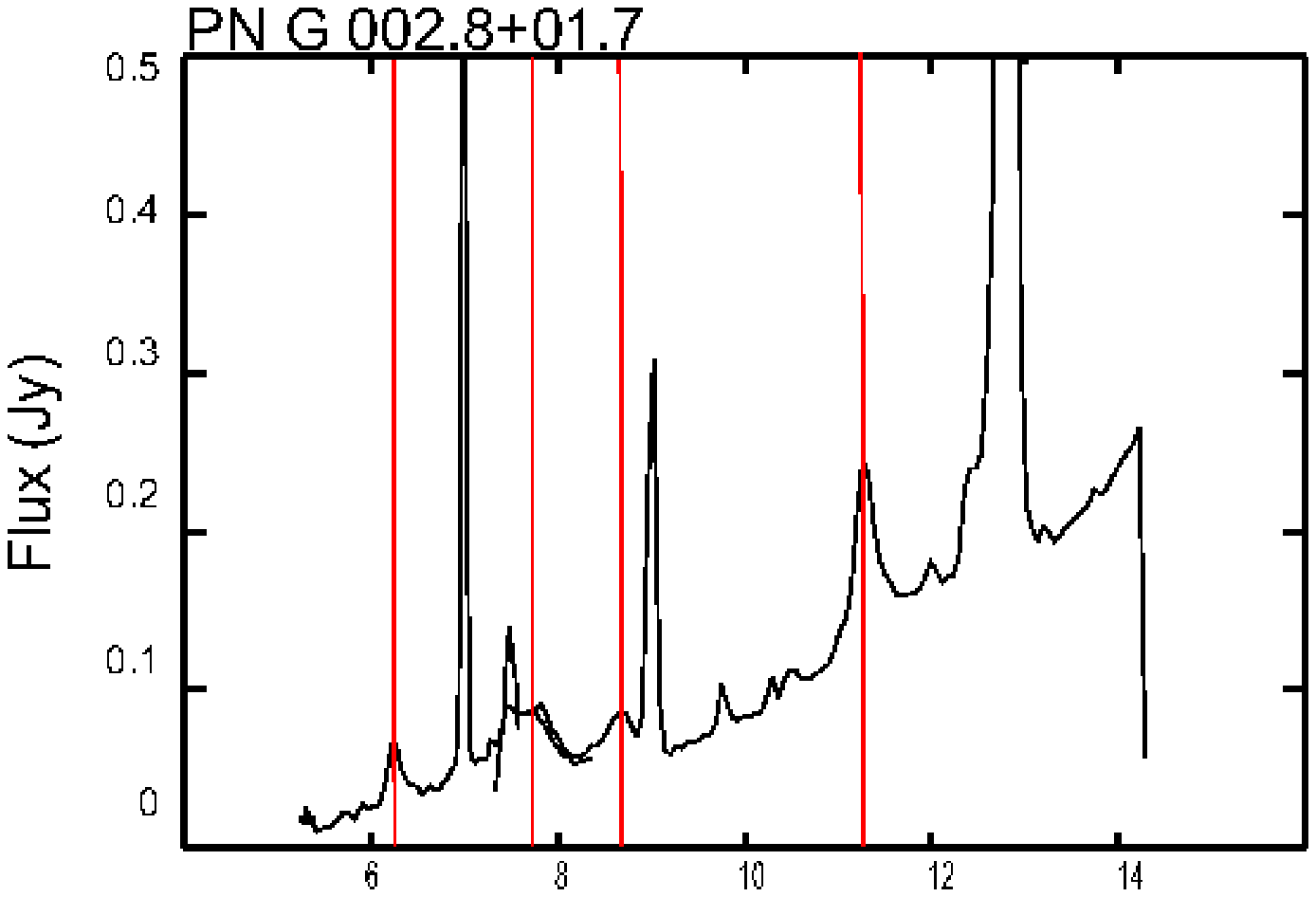}
          \hspace{3mm}
          \includegraphics[width=40mm]{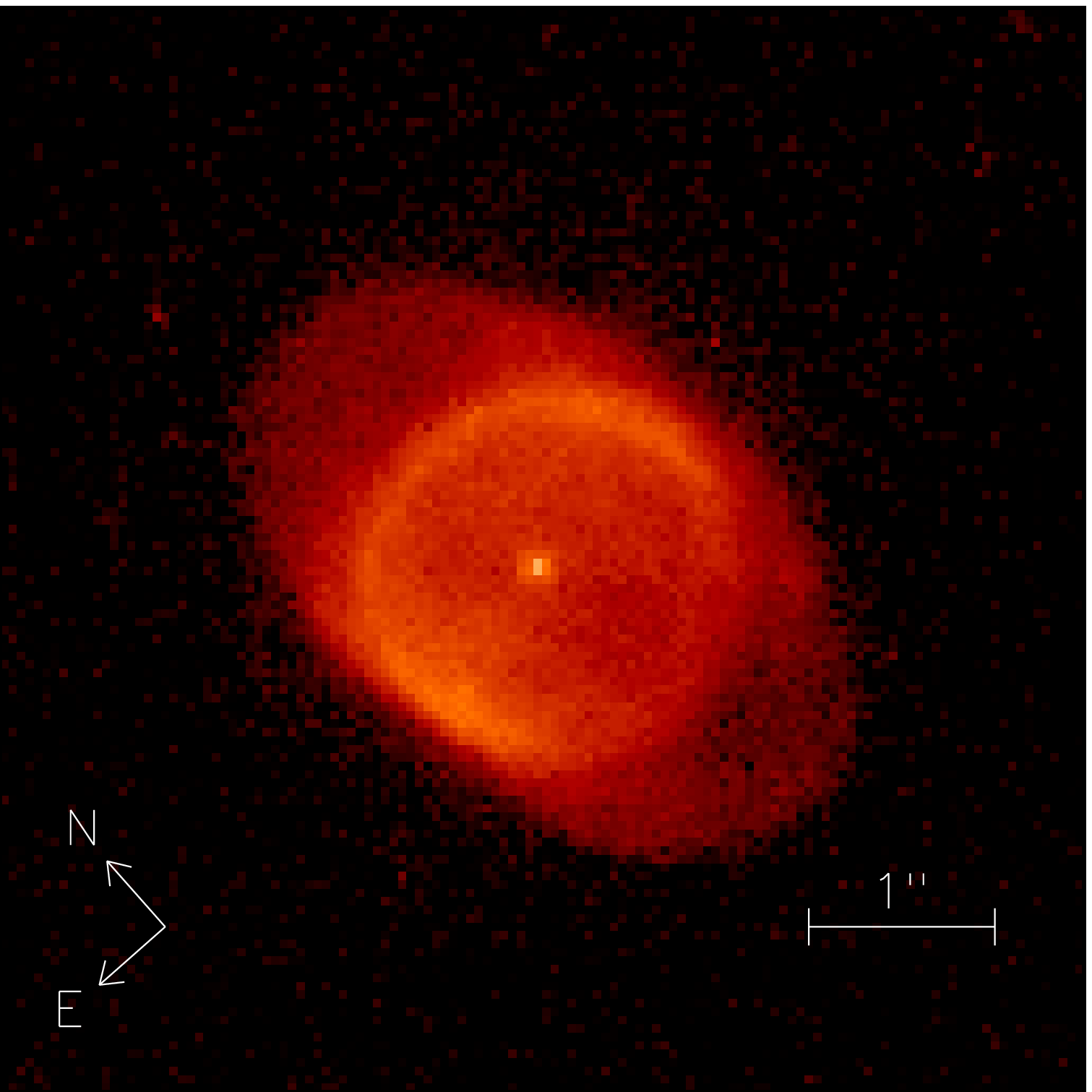}
          \hspace{3mm}
          \includegraphics[width=55mm]{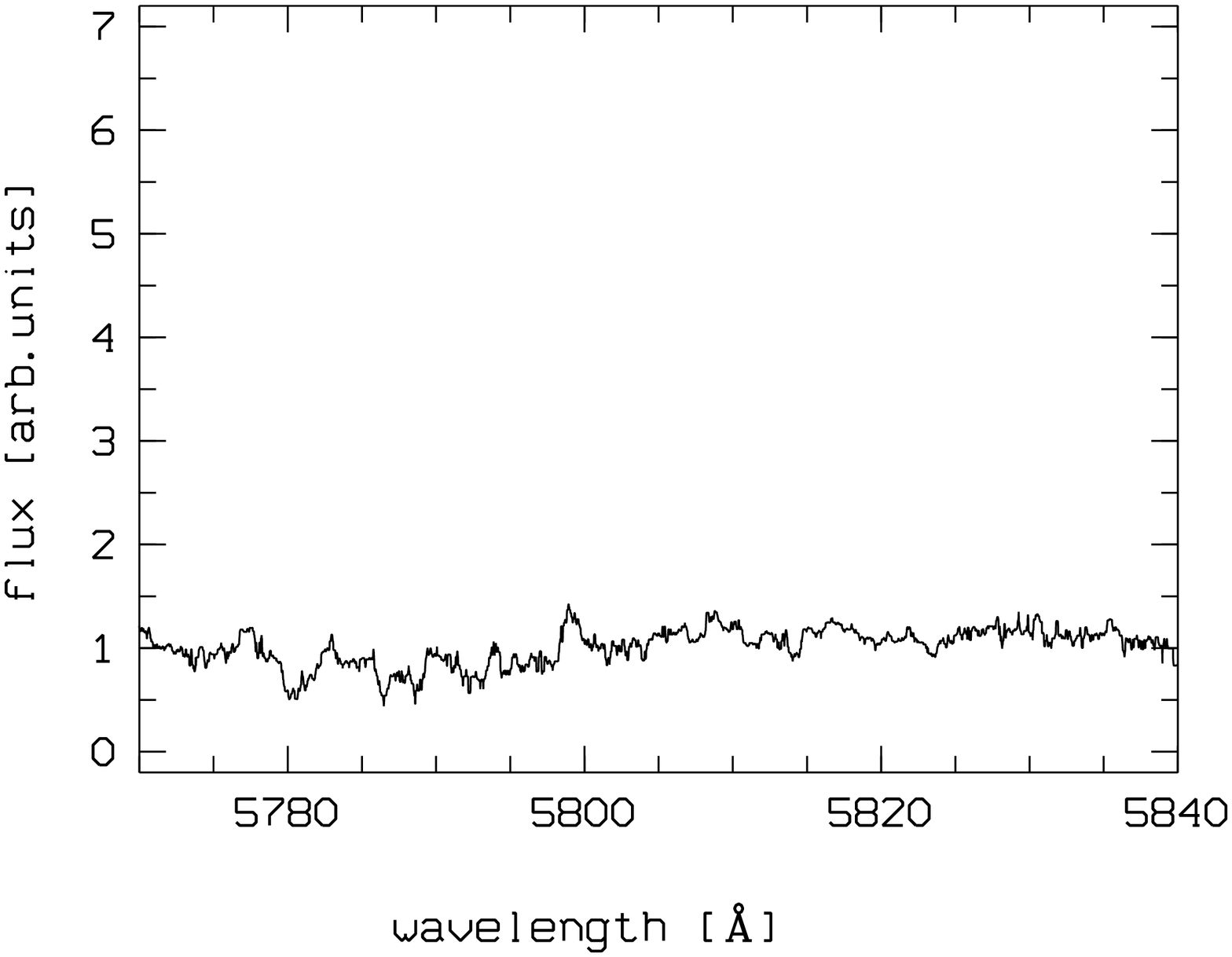}}
   \caption{(cont.) Correlation image showing in the first column the IR
       Spitzer spectrum, showing the short wavelength region, the
       vertical red lines show the PAH bands. In the second column is the
       corresponding HST image and in the third column we show a part
       of the UVES spectrum. }
     \label{correl2}
\end{figure*}      

\begin{figure*}
     \hbox{\includegraphics[width=50mm]{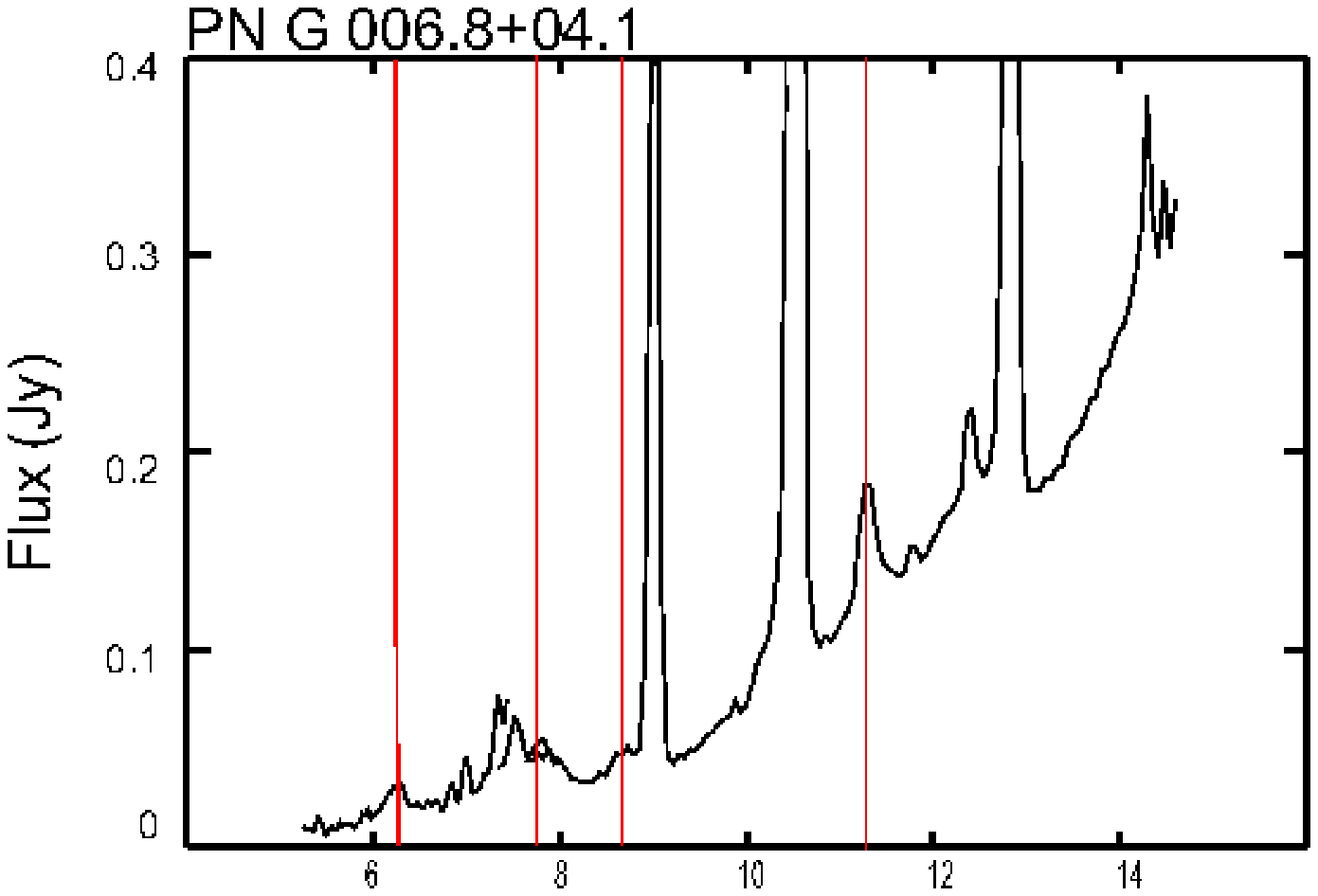}
          \hspace{3mm}
          \includegraphics[width=40mm]{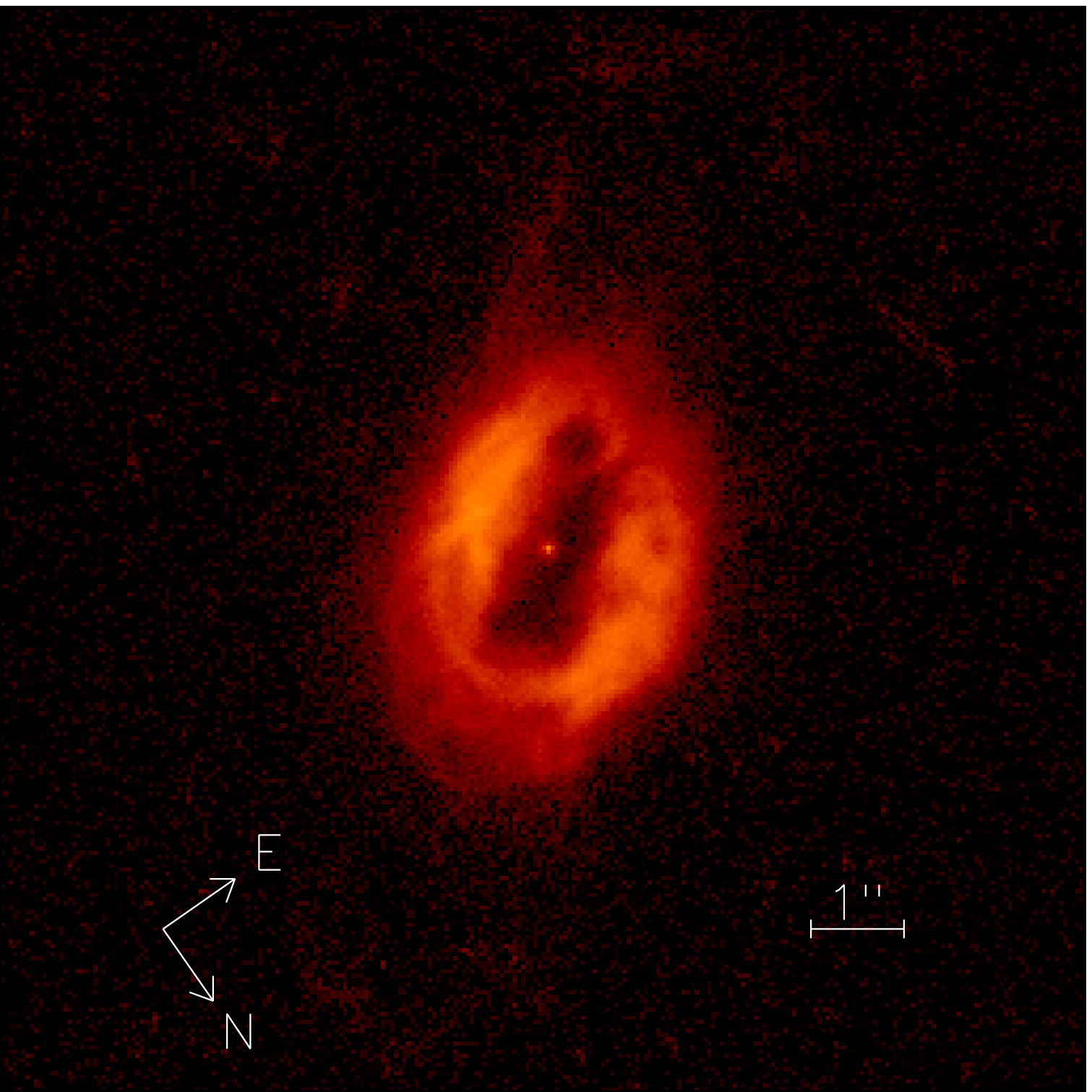}
          \hspace{3mm}
          \includegraphics[width=55mm]{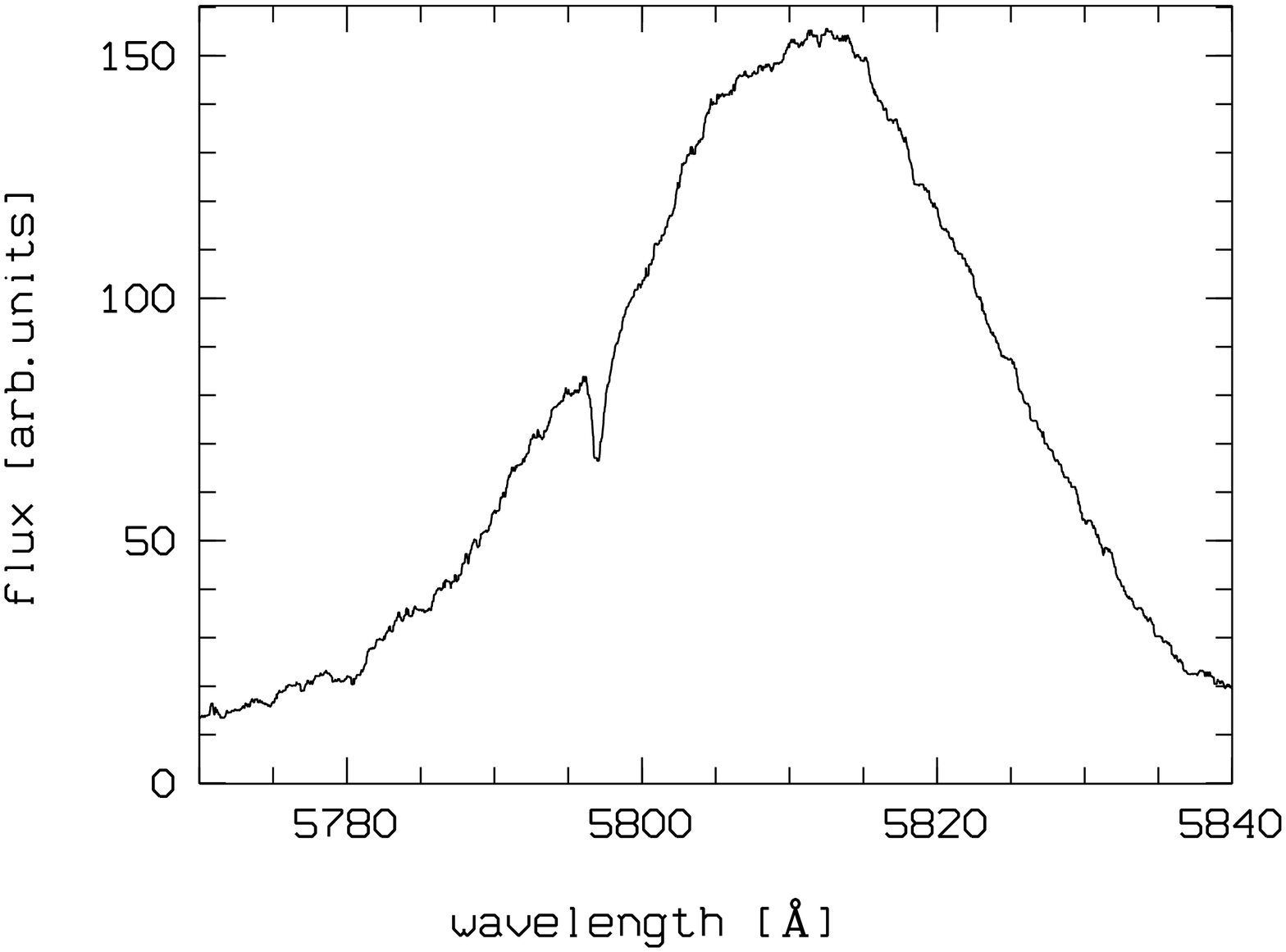}}
       \vspace{1mm}
   \hbox{\includegraphics[width=50mm]{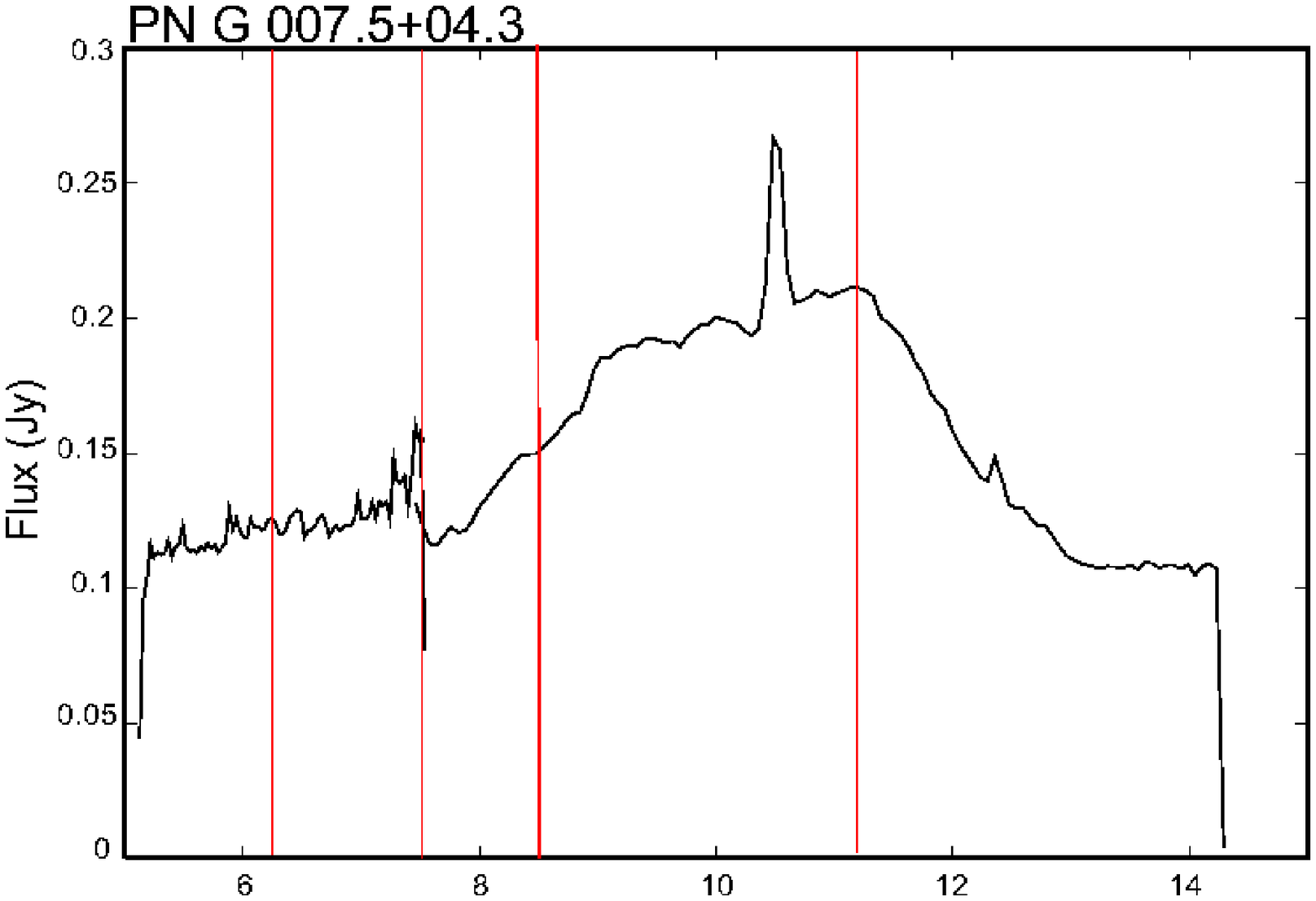}
          \hspace{3mm}
          \includegraphics[width=40mm]{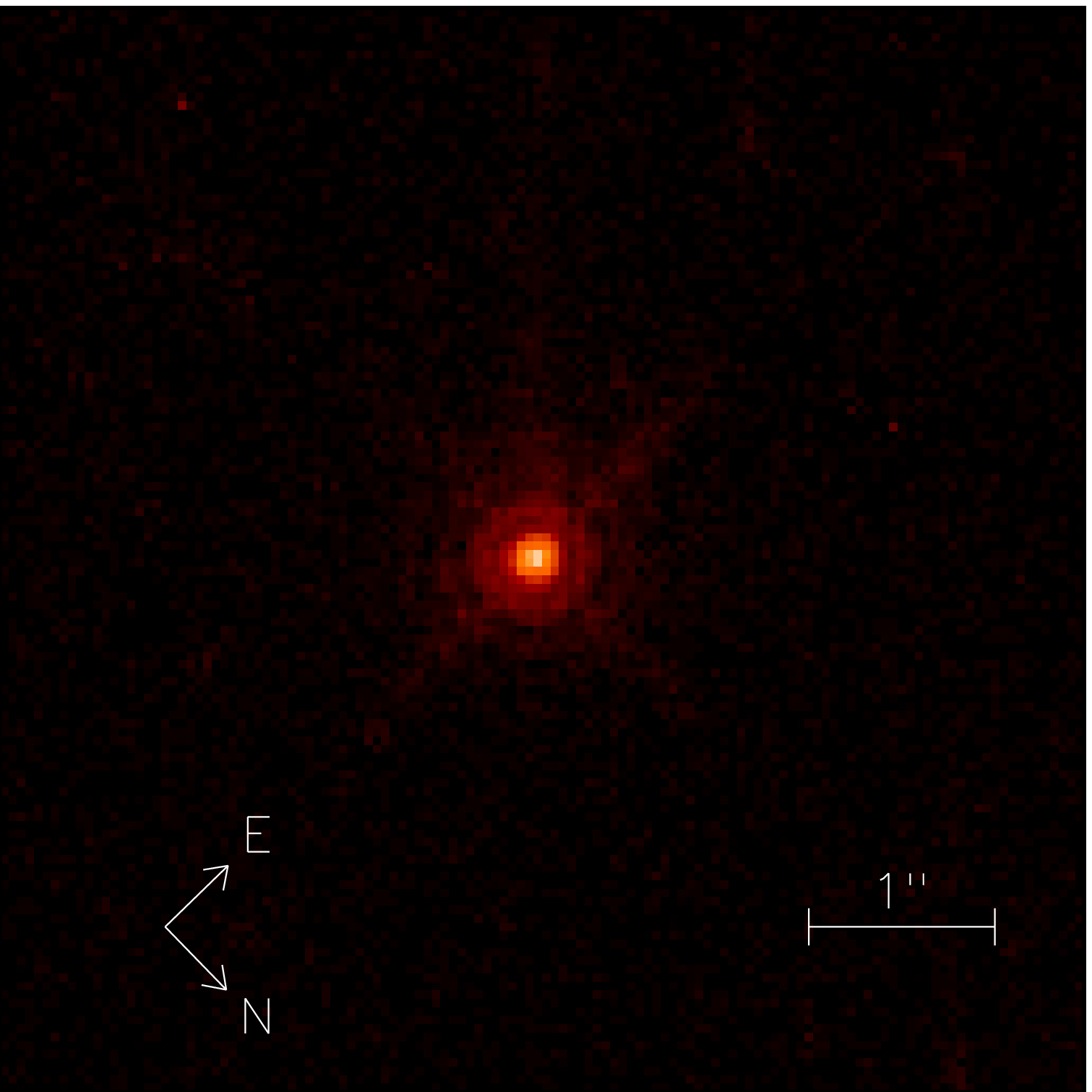}
          \hspace{6mm}
          \includegraphics[width=55mm]{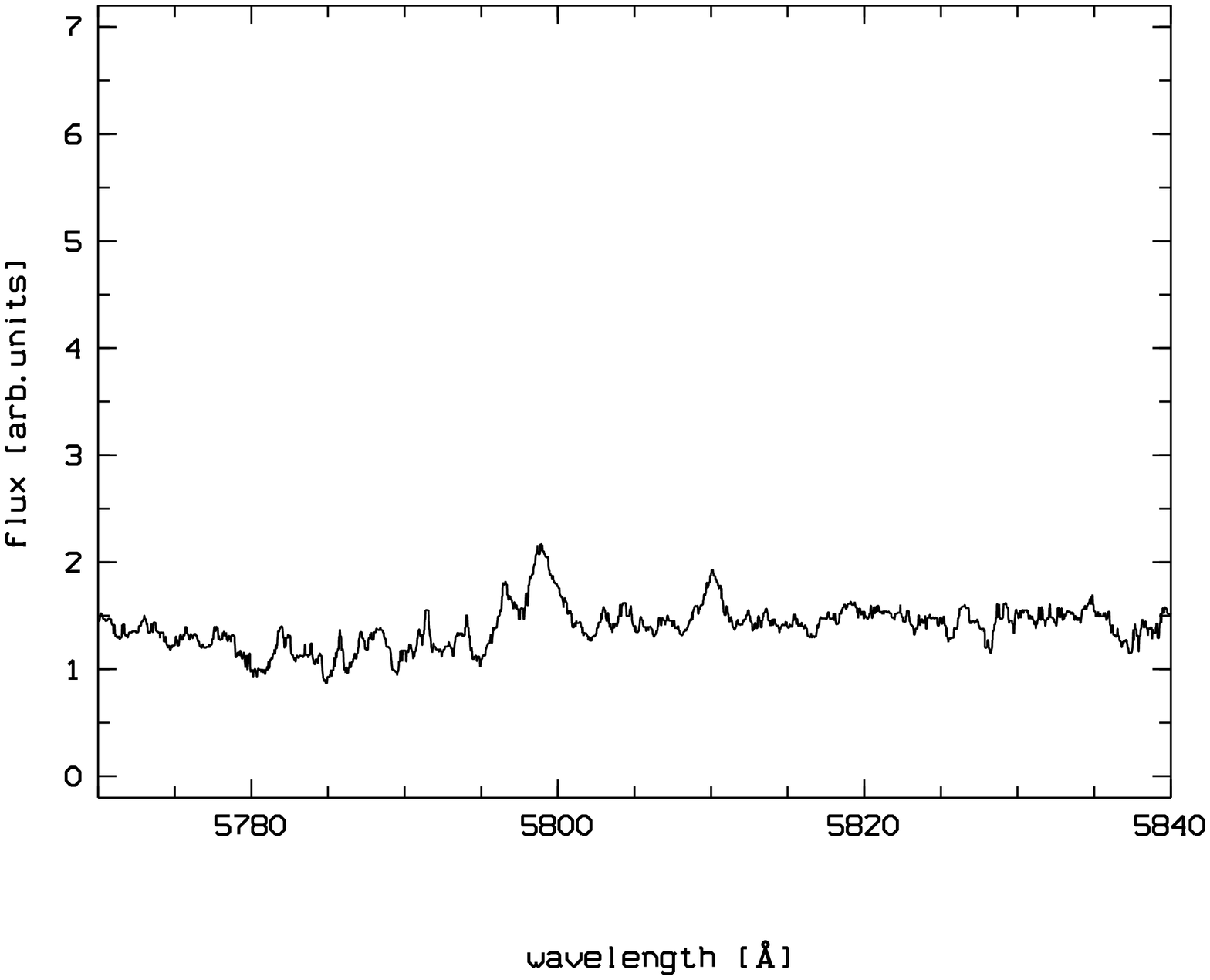}}
      \vspace{1mm}
    \hbox{\includegraphics[width=50mm]{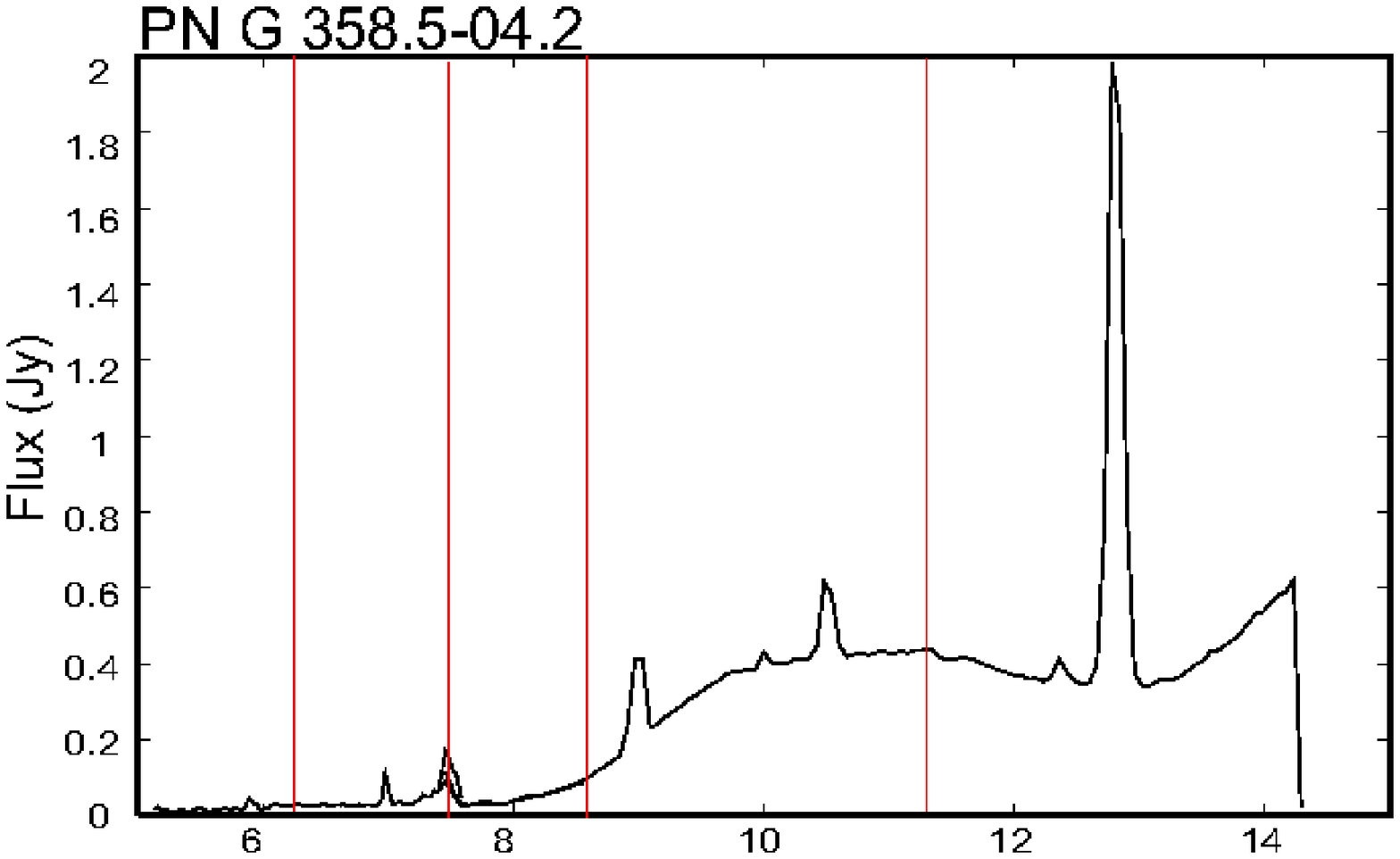}
          \hspace{3mm}
          \includegraphics[width=40mm]{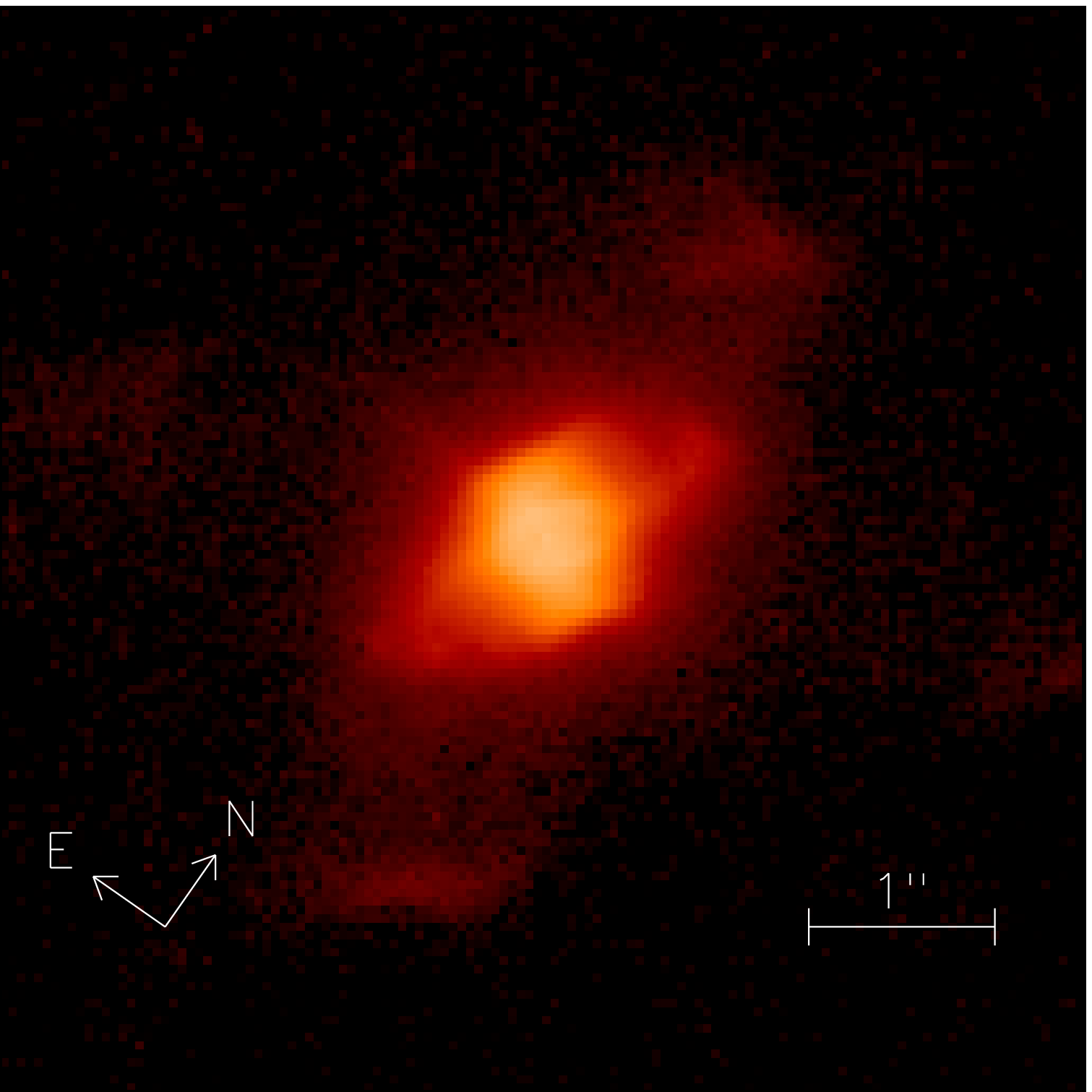}
          \hspace{6mm}
          \includegraphics[width=55mm]{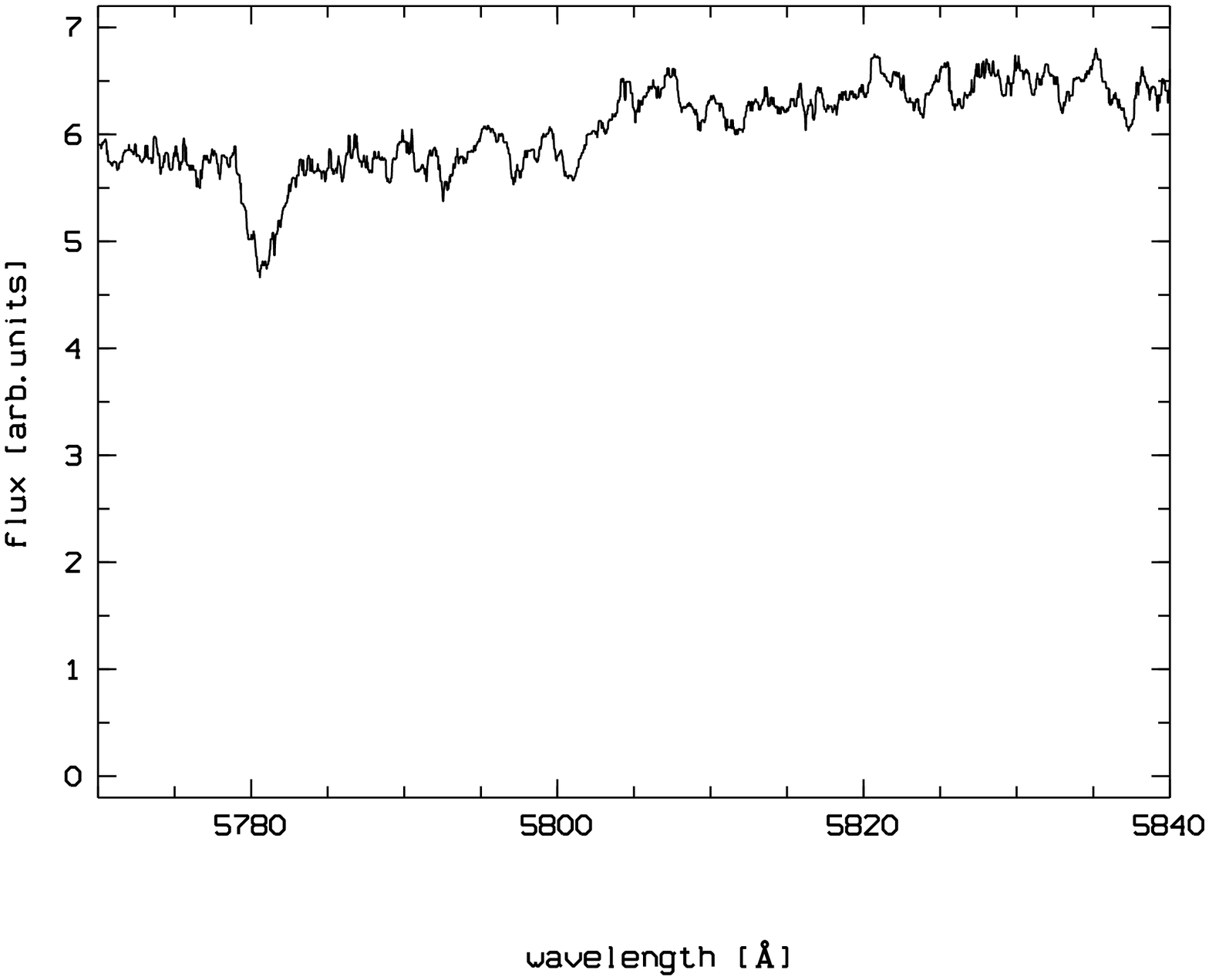}}
      \vspace{1mm}
       \hbox{\includegraphics[width=50mm]{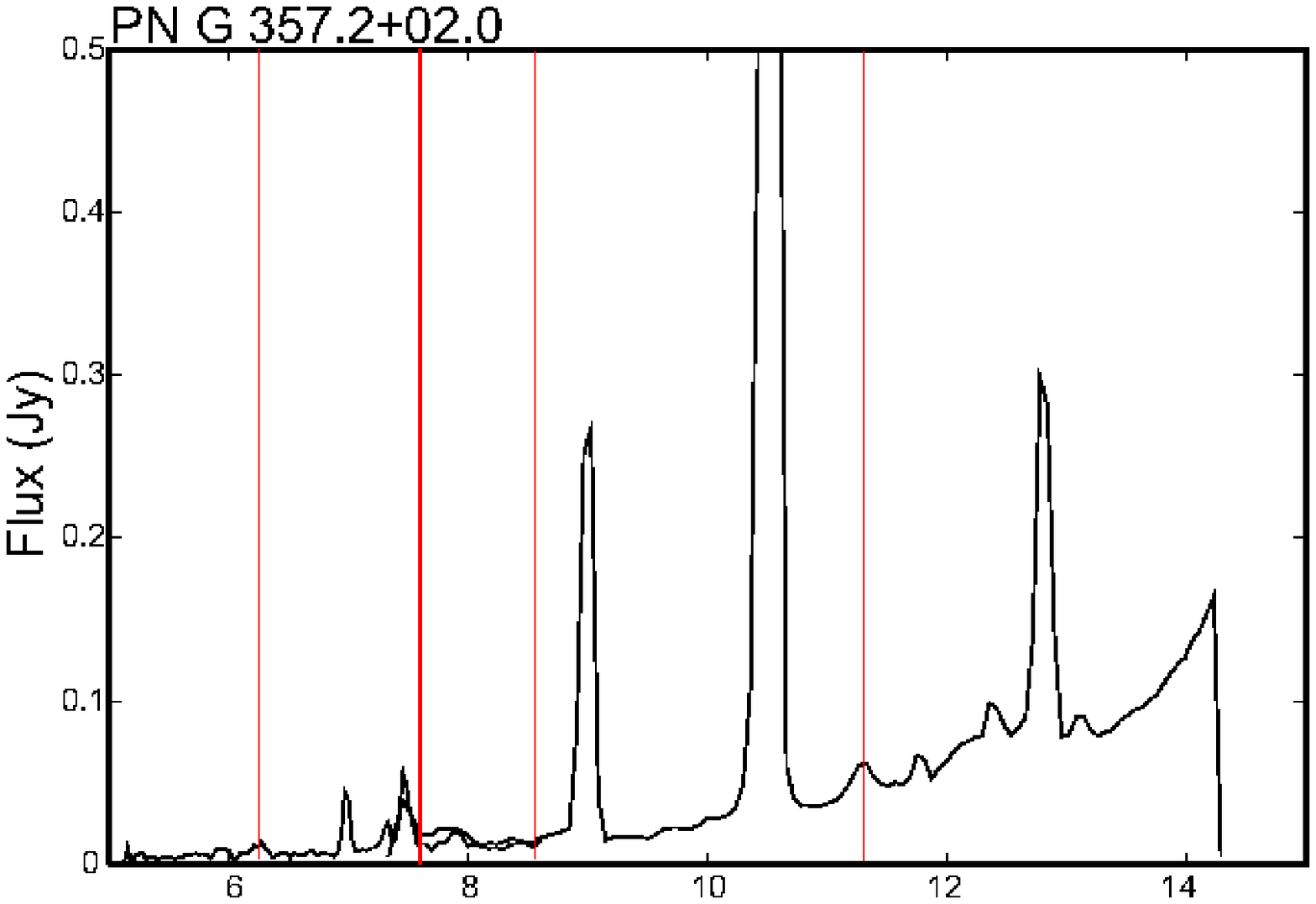}
          \hspace{3mm}
          \includegraphics[width=40mm]{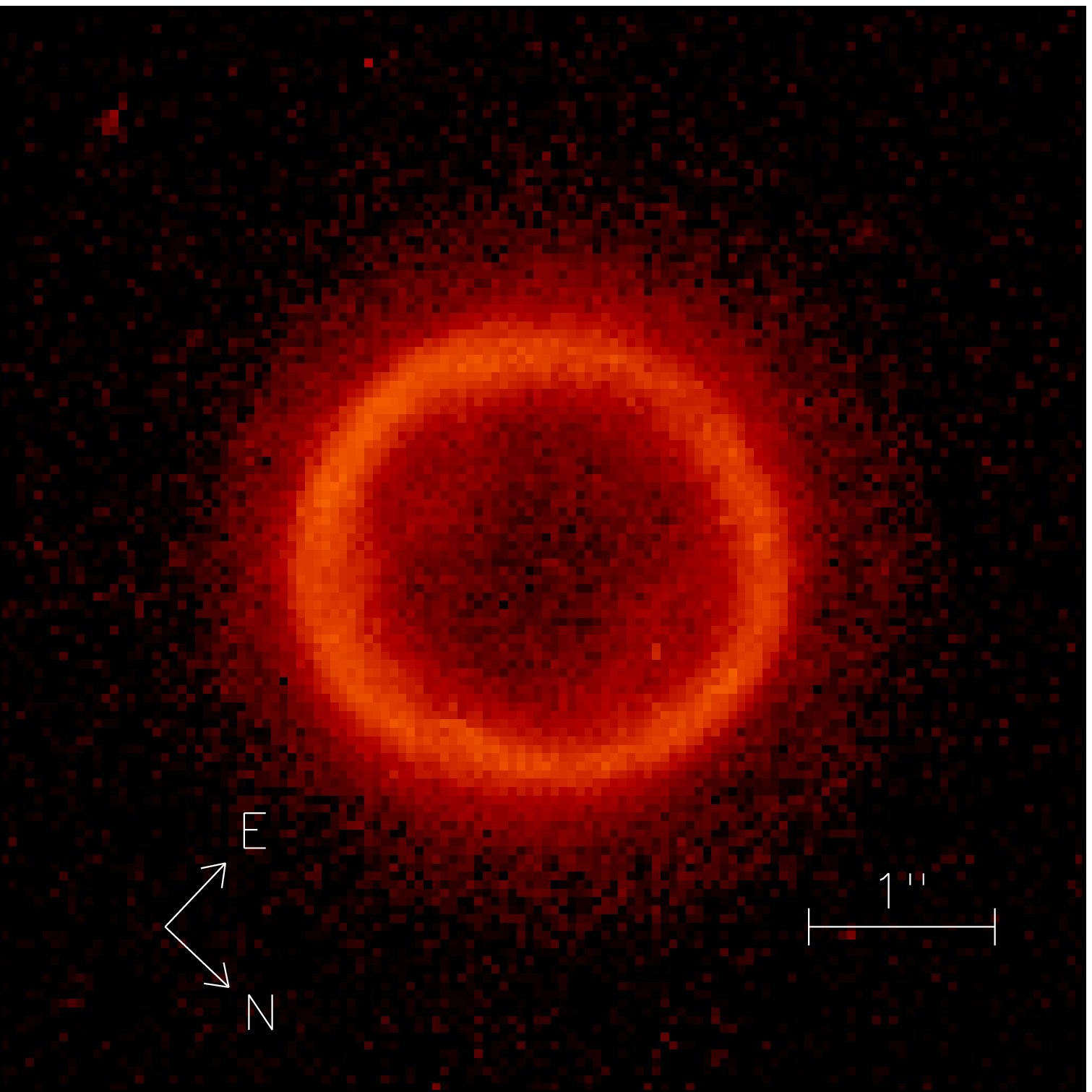}
          \hspace{6mm}
          \includegraphics[width=55mm]{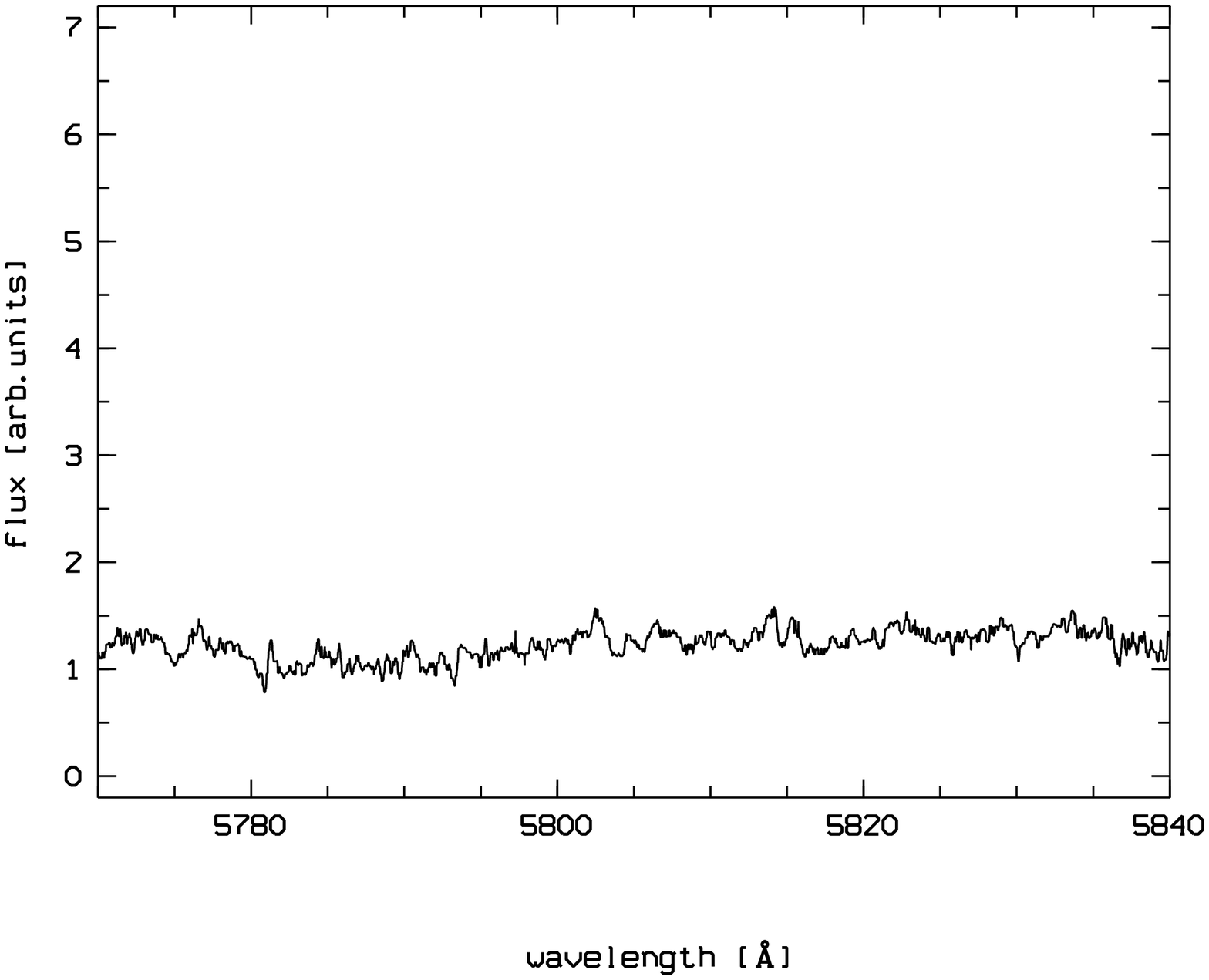}}
        \vspace{1mm} 
   \hbox{\includegraphics[width=50mm]{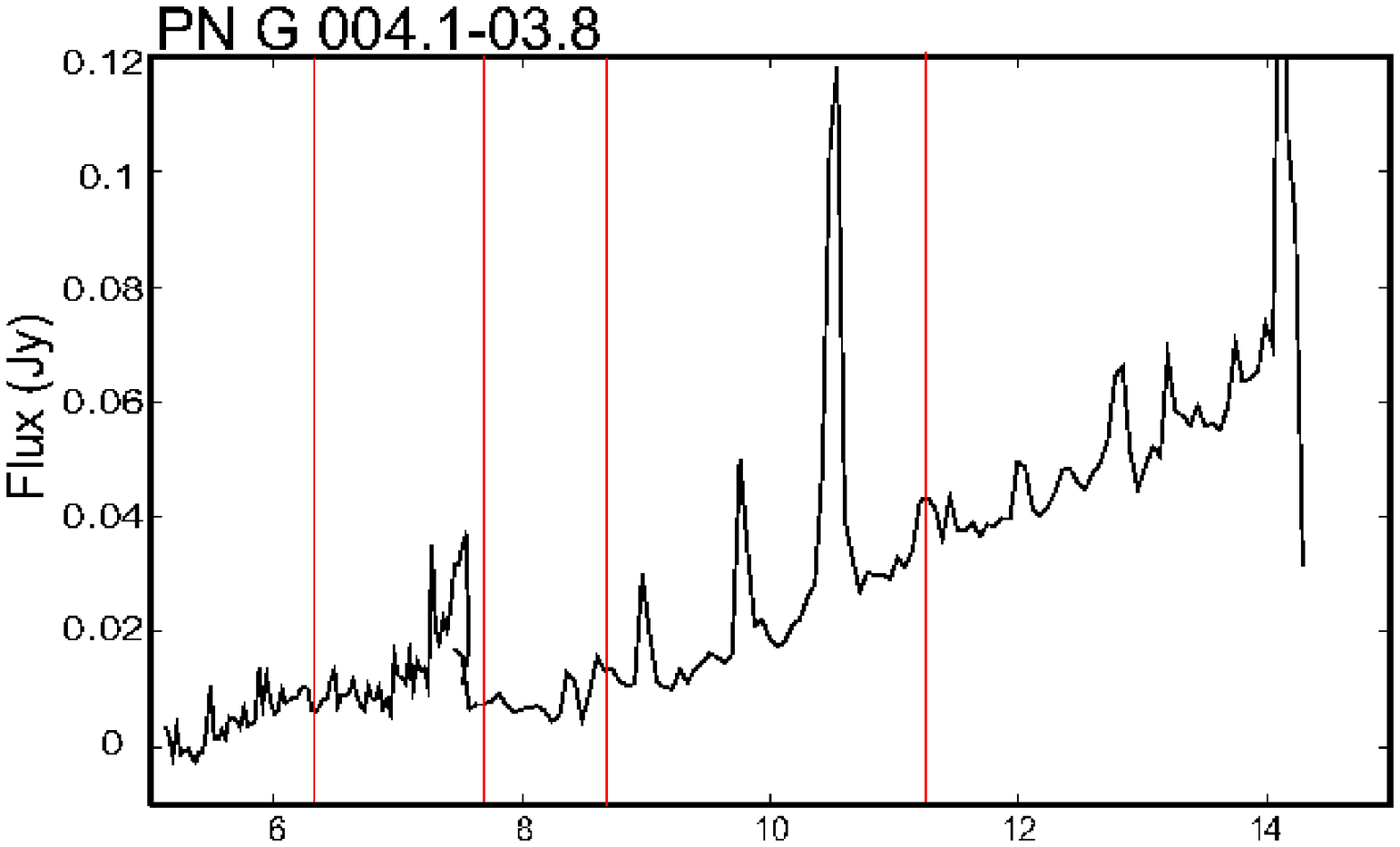}
          \hspace{3mm}
          \includegraphics[width=40mm]{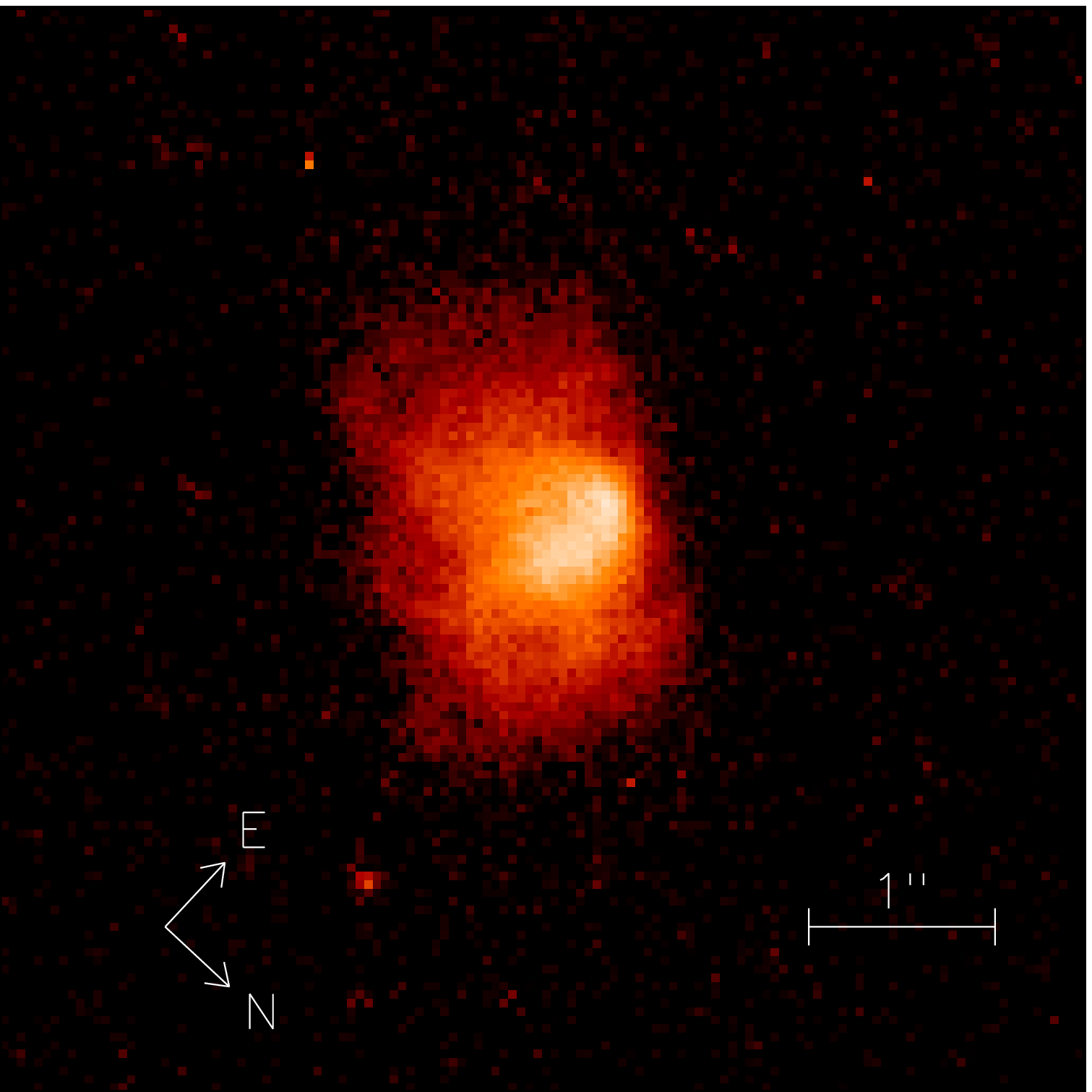}
          \hspace{6mm}
          \includegraphics[width=55mm]{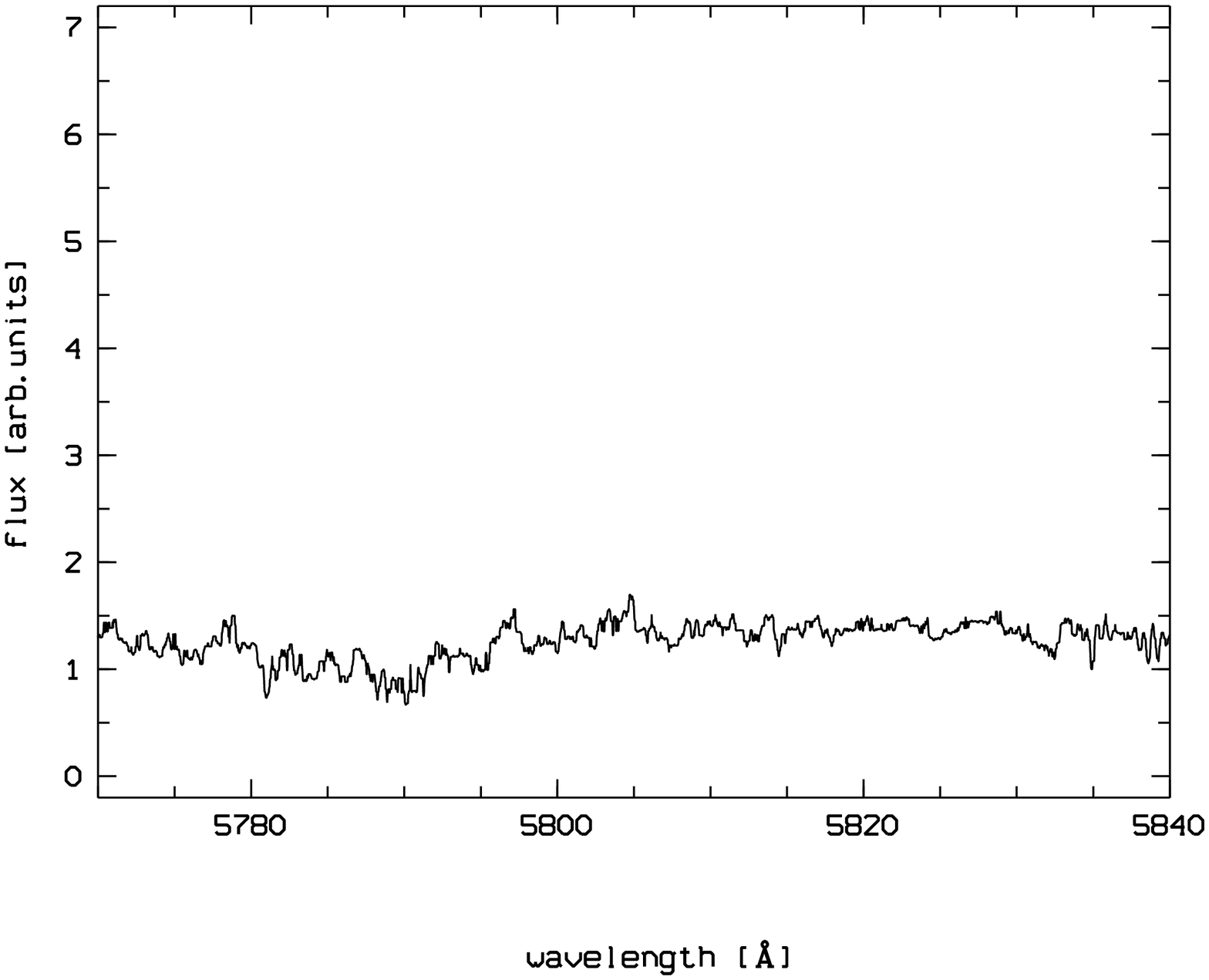}}
      \vspace{1mm}
      \hbox{\includegraphics[width=50mm]{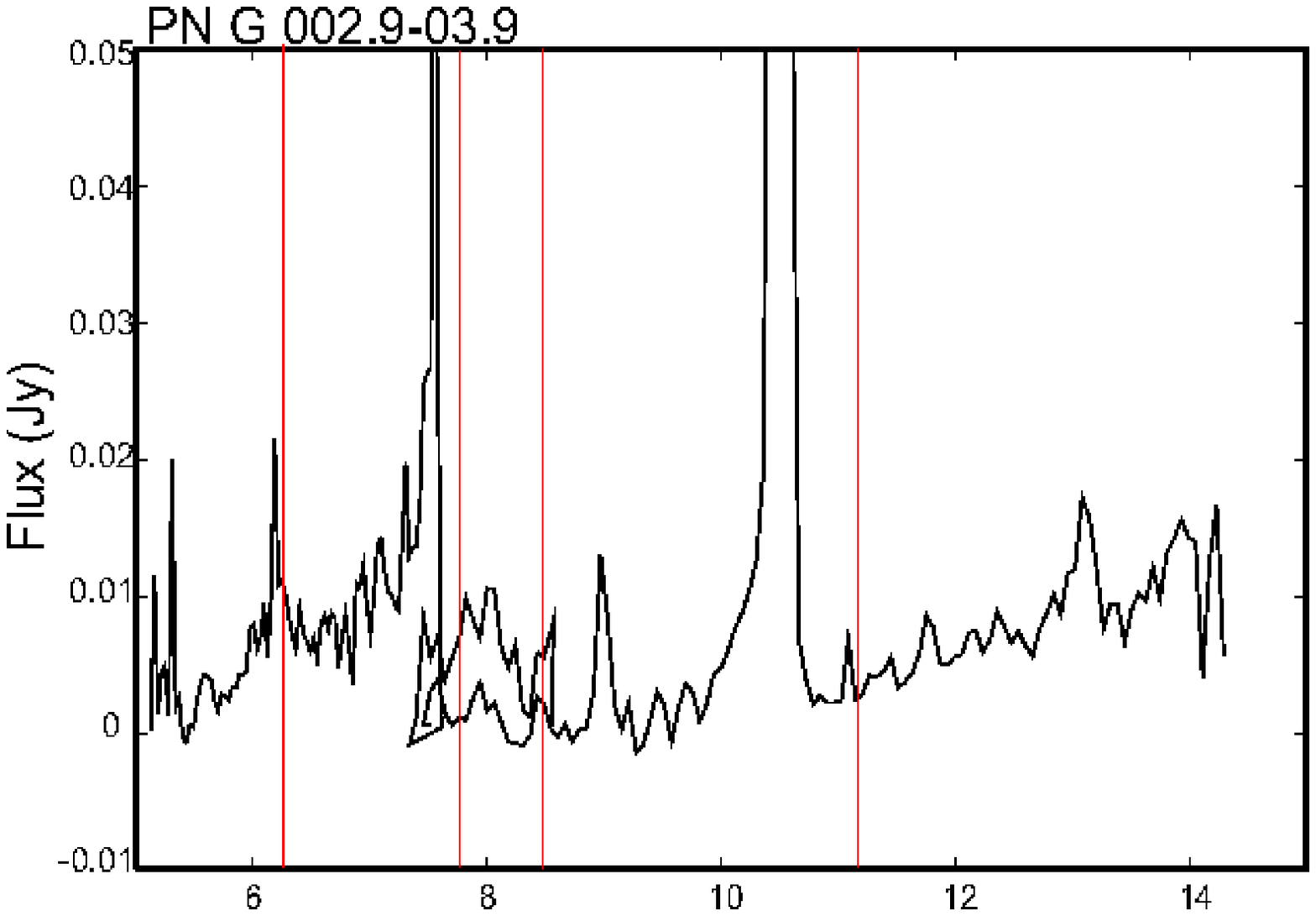}
          \hspace{3mm}
          \includegraphics[width=40mm]{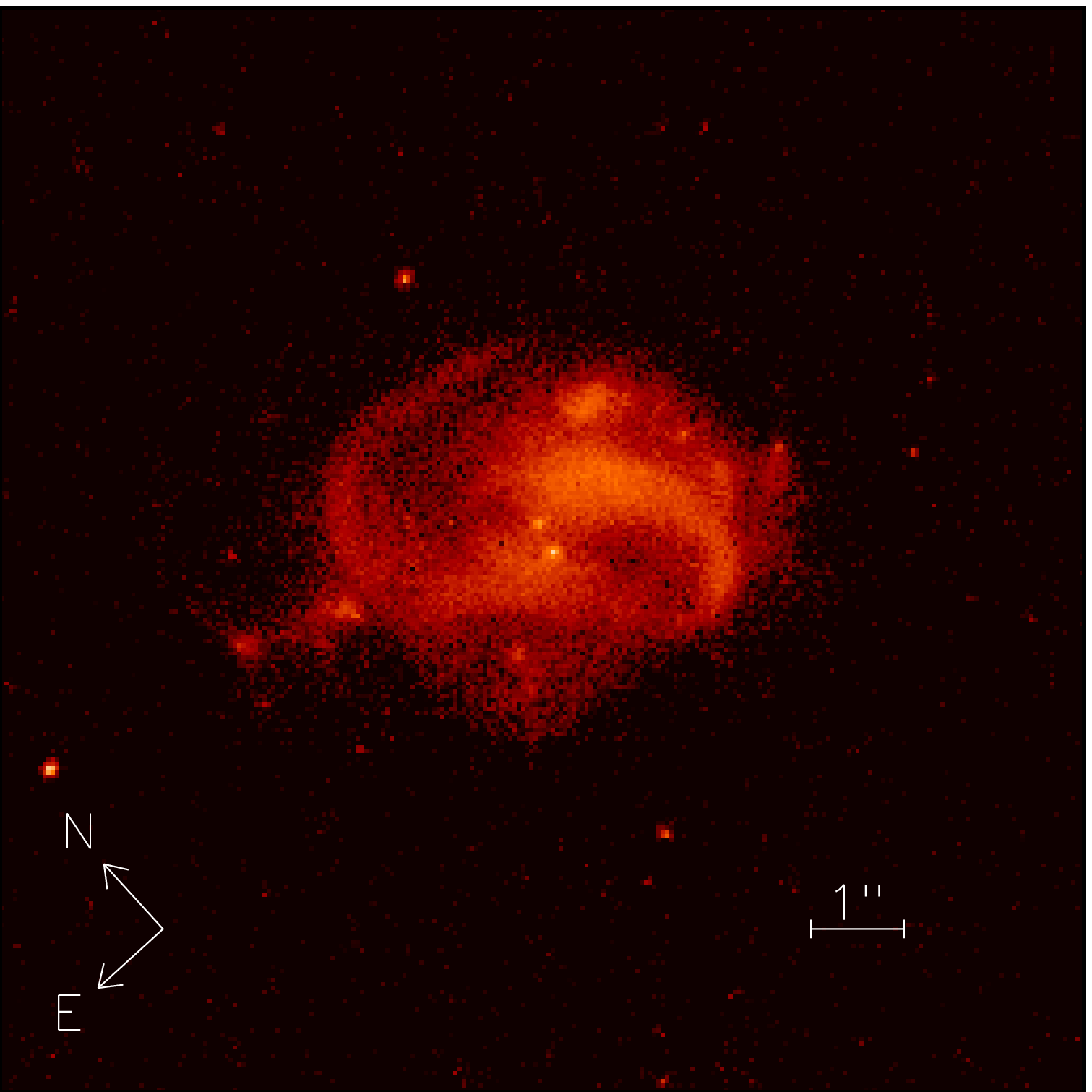}
          \hspace{6mm}
          \includegraphics[width=55mm]{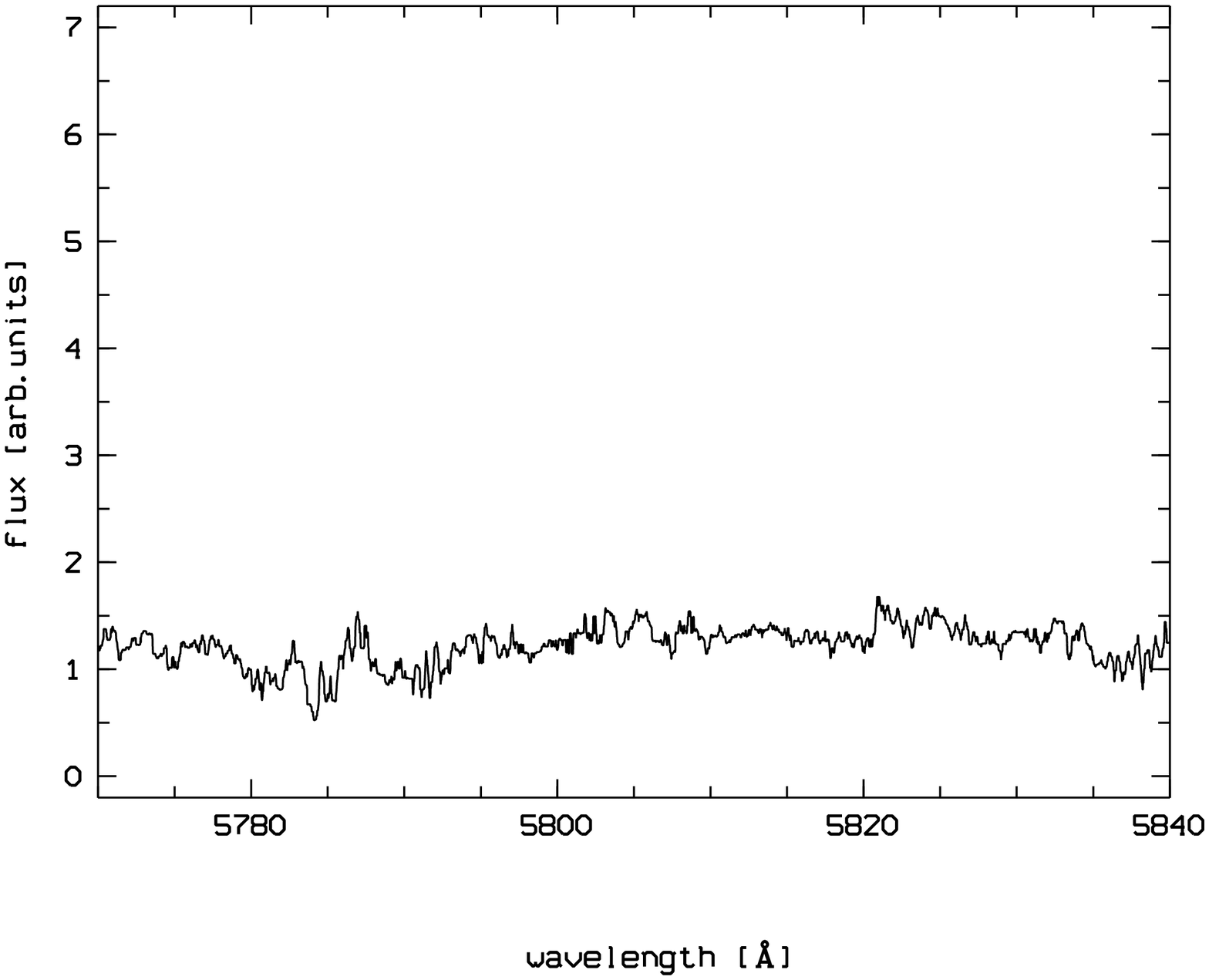}}
     \caption{(cont.) Correlation image showing in the first column the IR
       Spitzer spectrum, showing the short wavelength region, the
       vertical red lines show the PAH bands. In the second column is the
       corresponding HST image and in the third column we show a part
       of the UVES spectrum. }
     \label{correl3}
\end{figure*}

\begin{figure*}
    \hbox{\includegraphics[width=50mm]{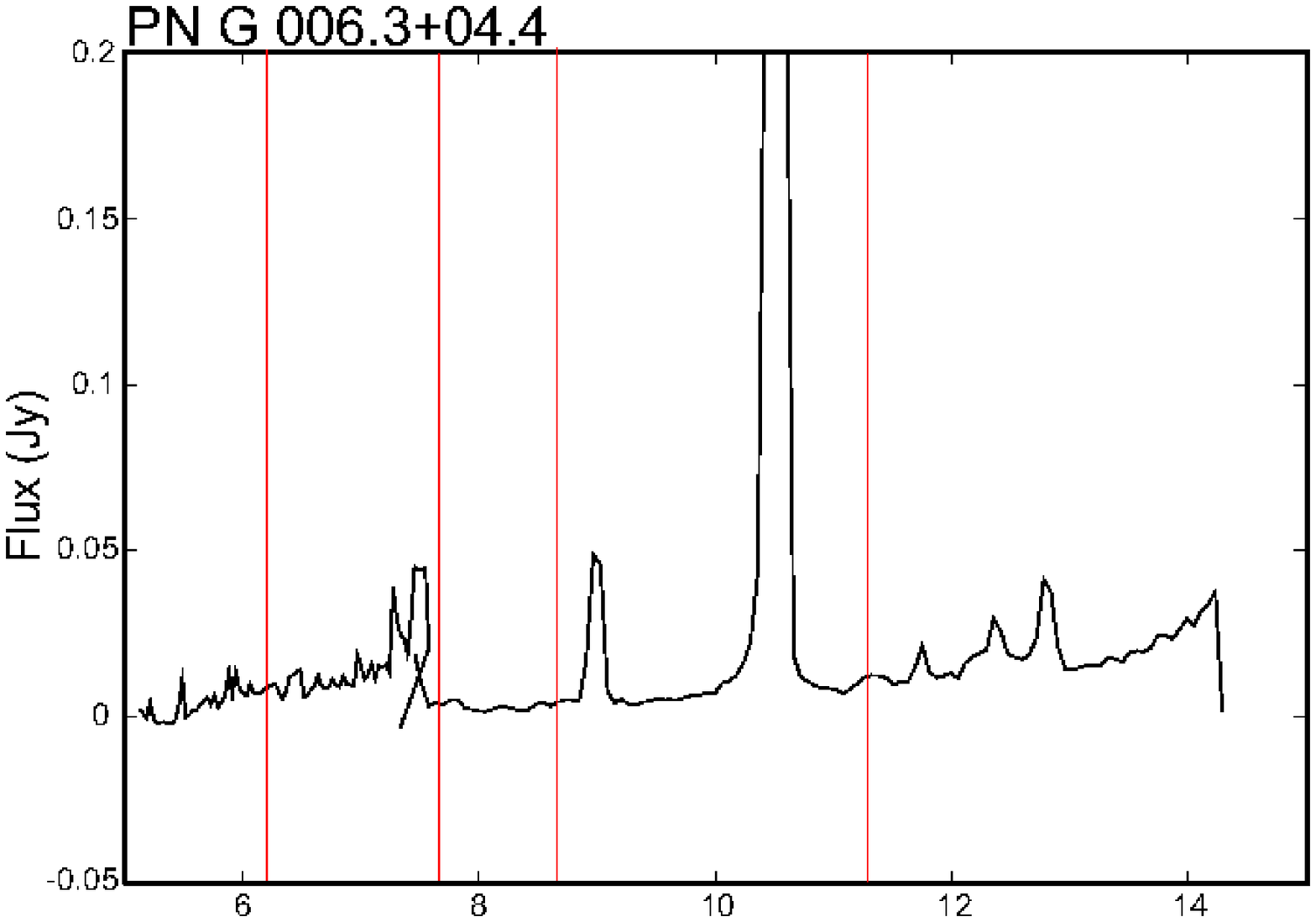}
          \hspace{3mm}
          \includegraphics[width=40mm]{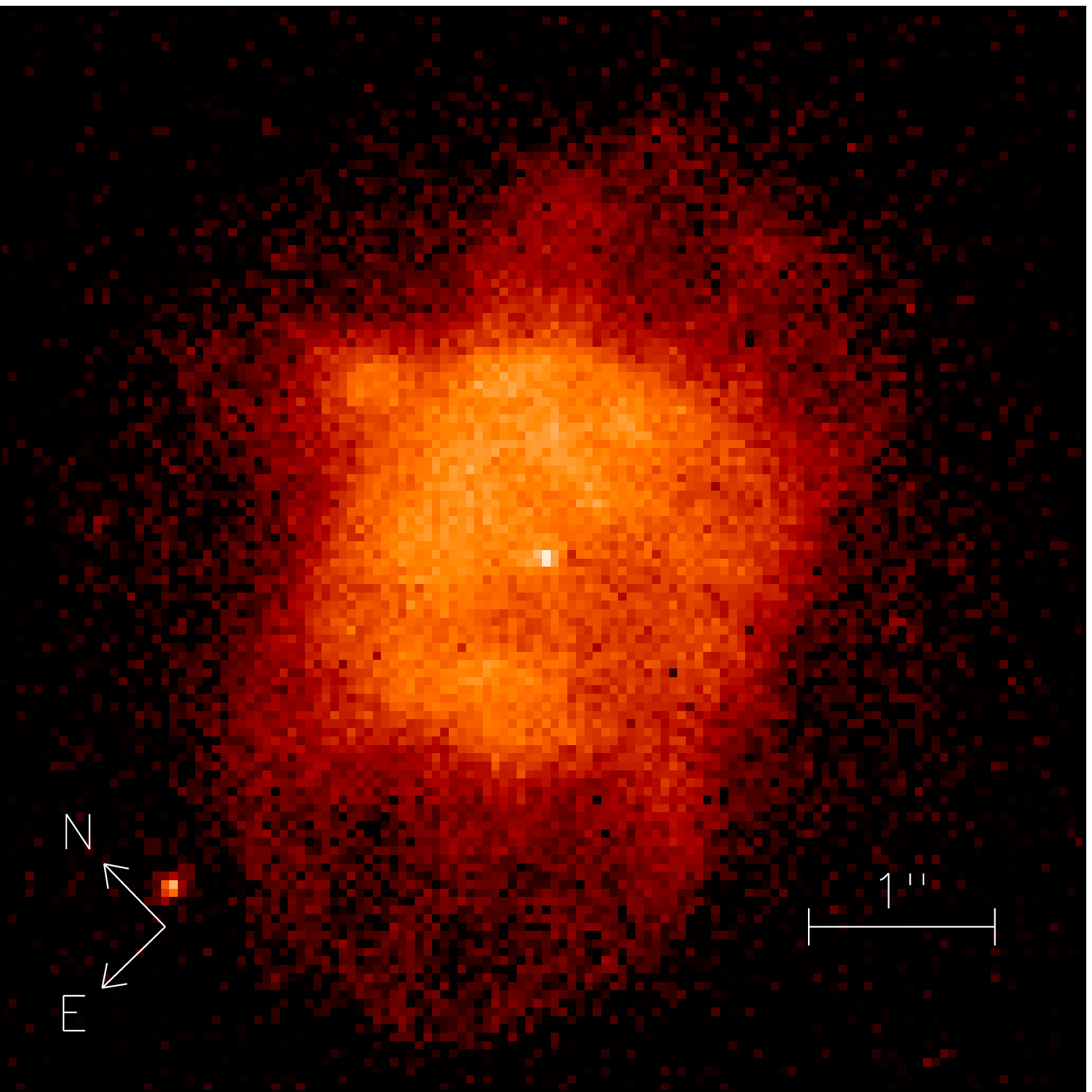}
          \hspace{6mm}
          \includegraphics[width=55mm]{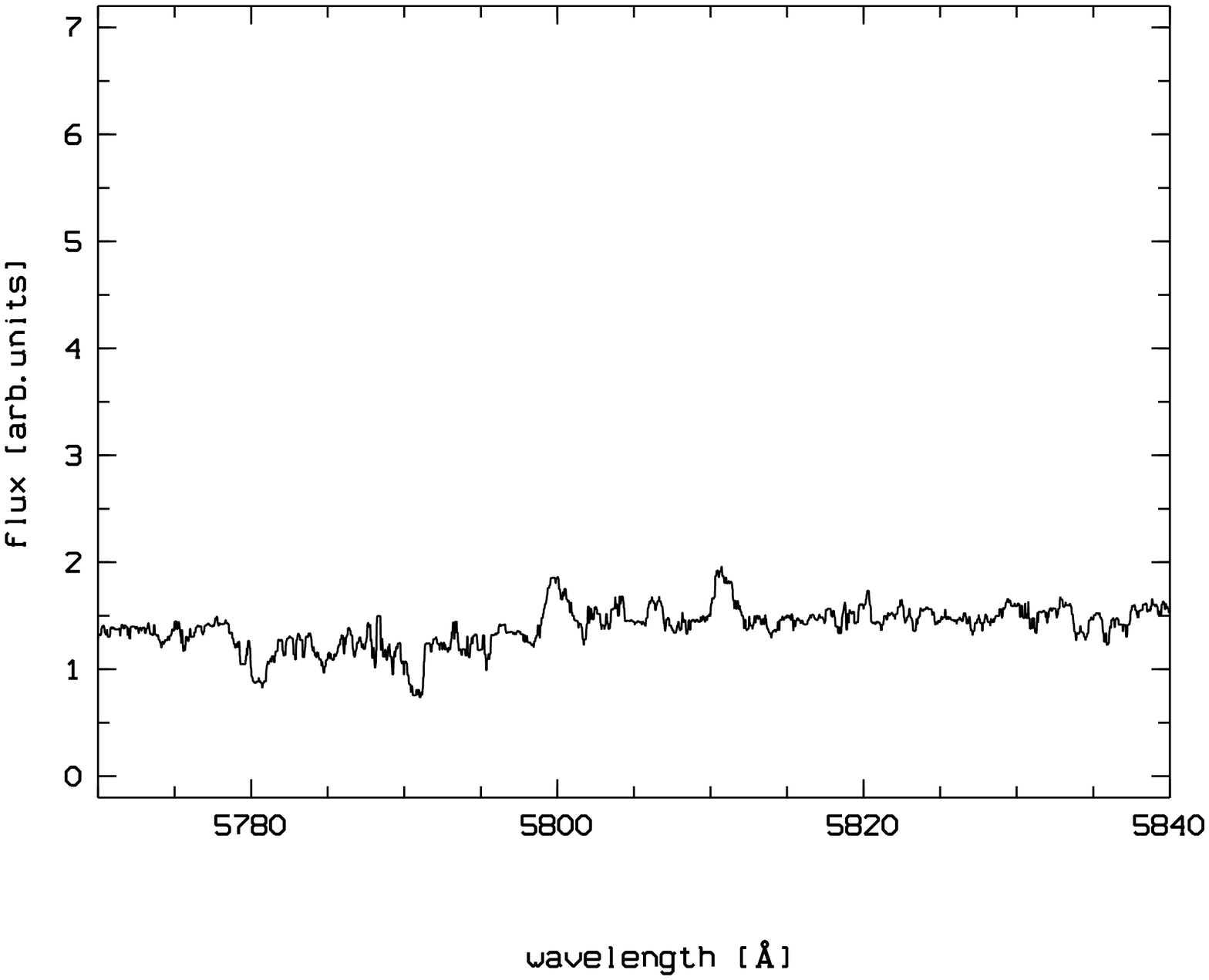}}
      \vspace{1mm}
     \hbox{\includegraphics[width=50mm]{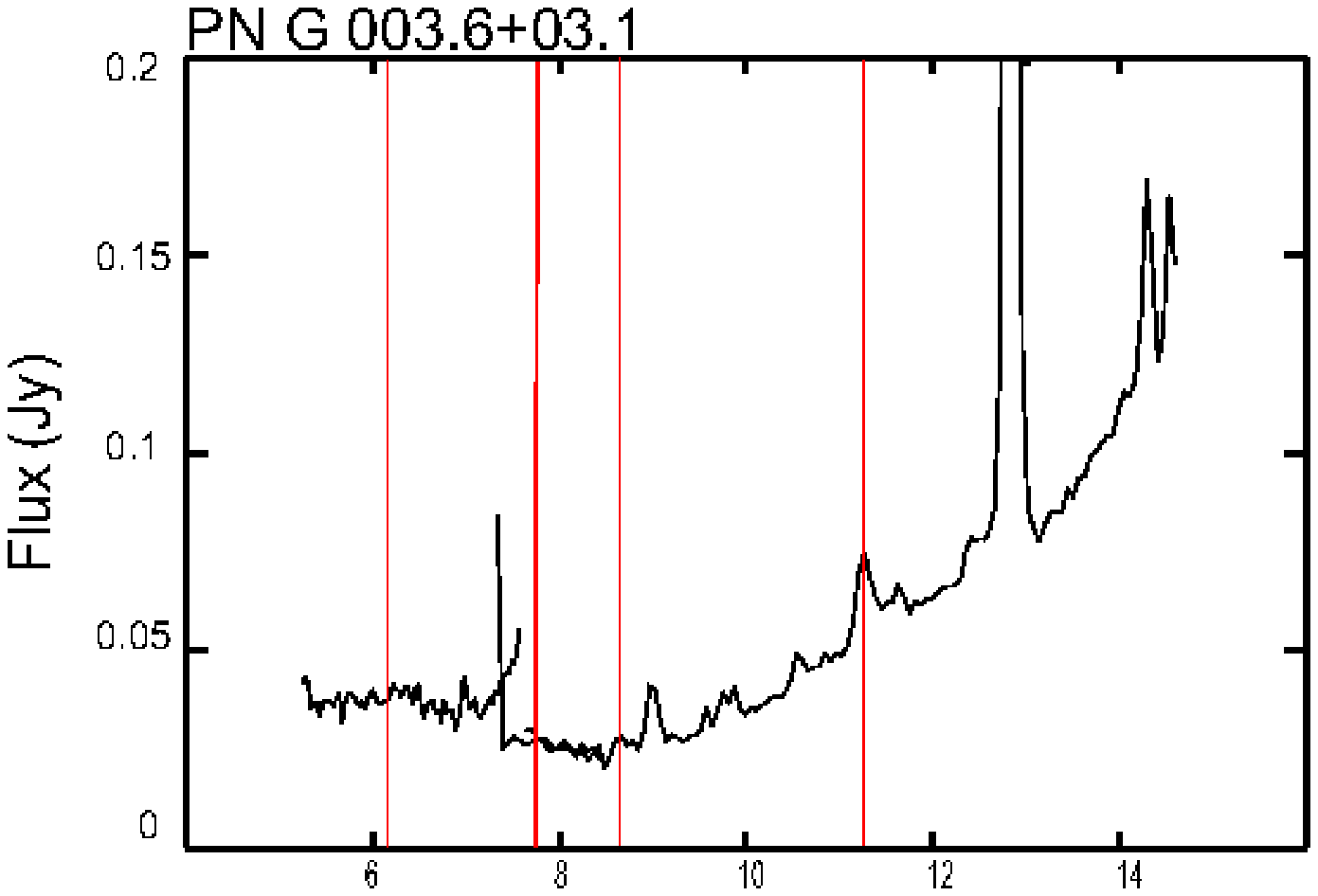}
          \hspace{3mm}
          \includegraphics[width=40mm]{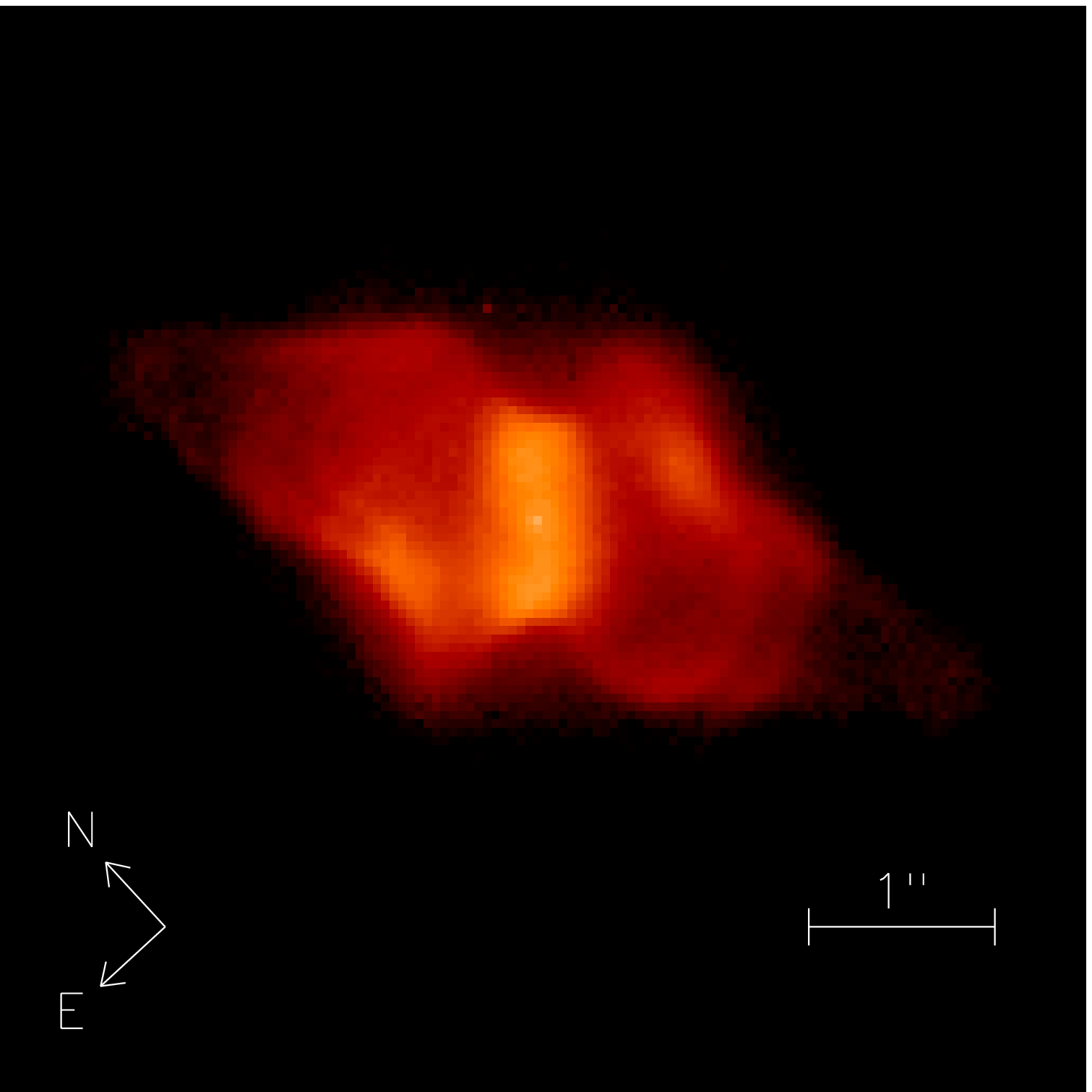}
          \hspace{3mm}
          \includegraphics[width=55mm]{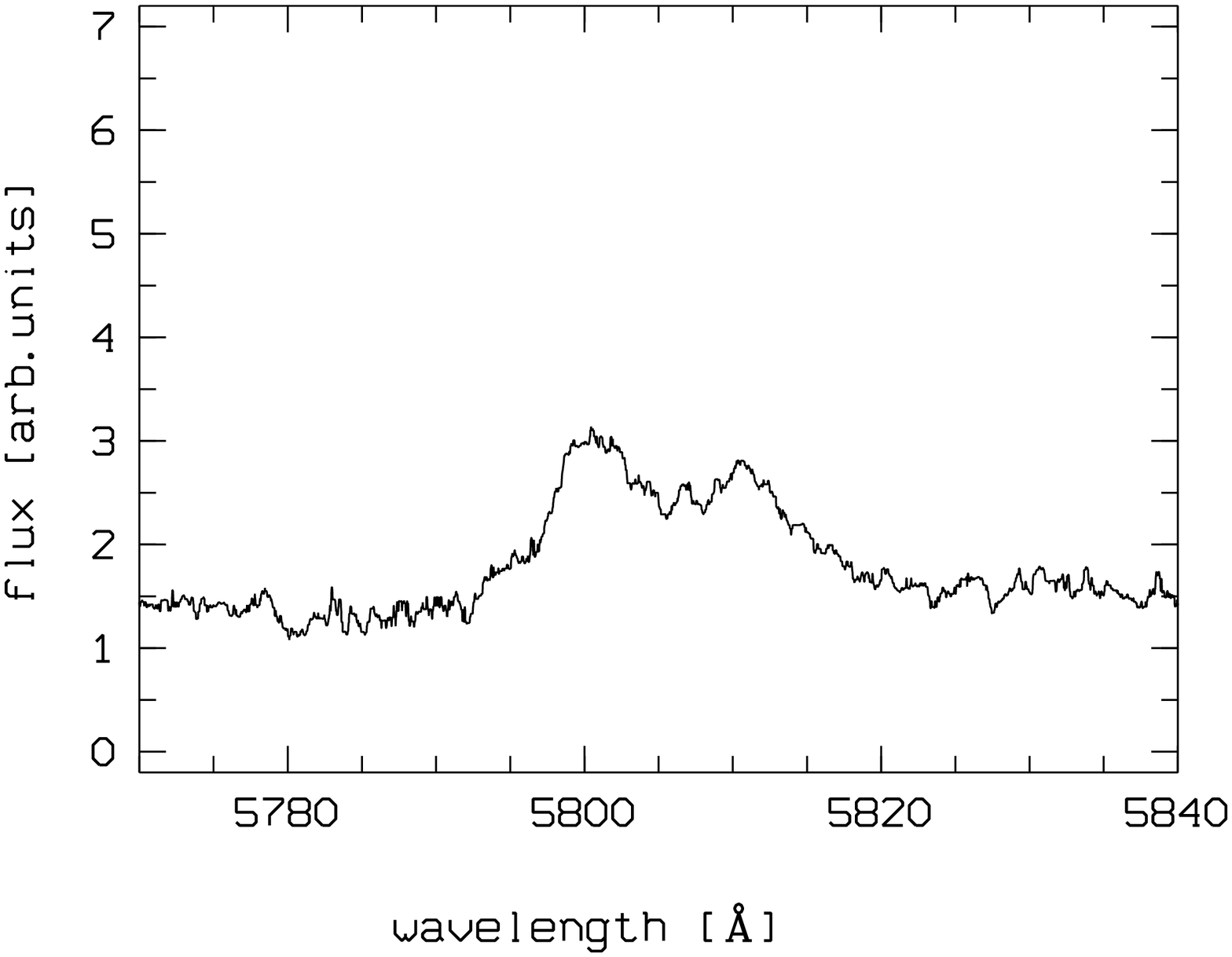}}
      \vspace{1mm}
      \hbox{\includegraphics[width=50mm]{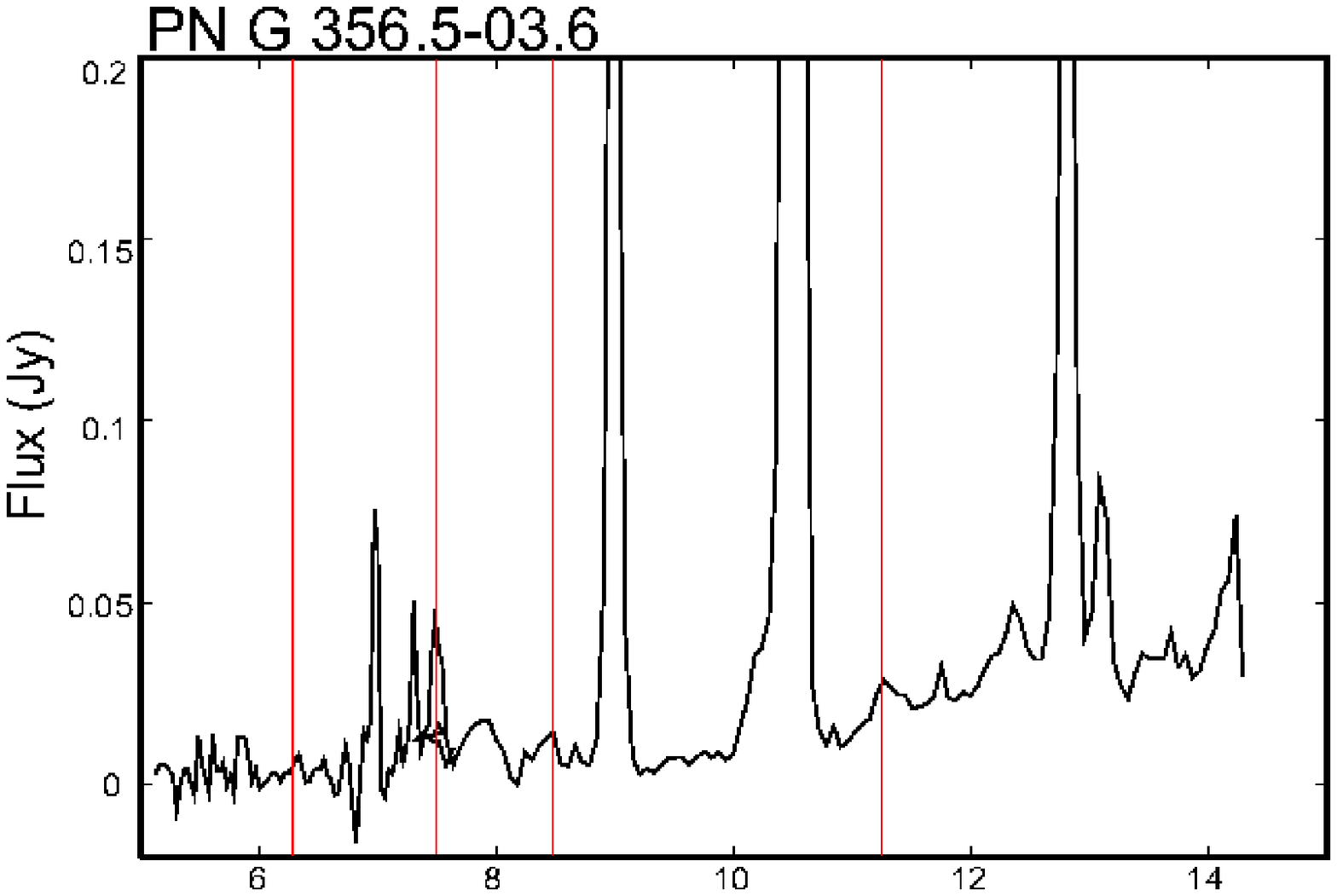}
          \hspace{3mm}
          \includegraphics[width=40mm]{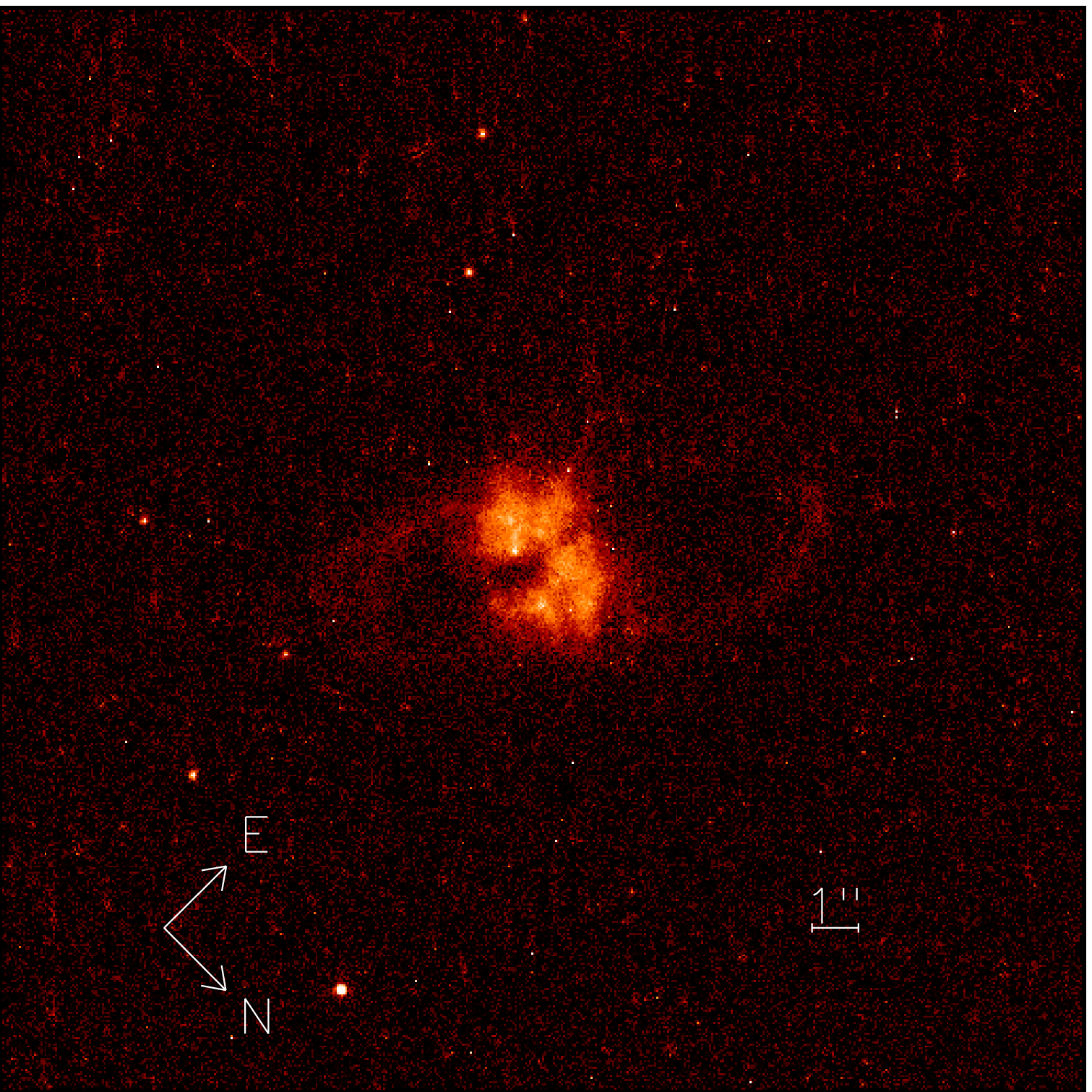}
          \hspace{6mm}
          \includegraphics[width=55mm]{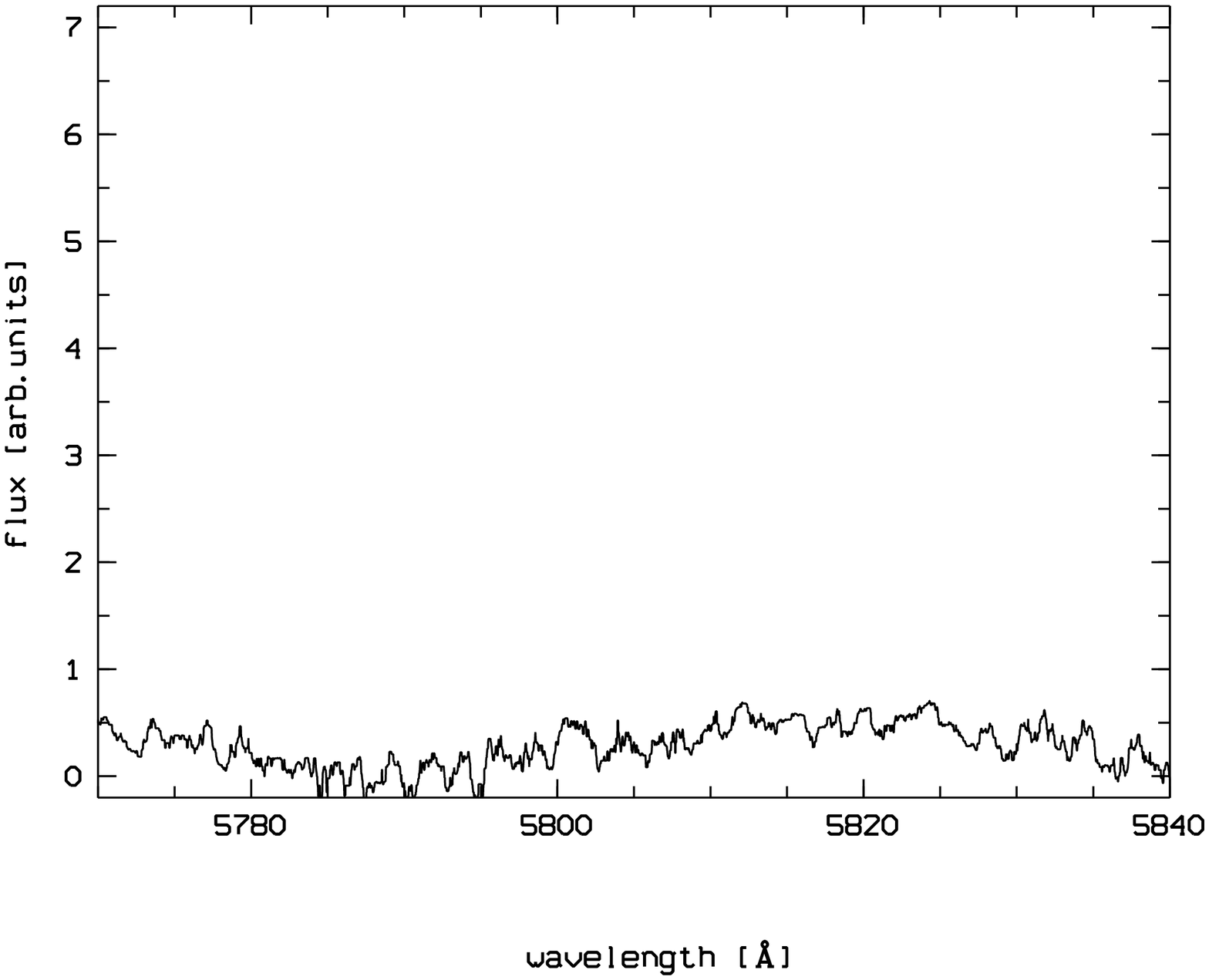}}
      \vspace{1mm}
     \hbox{\includegraphics[width=50mm]{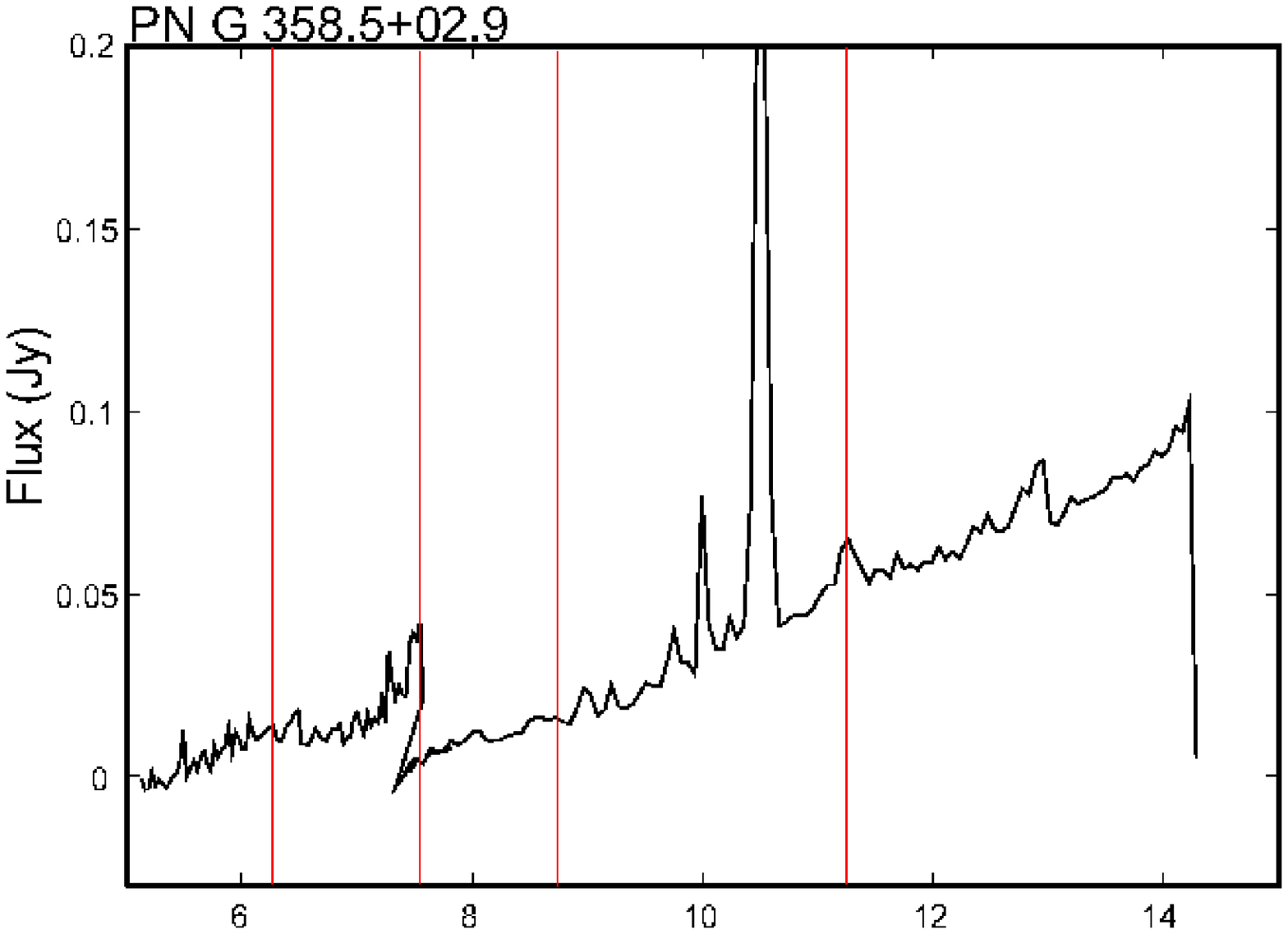}
          \hspace{3mm}
          \includegraphics[width=40mm]{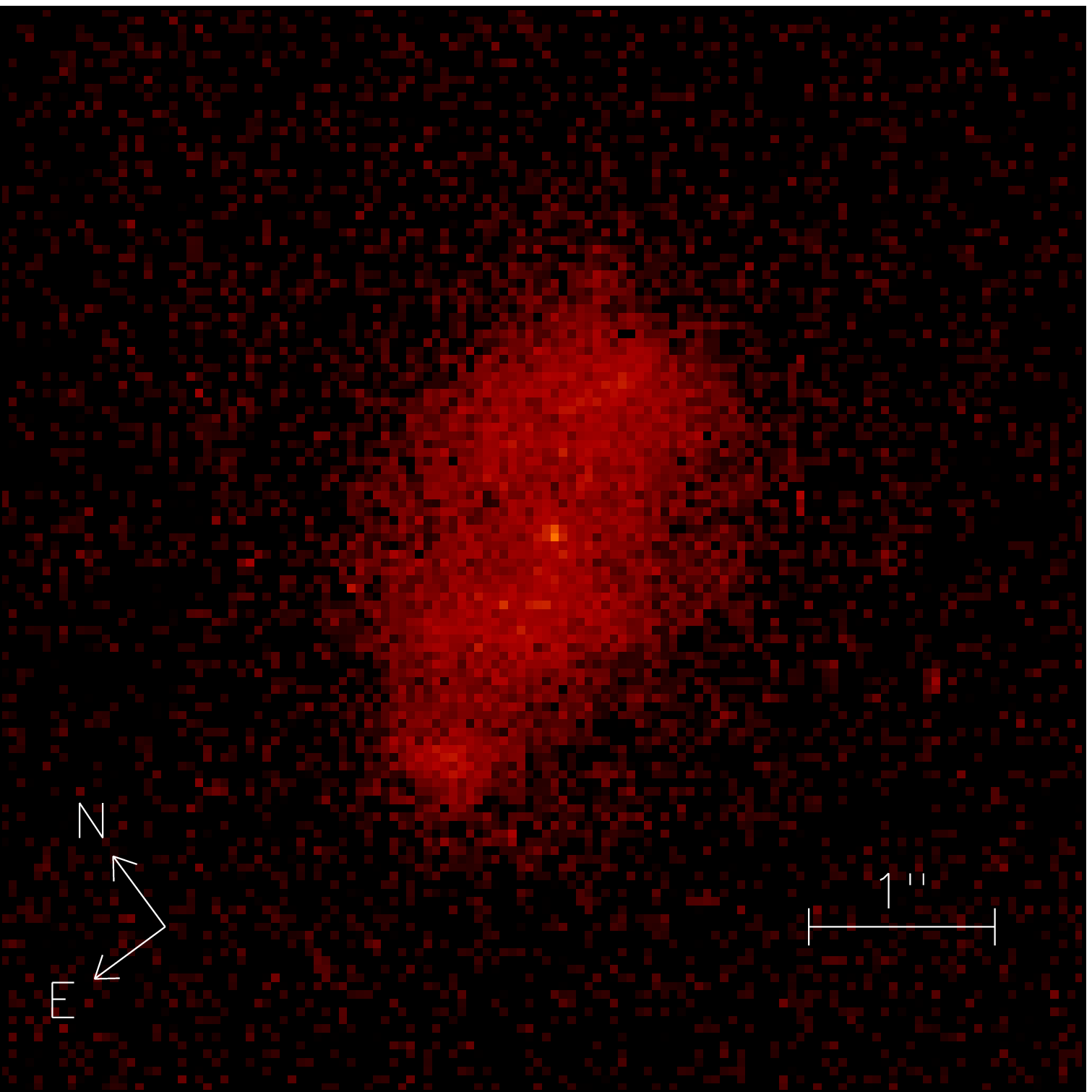}
          \hspace{6mm}
          \includegraphics[width=55mm]{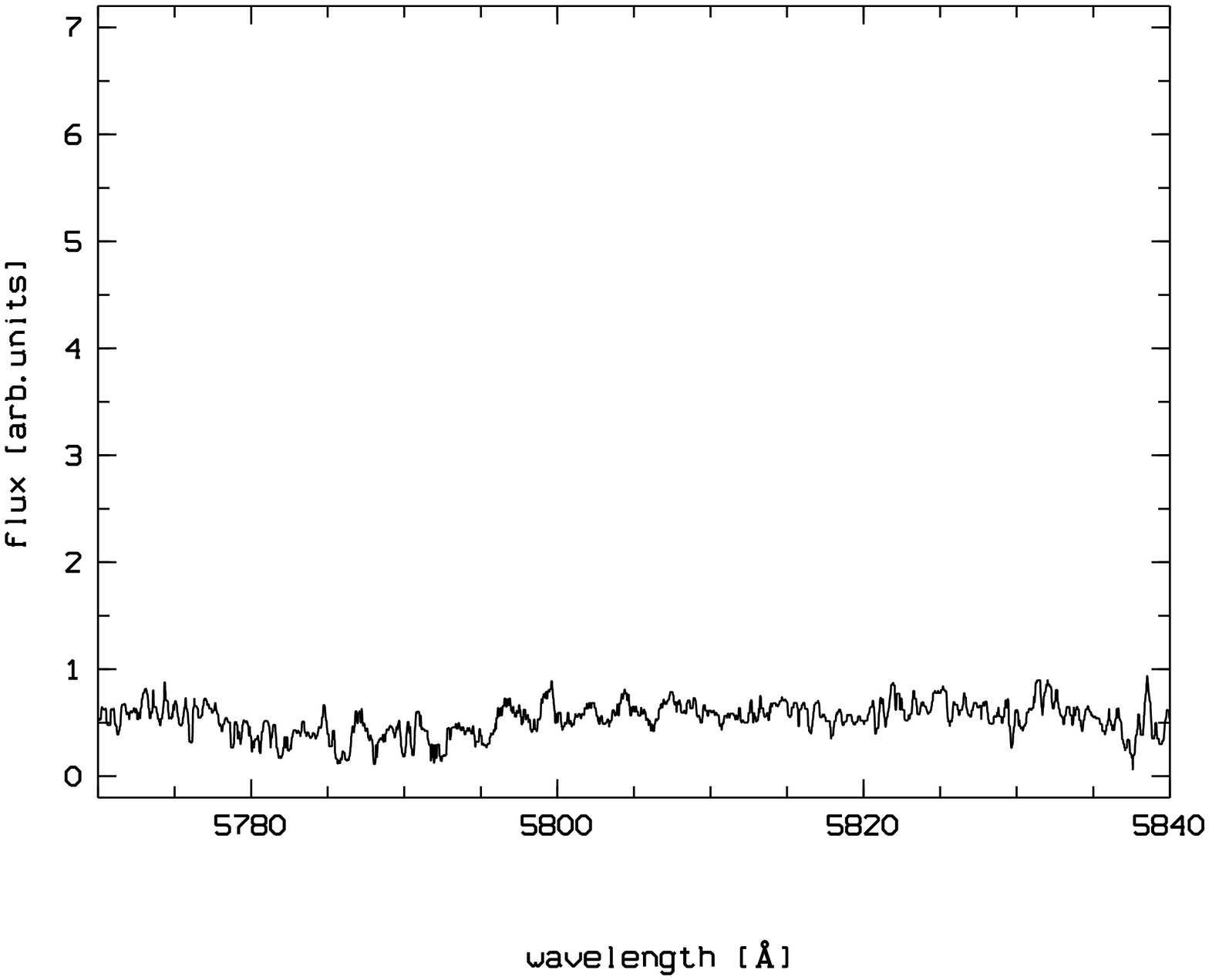}}
      \vspace{1mm}
       \hbox{\includegraphics[width=50mm]{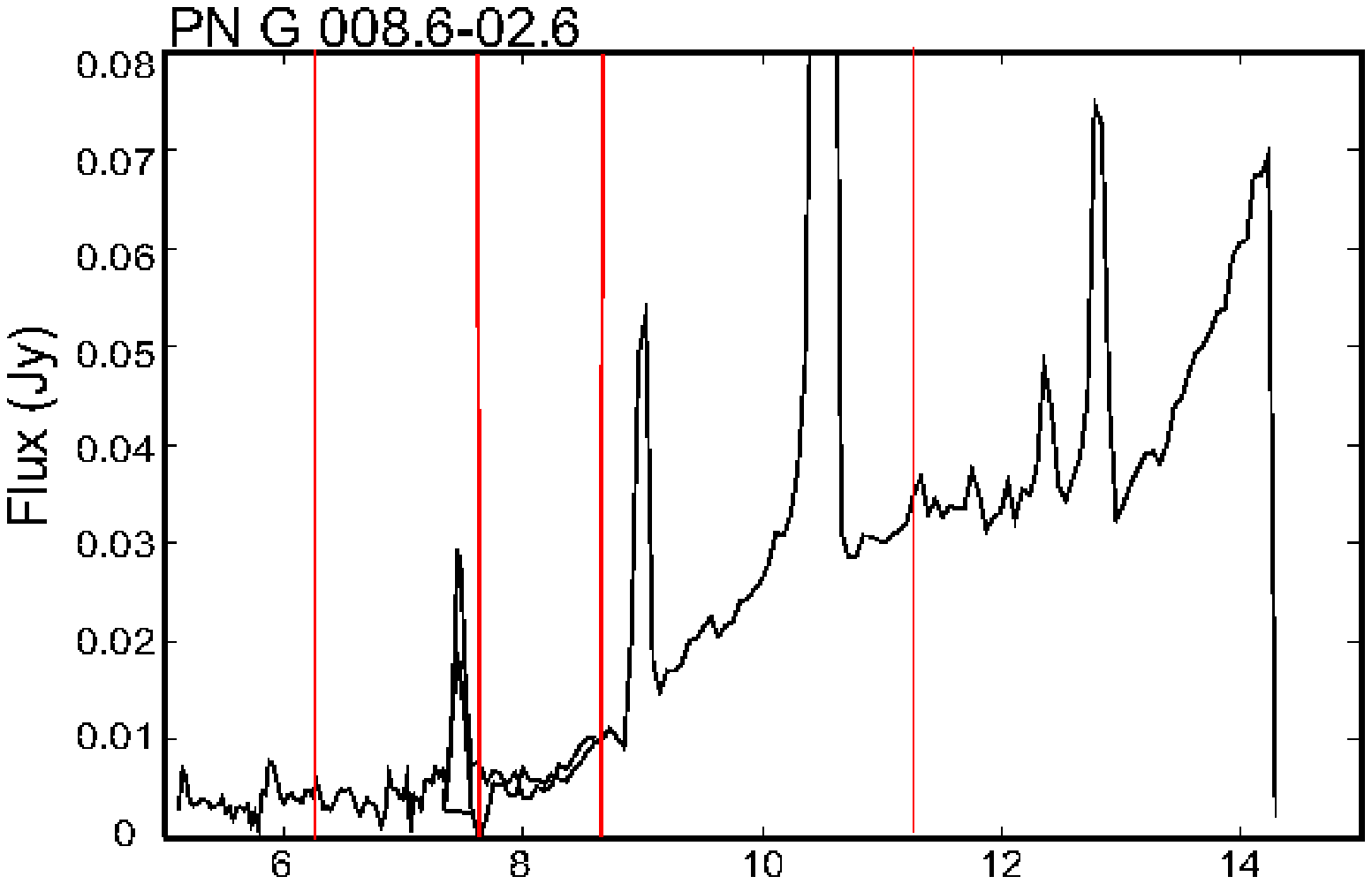}
          \hspace{3mm}
          \includegraphics[width=40mm]{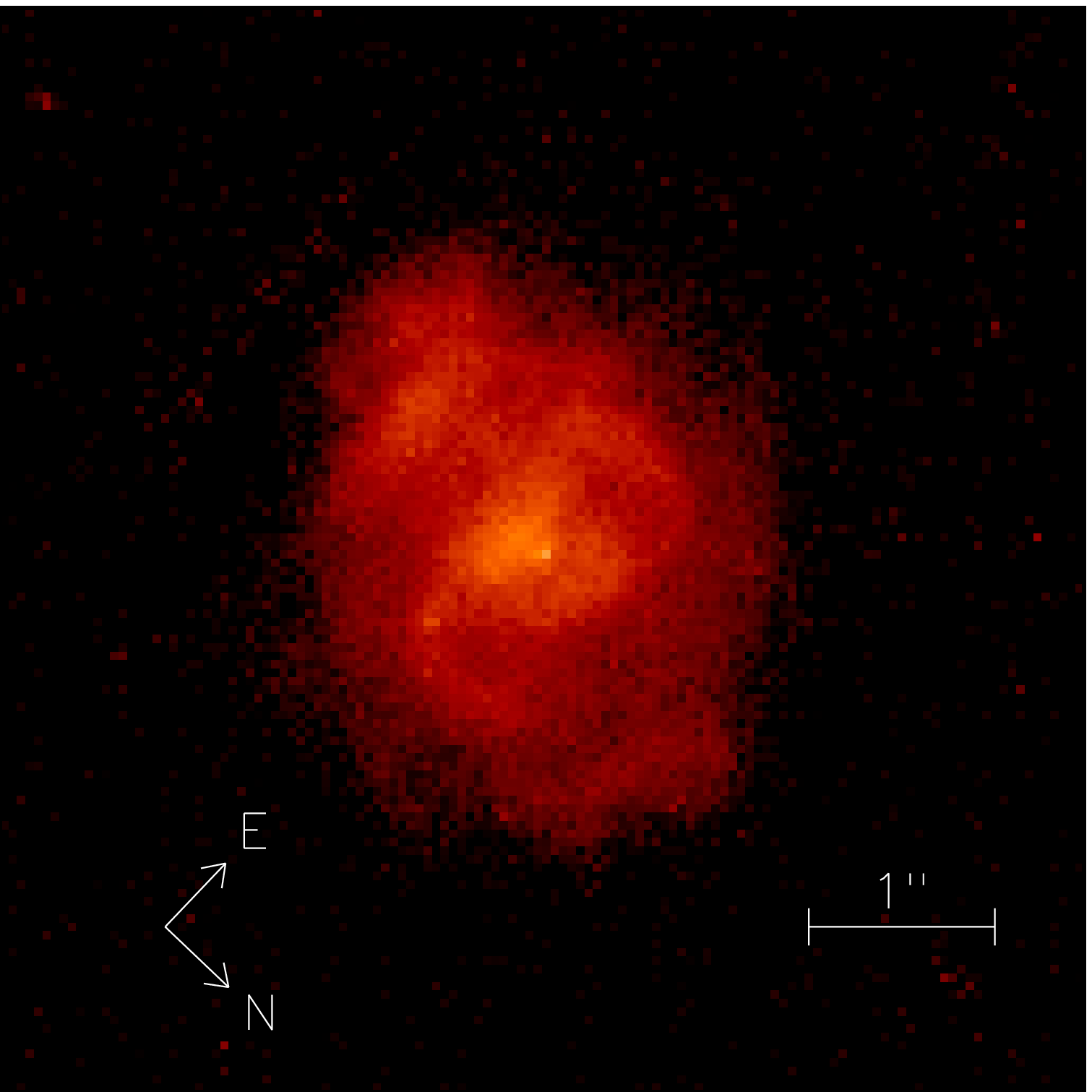}
          \hspace{6mm}
          \includegraphics[width=55mm]{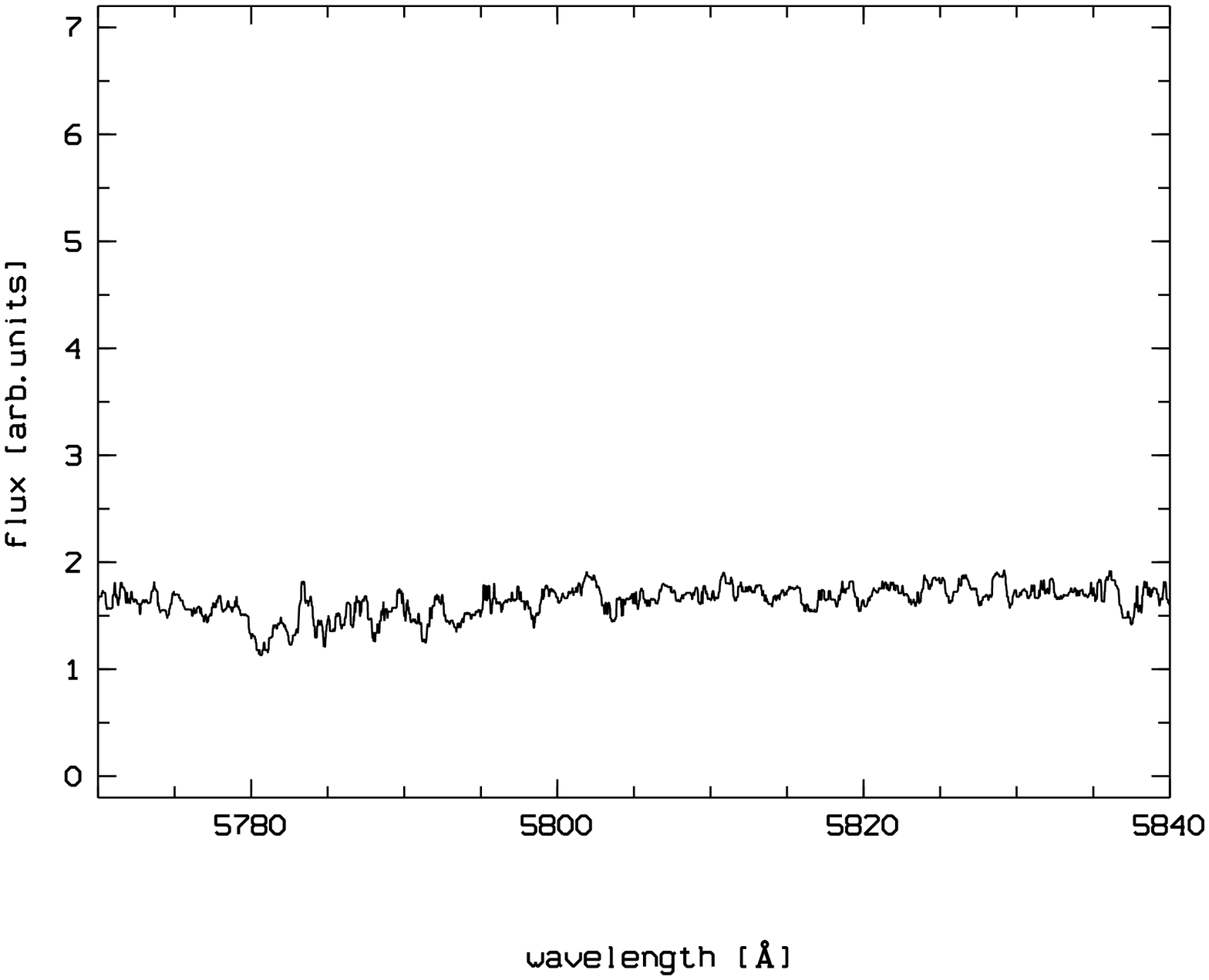}}
     \caption{(cont.) Correlation image showing in the first column the IR
       Spitzer spectrum, showing the short wavelength region, the
       vertical red lines show the PAH bands. In the second column is the
       corresponding HST image and in the third column we show a part
       of the UVES spectrum. }
     \label{correl4}
\end{figure*}

The most important aspect of Figs. \ref{correl1}, \ref{correl2},
\ref{correl3} and \ref{correl4} is the predominance of a very specific
morphology: a dense torus with bipolar or multipolar flows. This appears to
correlate well with the PAH band strength. In a few cases the torus appears to
be seen pole-on (PN G358.7+05.2 and PN G002.8+01.7), but in other cases the
structure is easily visible. The confining tori have a range of sizes but are
always well defined. The objects without 7.7$\mu$m feature, towards the end of the
sequence, tend to show less structured morphologies. The clearest exceptions to
the rule are PN G003.6+03.1 which has no 7.7$\mu$m detection (and has a
relatively large torus) and PN G003.1+03.4 which does not show elongated
lobes, possibly because of viewing angle. We note that one of the three
silicate objects (PN G007.5+04.3) is unresolved by HST and is extremely
 compact.

This type of morphology is not the most common in the Galaxy -- most PNe are
round or elliptical.  In the IAC morphological catalogue (\citealp{iac}), only
10 bipolar/quadrupolar nebulae are present among 240 objects.  Corradi
\&\ Schwarz (1995) \nocite{corradi} find that 15\%\ of Galactic PNe are
bipolar. \cite{zijltra07} finds 40\%\ of compact Bulge PNe are bipolar or
related. This is in contrast to our 70\%. 

We conclude that the mixed chemistry phenomenon is strongly related to
morphology, or the presence of a torus. There is no clear relation to the
central star, apart from a possible amplification of the PAH bands by emission-line stars.

\subsection{Modelling the Torus}

To test if the dust spectra can be related to the presence of a torus, we
model the Spitzer spectrum for the representative PN M1-31 (PN G006.4+02.0 ), using the MC3D
(MonteCarlo 3-D) program (\citealp{wolf}). We did not incorporate PAHs in this
model but only used astronomical silicate as the dust component as our aim is
to reproduce the shape of the continuum due to emission from these dust
grains.

The model has 6 main parameters that can be adjusted to arrive at the
best fit.  Other parameters were kept fixed: we used an  effective
temperature of the star of 57,000 K, a luminosity of 6,500L$_\odot$ (\citealp{gesicki07}),
an inclination of the disk of 90 degrees (edge-on), with the distance
to the Galactic Bulge taken as 8.5 kpc \citep{schneider96}. We did test calculations
for a range of inclinations, but found that this parameter has a limited
effect on the result, because of the limited opacity at 10$\mu$m of
the torus.

The Monte Carlo simulation calculates the radiative transfer for 100
wavelength points between $10^{-3}$ and $10^3\mu$m, to calculate the
dust temperature.  It then uses 29 wavelengths between 1 and 53$\mu$m
to calculate the SED.

\begin{figure}
\centering \includegraphics[width=9cm, clip=]{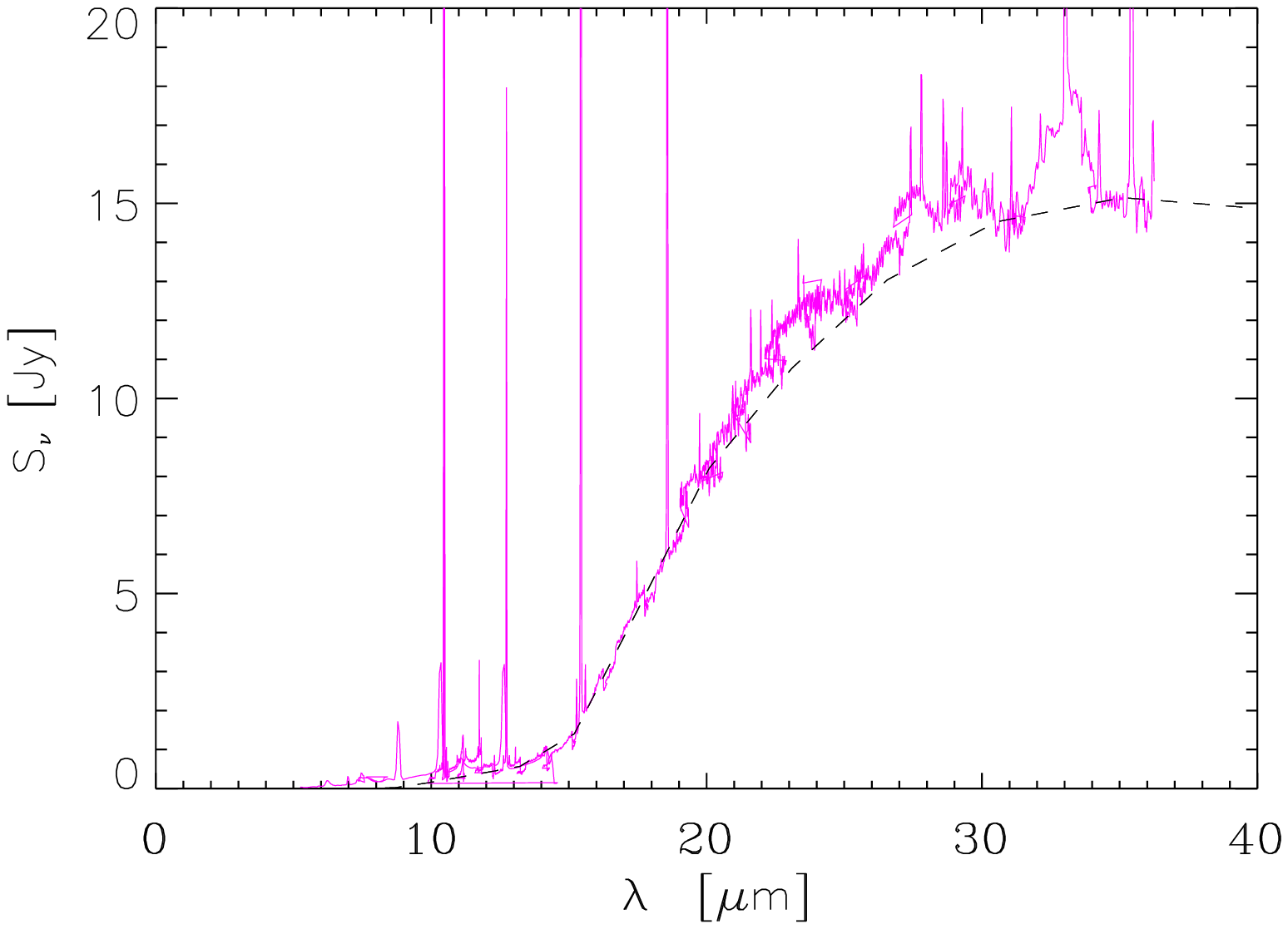}
\caption{The Spitzer spectrum of the PN M1-31.  The dashed
           black line is the model calculated using the MC3D code.}
\label{model}
\end{figure}

We obtained a best fit for the parameters listed in Table \ref{MC3D}.
The main parameters are: the inner radius of the torus ($R_{in}$), the
outer radius ($R_{out}$), the dust mass ($M_{dust}$), the exponent of
the power law describing  the density in the midplane ($r^{-\alpha}$),
flaring ($r^\beta$) and the scale height at 100 AU from $R_{in}$,
($h_0$) (\citealp{wood}). The fitted inner radius of the torus, of
5000\,AU, is in good agreement with that measured from the HST image
(0.55$^{\prime\prime}$ which at 8.5 kpc is 4600 AU).  In Fig \ref{model}, we
plot the Spitzer spectrum and the MC3D model. A good fit is obtained,
with the model reproducing the steep onset of the continuum at
14$\mu$m. The model is a little below the measured flux at shorter
wavelengths, which may be due to a contribution from PAH or very small
grain continuum emission. We also calculated a model with a mixture
of silicate and carbon dust, but it did not reproduce the longer
wavelength continuum as well, predicting too much flux longward of
24$\mu$m.
 
\begin{table}
\caption{Model parameters for the PN M1-31}
\label{MC3D}
\centering
\begin{tabular}{cccccc} 
\hline\hline     
$R_{\rm in}$ [AU] & $R_{\rm out}$[AU] & $M_{\rm dust}$(M$_{\odot}$)& $\alpha$ & $\beta$ & $h_0$ [AU]   \\ 
\hline
5000 & 50000  & 5.5$\times$10$^{-3}$ & 2.5  & 2 & 500 \\
 \hline
  \end{tabular}
\end{table}

The inner radius fits the HST image well but is relatively far from
the star. This indicates that the fitted component is the expanding
torus, and not a Keplerian disc. The model scale height is derived
very close to the inner edge of the torus (100 AU outside of the inner
radius of 5000 AU) where it is 10\%\ of the radius, i.e. a thin
torus. The flaring increases as $r^2$ and at the outer edge of the
model, the height is equal to the radius. The HST image suggests a
thickness of the torus which is intermediate between these values.
The dust mass indicates a gas mass of around 0.1\,M$_\odot$ (for an
assumed gas-to-dust ratio of 200)  The ionized mass has been estimated
as 0.29\,M$_\odot$ (\citealp{gesicki07}).  This suggests that the
modeled torus accounts for a large fraction of the nebular mass.

Our calculation shows that the dust continuum emission can be explained as
arising from the dense torus using oxygen-rich dust (silicate) to explain the
SED: there is no significant carbonaceous component. The fitted torus is large
and massive. This is a significant finding, as the traditional model for mixed
chemistry poses a long-lived, compact oxygen-rich disc. Such a disc is in
Keplerian rotation. \cite{waters} observed in the Red Rectangle a low-mass
disc of 500-2000 AU in size. \cite{oliv} observed a disc size of 9-500 AU in
Mz-3 with a dust mass of 1$\times$10$^{-5}$M$_{\odot}$. We do not find evidence for
such a component.
 
In the traditional stellar evolution model, the silicates were ejected
$>~10^5$yr ago, with the star subsequently becoming carbon-rich prior to the
ejection of the main nebula \citep{Habing96}. In contrast, the torus of M1-31 is part of the
main nebula and ejected $<10^4$ yr ago: at this time the ejecta were
oxygen-rich. The mixed chemistry phenomenon is therefore unlikely to be
related to the star becoming carbon-rich, as the chances of this occurring
over such a short time scale are low, whereas the occurrence of mixed
chemistry appears to be relatively common.

Interestingly, the two brightest PAH emitters in the mixed-chemistry sample,
M1-40 and Cn1-5, are foreground nebulae. These are emission-line stars for
which the traditional model of a long-lived disc may hold, similar to objects
such as IRAS 07027-7934 and the Red Rectangle.  The mixed chemistry nebulae
in the Bulge are a population which is distinct from these disc objects, with a
different origin of the mixed chemistry.

\section{Origin of the PAH in Bulge PNe}

The previous analysis shows that the mixed-chemistry Bulge nebulae are likely
oxygen-rich. This is consistent with expectations based on the stellar
population in the Bulge. The Bulge is believed to contain a 10Gyr old
population, with little or no trace of younger stars
(\citealp{zoccali}), although recently, some indications have been found
for a younger, highly metal-rich population (\citealp{bensby}).

Carbon stars originate from third dredge-up, which requires a minimum
stellar mass of $M_i >\sim 1.5\,M_\odot$ (\citealp{vassi}). It is not
expected to occur in 10Gyr old populations. Furthermore, at high
metallicity the formation of carbon stars is inhibited by the large
oxygen abundance which needs to be overcome by the primary carbon.
Carbon stars are rare in the inner Galaxy (\citealp{bertre}) and
absent in the Galactic Bulge (\citealp{feast}), consistent with this
expectation.

This strengthens the case that the mixed chemistry is occurring in
oxygen-rich gas, expelled by oxygen-rich stars.

We find a strong correlation between mixed chemistry and morphology,
specifically, the presence of a massive torus. This indicates that the
chemical pathway to the PAH molecules occurs within their high density
regions.  These tori remain molecular for some time after the ionization of
the nebula has started, due to the trapping of the ionization front and due to
the shielding effects of the dust column density (\citealp{woods05}). A mini PDR (photon-dominated
region) is expected in these molecular regions (\citealp{phillips}), slowly
being overrun by the advancing ionization front.

A mini-PDR within a dense torus is a plausible location for the PAH
formation.  In interstellar clouds, PAH emission is often associated with PDR
regions (\citealp{joblin,kassis}).  It is unclear whether PAHs form in these
regions or whether they become visible due to UV excitation. In our objects,
because of the close relationship between morphology and PAHs, it seems likely
they form in the regions where they are observed.

In AGB stars, the dichotomy between O-rich and C-rich chemistry can be
broken, as is shown by the observation of hot water in the carbon
stars IRC +10\,216 (\citealp{decin,neufeld,cherch11}). This may be due to collisional destruction of CO in the shocks (\citealp{duari}) or interstellar UV photons (\citealp{agun}).  The
latter authors find that for O-rich circumstellar envelopes, the
penetration of interstellar UV photons into the inner layers leads to
the formation of molecules such as CH$_4$ and HCN with abundances
relative to H$_2$ in the range of 10$^{-8}$ to 10$^{-7}$. This
chemistry is initiated by CO dissociation; the dissociation energy of
CO is 11.1 eV.

\subsection{Chemical reactions}

Here we investigate whether a C-rich chemistry can be similarly driven by UV
photons penetrating the dense, oxygen-rich torus. The penetration is limited
by the dust extinction: in our HST image, internal extinction is clearly
visible in M3-38 (PN G356.9+04.4) which has the most compact torus.

\begin{figure}
\centering \includegraphics[width=8cm]{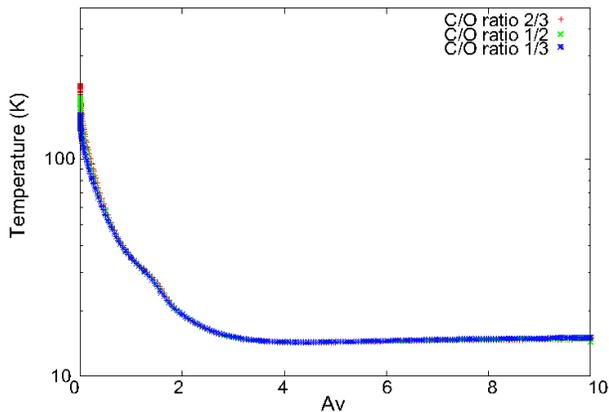}
\caption{Temperature as a function of visual extinction in our three models. The temperature is calculated self-consistently in the constant-density slab.}
\label{temp}
\end{figure}

Our model, which contains gas-phase chemistry only, is based on that of Ni
Chuimin (2009) (see also \citealp{tom07}), developed for conditions in PDR regions. It uses an
extrapolation of the Meudon 2006 PDR chemistry code (\citealp{petit}), and assumes
a semi-infinite and constant-density slab with a density of 2 $\times 10^4$ cm$^{-3}$, UV
radiation enhanced by a factor of 60 over the average interstellar field,
treats thermal balance self-consistently, and contains a mixture of reactions
extrapolated from the Meudon chemical files and also the UDFA 2006
(\citealp{woodall})\footnote{\url{http://www.udfa.net}}, Rate 99 (\citealp{teuff})
and Ohio State OSU-03\footnote{\url{http://www.physics.ohio-state.edu/~eric/research.html}} databases. In Fig. \ref{temp} we show the behaviour of the temperature against A$_v$. As expected, the temperature does not vary much between the three models with each displaying an increase in cooling by atomic O as its abundance increases.
Formation and destruction of large hydrocarbon
chains, that is with more than 4 carbon atoms, of the form C$_n$H$_m$, (n $\leq$ 23, m $\leq$ 2) and their associated ions were added. This resulted in a chemical network of 269 species connected by about 3500 reactions involving ion-neutral and neutral-neutral collisions, photodissociation and photoionization, dissociative and radiative recombination with electrons and radiative association reactions. The goal of the models is to study
the abundances of large hydrocarbons. The specific formation of benzene rings
(\citealp{paul,paul03}) and PAHs is not included, but cyclisation is assumed to be
possible for chains of 23 carbon atoms. Laboratory spectroscopy has shown that
this does not yet happen spontaneously for chains of 13 carbon atoms
(\citealp{giesen}).

The elemental carbon abundance is kept constant at the standard value of
1.32$\times$10$^{-4}$ relative to the total hydrogen abundance (\citealp{ss})
while the elemental oxygen abundance is varied to give a range of C/O values
of 2/3, 1/2 and 1/3.

The formation pathways for the carbon chain species in the chemical model
include neutral-neutral C atom reactions, radiative association reactions
with carbon atoms, and ion-molecule reactions with C$^+$. Destruction of these
chains occur in reactions with oxygen atoms (O + C$_n$), photodissociation
and photoionization, dissociative recombination and electronic recombination
reactions (Table \ref{chem}). The oxygen-destruction reaction rates are adopted and extrapolated
from the Meudon 2006 chemistry (\citealp{petit}).  The oxygen destruction rates 
for C$_2$ and C$_4$ are assumed to be reasonably fast ($1.5 \times 10^{-11}$
and $10^{-10}cm^3s^{-1}$, respectively), while the reaction with C$_3$ is assumed to have
an energy barrier (\citealp{woon}).
 
Example reactions of various types included in the model are listed in Table \ref{chem}. The last column shows the source of the adopted rate coefficients, either from the Meudon model, the UDFA, or OSU-03 databases.

\begin{table}
\caption{Example of model reactions of different types.}
\label{chem}
\centering
\begin{tabular}{lcr} 
\hline\hline              oxygen destruction     & $O + C_n
 \rightarrow C_{n-1} + CO$     & Meudon \\ neutral-neutral        &
 $C + C_nH \rightarrow C_{n+1} + H$      & UDFA \\ neutral-neutral
 & $C + C_nH_2 \rightarrow C_{n+1}H + H$   & UDFA \\ radiative
 association  & $C + C_n \rightarrow C_{n+1} + h\nu$    & Meudon \\
 radiative association  & $C + C_nH \rightarrow C_{n+1}H + h\nu$  &
 Meudon\\ ion-molecule           & $C^+ + C_nH \rightarrow
 C_{n+1}^+ + H$  &
 OSU-03\\
 \hline
  \end{tabular}
\end{table}

\subsection{Results}

If PAH emission does originate from the torus and not swept-up interstellar gas as is indicated by our observations, then it is valid to consider whether the PAH themselves can be formed in this oxygen-rich, UV-irradiated environment. Since there is no detailed chemical kinetic model for PAH formation, we may use our hydrocarbon chemistry as a proxy for PAH formation, noting that even the presence of abundant hydrocarbons in a PDR would not necessarily prove that PAHs are formed in the same environment.

\begin{figure}
\centering \includegraphics[width=8cm]{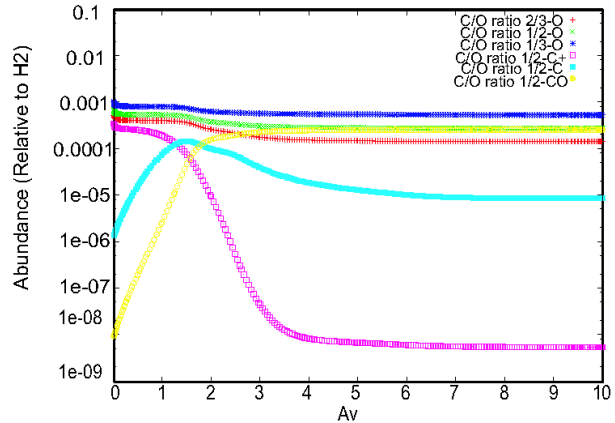}
\caption{Calculated abundances for O, C, C+ and CO for the three models with varying C/O ratios. Only the O atom abundance varies, other abundances are almost identical in the models and are shown here for the case of C/O = 1/2.}
\label{molecules}
\end{figure}

The PDR model calculations show that the chemistry first tends towards
complexity around the C$^+$/C/CO transition at $A_V \sim 1.4$ mag. This transition is seen in Fig. \ref{molecules} which plots the abundances of O, C, C+ and CO as a function of A$_v$. As expected, only O shows a difference between the three models, for the other species we simply plot the results for the model with C/O = 1/2.

\begin{figure}
\centering \includegraphics[width=8cm]{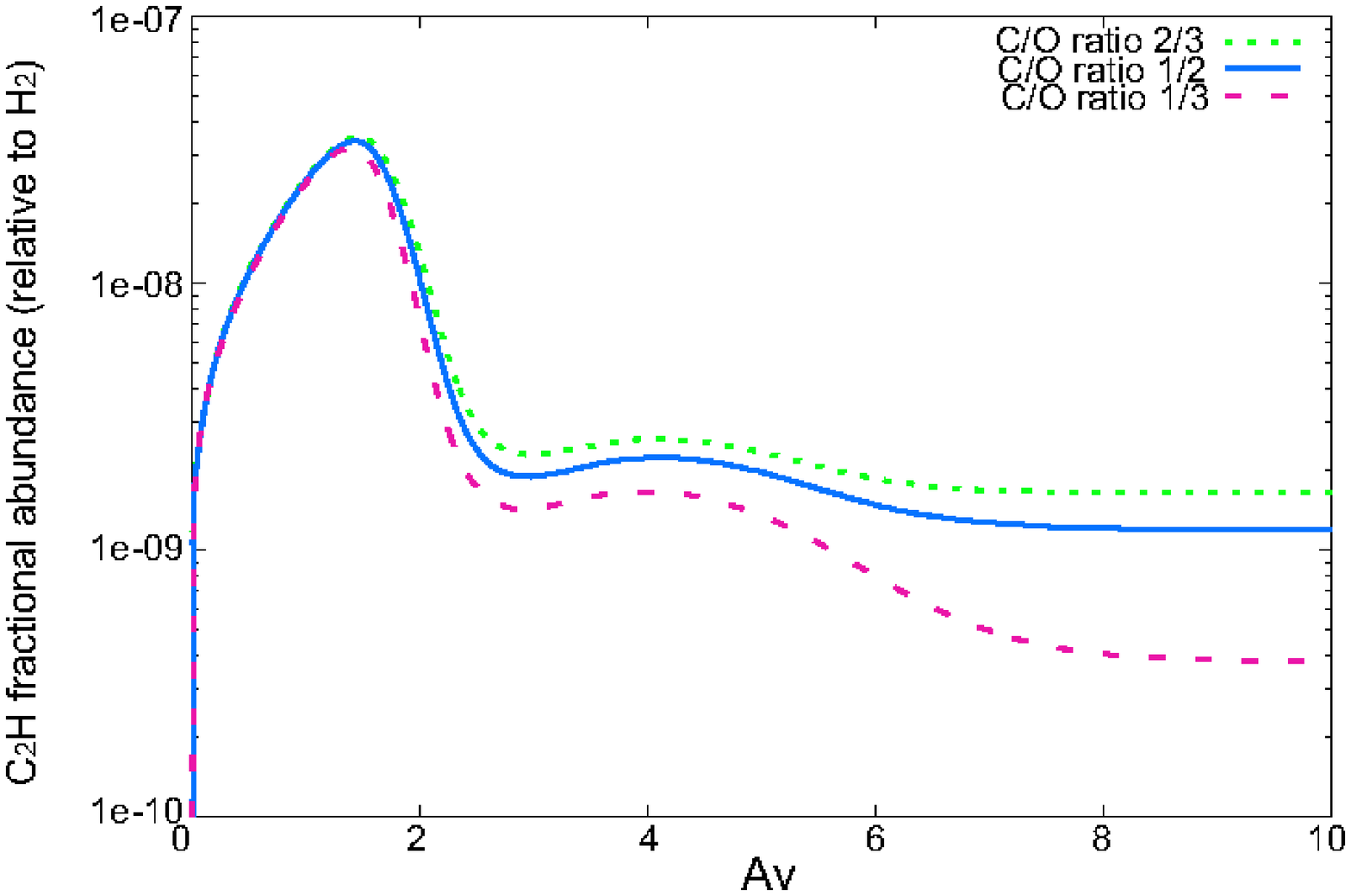}
\caption{Calculated abundances for varying C/O ratios for C$_2$H.}
\label{model-1}
\end{figure}

\begin{figure}
\centering \includegraphics[width=8cm]{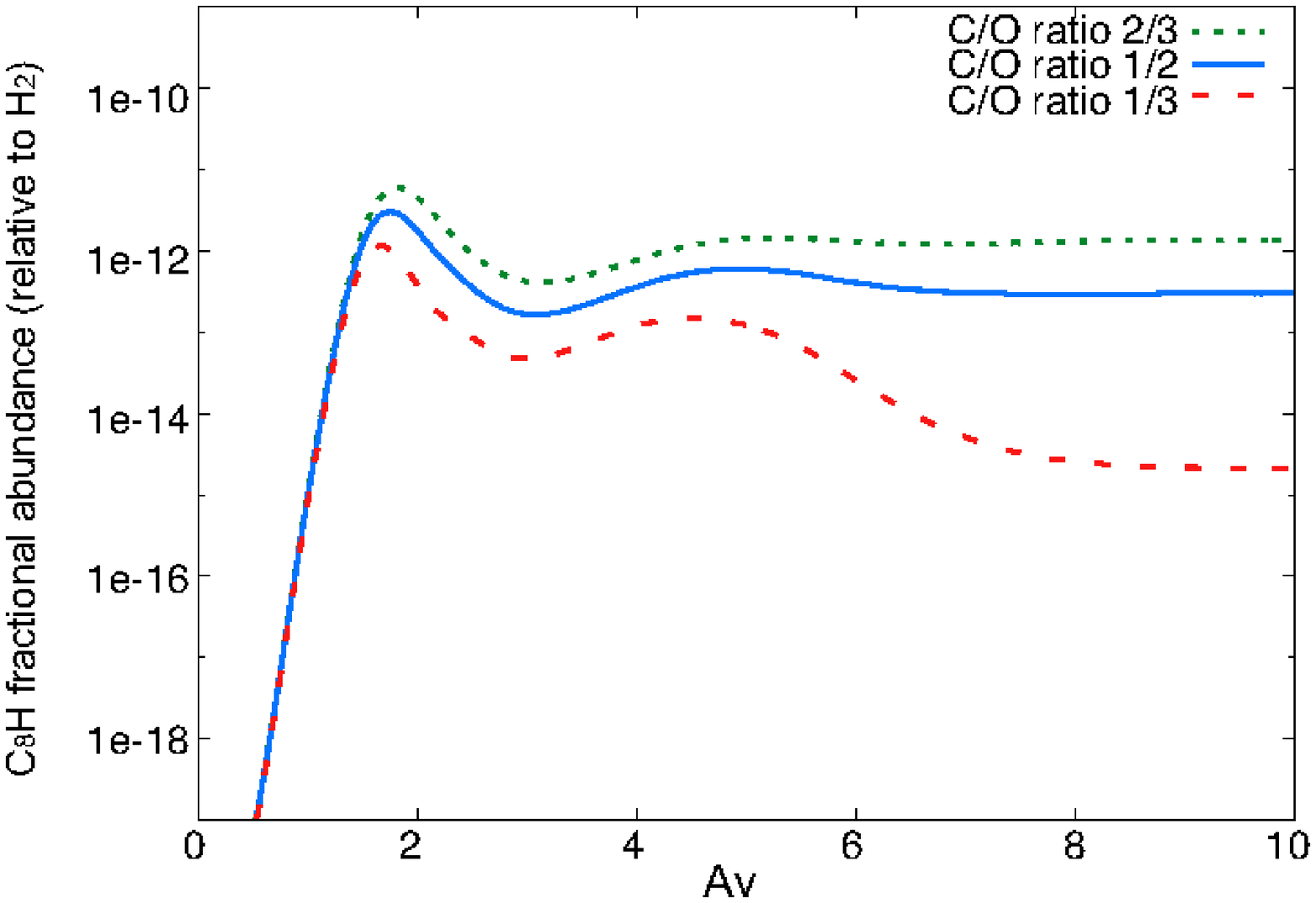}
\caption{Calculated abundances for varying C/O ratios for C$_8$H.}
\label{model-2}
\end{figure}

\begin{figure}
\centering \includegraphics[width=8cm]{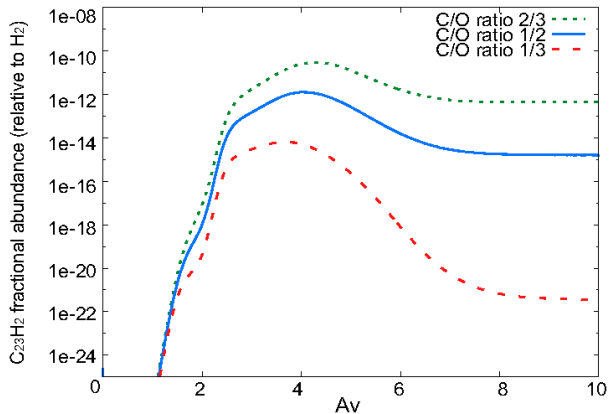}
\caption{Calculated abundances for varying C/O ratios for C$_{23}$H$_2$.}
\label{model-3}
\end{figure}

Figs. \ref{model-1}, \ref{model-2} and \ref{model-3} show the calculated
fractional abundances (relative to H$_2$) for the molecules C$_2$H, C$_8$H and
C$_{23}$H$_2$, respectively, as function of $A_v$, which measures depth into
the slab, and C/O ratio.

Two hydrocarbon peaks are seen, one at $A_V \sim 1.5$, and one at $A_V \sim
4$--6. The first peak occurs in the region in which C$^+$ and H$_2$ are
abundant with the hydrocarbon chemistry initiated by the C$^+$--H$_2$
radiative association reaction. Subsequent reactions with C and C$^+$ form
C$_2$H.  C$_2$H is the main seed molecule for the formation of larger
hydrocarbons, as it is in the carbon-rich envelope of IRC+10216
\citep{millar}, and reactions continue to build up to C$_8$H, although
large(r) hydrocarbons are not highly abundant at low extinction.

The second peak, at higher $A_V \sim 4-6$, occurs where atomic carbon becomes
abundant and drives chemistry through proton transfer with H$_3^+$ to form
CH$^+$ which subsequently reacts with H$_2$ to form simple hydrides. Here the chemistry produces a surprisingly large abundance of the biggest hydrocarbons in the model.

Oxygen atom destruction reactions are the dominant loss of larger hydrocarbons
at high $A_V$ where there is little C$^+$. The reaction rates of C$_2$H and C
with O are rather uncertain: they differ by an order of magnitude between the
UDFA and the Meudon chemical models. However, because of the importance of O,
the abundances depend on the C/O ratios: models with less O produce higher
abundances of large hydrocarbons. The difference amplifies for the larger
molecules: whereas for C$_2$H and C$_8$H the difference in abundance between
C/O=1/3 and 2/3 is within an order of magnitude, for C$_{23}$H$_2$ it is four orders
of magnitude. Even a small increase in the C/O ratio makes a large difference
here.

\subsection{Discussion}

Under the assumptions of the chemical model, it is clear that larger  hydrocarbons can form within oxygen-rich environments. It is assumed here that a mechanism exists to convert these long linear hydrocarbon chains into PAH molecules. Such a discovery is surprising, but clues have been previously found in the inner regions of oxygen-rich stars, where small carbon-bearing  molecules have been observed, and explained by means of shock chemistry (Cherchneff 2006)\nocite{cherch} or penetration of UV photons (Ag\'undez et al. 2010). Furthermore, in the oxygen-rich environment of a protoplanetary disc, Woods \& Willacy (2007)\nocite{woods07} showed that benzene and large polyynes of up to C$_8$H$_2$\ can form through ion-molecule chemistry in suitably dense conditions, where cosmic- and X-rays provide a source of ionisation. In this latter model, and in the ISM (McEwan et al. 1999)\nocite{mcewan}, and in proto-PNe with carbon-rich tori (Woods et al. 2002, 2003) closure of the aromatic ring occurs through the reaction of small (C$_{2\ldots4}^{\phantom{(+)}}$H$_{3\ldots5}^{(+)}$) hydrocarbons, which are not included in our model due to the simple nature of the Meudon chemistry (Table \ref{chem}). However, these small hydrocarbons are typically formed by the destruction of larger polyynes, and through reaction with atomic carbon, which are abundant in the chemical model. Once this first aromatic ring closure has occurred, subsequent aromatic growth is rapid in comparison (Cherchneff et al. 1992, Wang \& Frenklach 1997)\nocite{cherch92,wang}. 

If we assume that the underlying abundance distribution of hydrocarbon chains is related to that of PAHs, then we can make some inferences on the properties of PAH emission in the Galactic Bulge PNe. First, a reasonable extinction is
needed. At low $A_V$ ($<$2), where carbon is ionized, the abundances of the longest
chains are strongly suppressed by photodissociation and photoionization. This
still allows for the formation of shorter linear hydrocarbons, but PAH
formation is not expected.  Efficient PAH formation requires sufficient
extinction to render carbon neutral and to minimize the effects of
photodestruction. The optimum extinction in our model is around $A_V \sim
4$. At even higher extinction, carbon is locked up in CO and fewer of the longest
hydrocarbon chains form, although intermediate chains (C$_8$H) still form.

In order to generate sufficient reactive C and C$^+$ UV photons are required
to dissociate CO contained in the dense molecular torus. Thus, the central
star needs to be hot enough to generate these photons. Our models use a
radiation field of 60 times the intestellar radiation field which corresponds to
a star of spectral type B, or temperature around 15\,000K.  This suggests that
the PAHs will not form until the star is hot enough to begin to ionize the
nebula. By this time, the nebula will have expanded and the extinction has
dropped considerably.

The need for the torus may therefore be to provide an irradiated region dense
enough to allow a photon-dominated chemistry, in a nebula old enough to allow
for an ionizing star. Such a requirement would explain the close relation
between PAH emission in oxygen-rich PNe, and the presence of a dense
torus. This same relation also holds in non-Bulge PNe, e.g. NGC 6302 and
Roberts 22 (\citealp{zijlstra01}).

The predicted abundances show a strong dependence on C/O ratio
(Figs. \ref{model-1}, \ref{model-2}, \ref{model-3}).  This may explain the
finding of \cite{cohenBarlow05} that PAH emission at lower C/O ratio is faint.
An increase in the C/O ratio, even if it remains below unity, would benefit
PAH formation. But such an increase is not expected in Bulge PNe which do not
experience carbon dredge-up.
 
Finally, even though this model goes some way to explaining the
observed mixed chemistry, it remains unexplained why PNe with [WC]
central stars would have stronger PAH emission. Part of this correlation may
be due to the fact that the brightest objects in Table \ref{GBPNe} contain 
some foreground nebulae. However, even excluding these, it still appears that
the emission-line stars dominate the top end of the Table.  

Two possible explanations can be envisaged. First, the stellar winds may inject
some carbon into the surrounding nebulae. This would locally enhance the C/O
ratio, and increase the PAH formation rate.

The second explanation could be related to the energy deposition by the stellar
wind into the nebula.  \cite{gesicki06} show that nebulae around [WC] stars
are turbulent indicating that some energy deposition into the nebulae
occurs, for instance from the stellar wind. The turbulence may aid in
dissociating CO, or more likely, causes compressed regions with a larger range
of extinctions. The second explanation appears more plausible. However, the
relation between stellar winds and PAH emission needs to be studied further. 

We finally note that the $A_V$ in Figs.  \ref{model-1}, \ref{model-2},
\ref{model-3}, \ref{molecules} is time-dependent. As the torus expands, the extinction drops.
The local conditions thus evolve from right to left in the Figures. This
predicts two layers within the torus, one with the smaller hydrocarbons at $A_V
\sim 1.5$ and one with PAHs at $A_V \sim 4$, which are both moving inward into
the torus as the nebula evolves. This makes the layer at lower extinction,
with shorter hydrocarbons, a product of the destruction of the PAHs by the
increasing UV radiation field. If the PAHs are more UV-stable than the largest
chains in the model, then PAHs could be present at lower extinction than
predicted here.

\section{Conclusions}

We summarize the main conclusions of this research in the following
list:
\begin{itemize}
\item We analyzed observations of 40 PNe towards the Galactic Bulge using the IRS
 instrument on board Spitzer. We found that 30 of
 them present a mixed chemistry phenomenon. From these 40 we used
 HST observations of 22  to analyze their morphology.

\item A strong correlation was found between strength of the PAH bands and
  morphology, in particular, the presence of a massive torus. Data from
  VLT/UVES was also used to study their central stars. We find that there is
  no correlation between the classification of the central star and the
  mixed chemistry phenomenon, although PNe with  emission
  line stars (wels or [WC]) tend to have the stronger PAHs
  emission. This could be related to carbon injection or energy deposition by
  the wind into the nebula, enhancing PAH formation.

\item We modelled the Spitzer spectrum for the representative PN M1-31. The SED
  can be explained using oxygen-rich (silicate) dust. The fitted torus is
  large and massive, with a size of 5000-50000 AU and a mass of 5$\times
  10^{-3}$M$_{\odot}$. Because of the age of the torus of M1-31
  ($<10^4$yr) the mixed chemistry phenomenon is unlikely to be related to a
  change of composition of the stellar ejecta over time. Such a change would
  be unexpected in the Bulge population (which does not show third dredge-up),
  and would be very rare over such a short period in any case,  whereas
  the occurrence of mixed chemistry appears to be relatively common.

\item For the chemical analysis of the PAH formation in an oxygen-rich environment, we
  used the gas-phase models from N\'i Chim\'in (2009)\nocite{rois}. These are an
  extrapolation of the Meudon 2006 PDR chemistry code to include the chemistry
  of large hydrocarbons. They cover a range of $A_v$ and C/O ratios of 1/3,
  1/2 and 2/3. Although the model does not include PAH formation, it shows that large hydrocarbons can form in oxygen-rich environments and that their abundances can be relatively large in moderately shielded regions. The model predicts
  two layers, one at $A_V\sim 1.5$ where small hydrocarbons form from
  reactions with C$^+$, and one at $A_V\sim 4$, where larger chains (and by
  implication, PAHs) form from reactions with neutral, atomic carbon.

\item The torus observed in all of the PNe that present the mixed chemistry
  phenomenon could provide the dense, irradiated environment needed to form the
  PAHs detected. 

\end{itemize}

\section*{Acknowledgments}
This work is based on observations made with the Spitzer Space Telescope,
which is operated by the Jet Propulsion Laboratory, California Institute of
Technology under a contract with NASA. TheVLT/UVES observations are from ESO
program 075.D-0104, and the HST images are from the proposal number 9356, PI
Zijlstra. LGR and AZ acknowledge the support of CONACyT (Mexico). Astrophysics
at QUB and at Manchester is supported by  grants from the STFC.

\bibliography{biblio}
\bibliographystyle{mn2e}

\label{lastpage}
\end{document}